\pdfoutput=1
\documentclass[aps,prx,reprint,superscriptaddress,amsmath,amssymb,longbibliography,nofootinbib,floatfix]{revtex4-2}
\usepackage[T1]{fontenc}
\usepackage{amsthm,mathrsfs,booktabs,tikz}
\usetikzlibrary{arrows.meta,positioning,calc,shapes.geometric}
\usepackage[ruled,vlined,linesnumbered]{algorithm2e}
\usepackage{xurl}
\usepackage[colorlinks=true,linkcolor=blue,citecolor=blue,urlcolor=blue]{hyperref}
\SetKwInOut{Input}{Input}\SetKwInOut{Output}{Output}
\newtheorem{theorem}{Theorem}
\newtheorem{proposition}{Proposition}
\newtheorem{lemma}{Lemma}
\newtheorem{corollary}{Corollary}
\newcommand{\restatedname}{}
\newtheorem*{restatedthm}{\restatedname}
\newenvironment{restated}[1]{\renewcommand{\restatedname}{#1}%
 \begin{restatedthm}}{\end{restatedthm}}
\newcommand{\F}{\mathbb F_p}
\newcommand{\Z}{\mathbb Z}
\newcommand{\im}{\operatorname{im}}
\newcommand{\coker}{\operatorname{coker}}
\newcommand{\Ext}{\operatorname{Ext}}
\newcommand{\Hom}{\operatorname{Hom}}
\newcommand{\rank}{\operatorname{rank}}

\newcommand{\diag}{\operatorname{diag}}
\newcommand{\TC}{\operatorname{TC}}
\newcommand{\Ccode}{C_{\mathrm{code}}}
\newcommand{\Cstack}{C_{\mathrm{stack}}}
\newcommand{\dcode}{d_{\mathrm{code}}}
\newcommand{\dstack}{d_{\mathrm{stack}}}
\newcommand{\Ltriv}{\Lambda_{\mathrm{triv}}}
\newcommand{\Lsuff}{\Lambda_{\mathrm{suff}}}
\newcommand{\GC}{\mathrm{GC}}
\newcommand{\ket}[1]{\lvert#1\rangle}

\renewcommand{\tilde}{\widetilde}

\renewcommand{\leq}{\leqslant}
\renewcommand{\geq}{\geqslant}
\definecolor{tcone}{RGB}{35,87,155}
\definecolor{tctwo}{RGB}{216,90,48}
\tikzset{
 chk/.style={star,star points=5,star point ratio=2.3,inner sep=0.95pt,
  draw=white,line width=0.3pt},
 zchk/.style={diamond,inner sep=1.25pt,draw=white,line width=0.3pt},
 xone/.style={chk,fill=tcone},xtwo/.style={chk,fill=tctwo},
 zone/.style={zchk,fill=tcone},ztwo/.style={zchk,fill=tctwo},
 xoneopen/.style={chk,fill=white,draw=tcone,line width=0.5pt},
 qone/.style={tcone,line width=1.6pt},
  regroup/.style={black!70,densely dashed,rounded corners=2pt,line width=0.5pt},
 cxdeg/.style={font=\scriptsize,text=gray!70!black},
 cxlab/.style={font=\scriptsize}}
\begin{document}
\title{Toric codes all the way down: decoupling translation-invariant topological CSS codes in higher dimensions}
\author{Mingxin He}
\affiliation{Chen-Ning Yang Institute for Advanced Study, Tsinghua University, Beijing, 100084, China}
\affiliation{Beijing Key Laboratory of Cold Atom Quantum Computation, Tsinghua University, Beijing, 100084, China}

\author{Yifei Wang}
\affiliation{Chen-Ning Yang Institute for Advanced Study, Tsinghua University, Beijing, 100084, China}
\affiliation{Beijing Key Laboratory of Cold Atom Quantum Computation, Tsinghua University, Beijing, 100084, China}

\author{Hao Song}
\email[E-mail: ]{songhao@itp.ac.cn}
\affiliation{Institute of Theoretical Physics, Chinese Academy of Sciences, Beijing 100190, China}

\author{Yingfei Gu}
\email[E-mail: ]{guyingfei@tsinghua.edu.cn}
\affiliation{Chen-Ning Yang Institute for Advanced Study, Tsinghua University, Beijing, 100084, China}
\affiliation{Beijing Key Laboratory of Cold Atom Quantum Computation, Tsinghua University, Beijing, 100084, China}

\date{\today}
\begin{abstract}
Topological Calderbank--Shor--Steane (CSS) codes in more than two spatial dimensions offer features useful for fault-tolerant quantum computation,
such as transversal non-Clifford gates and single-shot decoding, but a systematic understanding of them is still lacking. 
One approach is to \emph{decouple} a code: map it to a stack of higher-form toric codes by a finite-depth local CSS-preserving Clifford circuit, possibly after adding ancillas. 
For translation-invariant $p$-ary CSS codes, where $p$ is prime ($p=2$ for qubits), we require the circuit to commute with the translations of a superlattice. 
In two dimensions, such a decoupling exists once these translations act trivially on every anyon type. 
In higher dimensions, excitations are sorted by their dimension into superselection layers,
which can remain glued together even when translations act trivially on each layer. 
We show that this gluing is the only remaining obstruction:
a code can be decoupled on a superlattice exactly when, 
after coarse-graining to it, 
translations act trivially on every layer and the gluing vanishes. 
Moreover, for a code with finitely many excitation types in each layer and \(m\) nonzero layers,
once translations act trivially on every layer,
further coarse-graining by a factor of \(p^{m-1}\) in every direction removes the aforementioned gluing.
Physically, the same criterion determines when the code encodes the maximal number of logical qudits on a finite torus.
Finally, we give an explicit algorithm that constructs the decoupling as a circuit of SUM (qudit CNOT) and SWAP gates.
\end{abstract}
\maketitle

\section{Introduction}\label{sec:introduction}
Topological quantum codes are defined by local checks but store quantum information nonlocally, which protects it against local noise \cite{Kitaev2003,Dennis2002}. 
Codes in three or more dimensions offer capabilities beyond those of two-dimensional codes: 
some admit transversal non-Clifford gates \cite{BombinMartinDelgado2007c,KubicaBeverland2015,Kubica2015,VasmerBrowne2019},
and some support single-shot decoding \cite{Bombin2015,Jacob2025,Menon2026MagicTricycles,Mian2026}.
Such codes come in many forms, however, and relations among them are known only in special cases. 
For example, the three-dimensional color code is locally equivalent to three copies of the three-dimensional toric code \cite{Kubica2015}, 
so it can be decoded with toric-code decoders \cite{KubicaDelfosse2023}.

A systematic way to organize topological codes is to ask when two of them are related by local transformations, 
that is, 
when they realize the same topological phase \cite{ChenGuWen2010}. 
Toric codes are the simplest and best-understood representatives, so a basic question is when a given code is equivalent to a stack of toric codes. 
We address this question for translation-invariant CSS codes \cite{CalderbankShor1996,Steane1996}, a family that includes bivariate bicycle codes \cite{Bravyi2024}. 
We allow finite-depth local CSS-preserving Clifford circuits that commute with translations,
together with the addition of ancillas, 
and call an equivalence of this kind to a toric-code stack a \emph{decoupling}.

In two dimensions, a finite-depth local circuit that commutes with translations preserves how translations act on the anyons of a translation-invariant topological Pauli stabilizer code \cite{Song2026}. 
Translations act trivially on the anyons of a toric-code stack, 
but they can permute the anyons of a general topological code, 
a phenomenon known as weak breaking of translation symmetry \cite{Wen2003,Kitaev2006,RaoSodemann2021}. 
One must therefore coarse-grain, 
retaining only a finite-index subgroup of translations, or \emph{superlattice}, 
whose elements preserve every anyon type. 
Such a code can always be mapped to a toric-code stack, possibly after adding ancillas, 
by a finite-depth local circuit commuting with the translations of some superlattice \cite{BombinDuclosPoulin2012,Bombin2014,Haah2016,Haah2021}, 
although the known constructions may coarse-grain more than necessary. 
For topological CSS codes, Wang \emph{et~al.} showed that preserving every anyon type is also sufficient: 
a decoupling already exists on the superlattice of all anyon-preserving translations \cite{Wang2026}.

In \(D\geq3\) spatial dimensions, excitations of CSS codes can be pointlike, looplike, or higher-dimensional;
in each of the \(X\) and \(Z\) sectors, they are sorted by dimension into \emph{superselection layers} \cite{RubaYang2024}. 
The two-dimensional argument applies to every layer, 
so the retained translations must act trivially on each of them (this is possible only if every layer has finitely many excitation types, which excludes fracton models \cite{NandkishoreHermele2019,PretkoChenYou2020} such as Haah's cubic code \cite{Haah2011} and the X-cube model \cite{Vijay2016}). 
For codes with a single nonzero layer, Song~\cite{Song2026} proved that a decoupling exists after coarse-graining precisely when the retained translations act trivially on that layer. 
Li and Williamson~\cite{LiWilliamson2026} established toric-stack equivalence for qubit CSS codes obtained from three-term segments of length-$D$ free resolutions whose $ZX$ duals are also resolutions. 
Color codes \cite{BombinMartinDelgado2006,BombinMartinDelgado2007}, explicitly decoupled earlier by Kubica~\emph{et~al.}~\cite{Kubica2015}, are examples of such single-layer codes.

When several layers coexist, however, the full excitation structure, called the \emph{superselection profile} \cite{Song2026}, also records how the layers are glued together. 
This gluing is an additional obstruction to decoupling: Song constructed three-dimensional examples in which translations act trivially on every layer, yet the gluing prevents a decoupling that retains these translations. 
In this work, we show that for translation-invariant topological \(p\)-ary CSS codes in any spatial dimension \(D\), with \(p\) any prime, further coarse-graining always removes this obstruction, provided that every layer has finitely many excitation types. 
The code can then be decoupled by an explicit circuit of SUM (qudit CNOT) and SWAP gates.

Our main result is a decoupling criterion: a code can be decoupled on a superlattice exactly when, after coarse-graining to that superlattice, its profile is that of a toric-code stack.
This condition has two parts: the translations in the superlattice must act trivially on every layer, and the gluing must vanish.
Once the first holds, each further coarse-graining by a factor of \(p\) in every direction detaches one more layer from those below it.
Hence, for a code with \(m\) nonzero layers, a further factor of \(p^{m-1}\) suffices.
Coarse-graining in every direction can be more than necessary;
when at most two layers are nonzero,
as in every three-dimensional code, we also determine the minimal coarse-graining.
We turn these results into two algorithms: the first finds a sufficient superlattice,
and the second constructs the decoupling circuit on it,
in polynomial time for fixed spatial dimension once the profile has been computed (Proposition~\ref{prop:complexity}).
Finally, the criterion also applies to finite systems: a torus encodes the maximal number of logical qudits exactly when the code can be decoupled on its period lattice.

The rest of the paper is organized as follows.
Section~\ref{sec:framework} reviews the algebraic framework.
Section~\ref{sec:gluing} defines the inter-layer gluing and shows, for a glued code from Ref.~\cite{Song2026}, that it obstructs decoupling.
Section~\ref{sec:main} removes the gluing by coarse-graining, states the decoupling criterion (Theorem~\ref{thm:main}), turns it into algorithms, and applies it to tori.
Section~\ref{sec:example} removes the gluing of the glued code.
Section~\ref{sec:discussion} discusses applications to decoding and logical gates, and open questions. 
Some proofs, including the explicit construction of the decoupling circuit, are deferred to the appendixes.

\section{Preliminaries}\label{sec:framework}
We first describe decouplings and excitations in a common algebraic language. 
In this language, a decoupling becomes a commuting square of invertible maps, and the excitations form the superselection profile, which a decoupling cannot change.
We then introduce coarse-graining, which a decoupling generally requires because translations can act nontrivially on the excitations of a code but not on those of a toric-code stack.

\subsection{Polynomial representation and the decoupling problem}
\label{subsec:polynomial}
Following Haah \cite{Haah2013}, we use translation invariance to turn the decoupling problem into one about modules over a ring of Laurent polynomials. 
Consider the lattice \(\Z^D\), \(D\geq2\), with \(q\) \(p\)-ary qudits in each unit cell. 
The generalized Pauli operators act as \(X\ket a=\ket{a+1}\) and \(Z\ket a=\omega^a\ket a\), 
with \(a\in\F\) and \(\omega=e^{2\pi i/p}\) \cite{Hostens2005}. 
Let \(R=\F[x_1^{\pm1},\ldots,x_D^{\pm1}]\) be the Laurent polynomial ring,
write \(x^\alpha:=x_1^{\alpha_1}\cdots x_D^{\alpha_D}\) for \(\alpha\in\Z^D\), 
and let \(e_1,\ldots,e_q\) denote the standard basis of \(R^q\). 
A finitely supported \(X\)-type Pauli operator, modulo phases, is represented by an element \(u\in R^q\): 
the term \(\lambda x^\alpha e_i\), with \(\lambda\in\F\), represents \(X^\lambda\) on qudit \(i\) of the unit cell at \(\alpha\). 
We write \(X(u)\) for the operator represented by \(u\), and define \(Z(v)\) for \(v\in R^q\) in the same way. 
Multiplication of operators corresponds to addition of representatives, 
and translating an operator by \(\alpha\) multiplies its representative by \(x^\alpha\).

To express commutation, we use the antipode \(f\mapsto\overline f\),
the ring automorphism of \(R\) that sends each \(x_i\) to \(\overline x_i:=x_i^{-1}\). 
For a matrix over \(R\), \(\dagger\) denotes the transpose with the antipode applied entrywise.  
Commutation between \(X\)- and \(Z\)-type operators is then encoded by the sesquilinear symplectic form \((u,v)\mapsto u^\dagger v\in R\) \cite{Haah2013,Wang2026}. 
If \(\lambda\in\F\) is the coefficient of \(x^\alpha\) in \(u^\dagger v\), then \(X(x^\alpha u)\,Z(v)=\omega^{-\lambda}Z(v)\,X(x^\alpha u)\).  
In particular, \(X(u)\) commutes with every translate of \(Z(v)\) if and only if \(u^\dagger v=0\).

A translation-invariant CSS code is specified by its check matrices \(H_X\in R^{s\times q}\) and \(H_Z\in R^{r\times q}\). 
The \(a\)th column \(H_X^\dagger e_a\) of \(H_X^\dagger\), with \(e_a\) the \(a\)th standard basis vector of \(R^s\), represents the \(a\)th \(X\)-type check. 
Its translate by \(\alpha\) is represented by \(x^\alpha H_X^\dagger e_a\), and likewise for \(H_Z^\dagger\). 
The checks, taken with trivial phase, and their translates generate the stabilizer group. 
By the symplectic form above, the \(a\)th entry of \(H_Xv\) encodes the commutation exponents of \(Z(v)\) with the translates of the \(a\)th \(X\)-type check. 
In other words, \(H_X\) maps a \(Z\)-type operator to the \(X\)-check syndrome it creates, and \(\im H_X\) consists of the syndromes created by finitely supported \(Z\)-type operators.

Following Ref.~\cite{Wang2026}, we represent the code by the complex
\begin{equation}
\begin{tikzpicture}[baseline=(b.base),>=To,
 x=1.35cm,every node/.style={inner sep=2pt}]
\foreach \n in {-1,0,1} \node[cxdeg] at (\n,0.6) {\(\n\)};
\node[cxdeg,anchor=east] at (-1.3,0.6) {degree};
\node (a) at (-1,0) {\(R^r\)};
\node (b) at (0,0) {\(R^q\)};
\node (c) at (1,0) {\(R^s\)};
\draw[->] (a)--node[above,cxlab]{\(H_Z^\dagger\)}(b);
\draw[->] (b)--node[above,cxlab]{\(H_X\)}(c);
\end{tikzpicture}
\label{eq:physical-window}
\end{equation}
and its dual \(R^s\xrightarrow{H_X^\dagger}R^q\xrightarrow{H_Z}R^r\). 
This degree convention places the qudits in degree zero; 
we use it throughout the paper because it matches the superselection profile of Sec.~\ref{subsec:profile}.
We call the code \emph{topological} if \cite{Haah2013,Song2026}
\begin{equation}
 \ker H_X=\im H_Z^\dagger,\qquad
 \ker H_Z=\im H_X^\dagger.
 \label{eq:exactness}
\end{equation}
These equalities imply that the stabilizers commute, \(H_XH_Z^\dagger=0\), and they exclude finitely supported logical operators on the infinite lattice.

In the polynomial representation, a translation-invariant CSS-preserving Clifford circuit maps each \(Z\)-type operator \(Z(v)\) to \(Z(Uv)\) for an invertible matrix \(U\) over \(R\). 
Since it preserves the symplectic form, it maps each \(X\)-type operator \(X(u)\) to \(X\bigl((U^\dagger)^{-1}u\bigr)\) \cite{Wang2026}.
We call the complex~\eqref{eq:physical-window} of the code \(\Ccode\) and compare it with the complex \(\Cstack\) of a toric-code stack (Sec.~\ref{subsec:toric}). 
We write \(\dcode\) and \(\dstack\) for their differentials and \(\dcode^n\), \(\dstack^n\) for the components from degree \(n\) to degree \(n+1\), so that \(\dcode^0=H_X\).
After ancillas and identity \(X\)-type checks (zero rows of \(H_X\)) are adjoined (Appendix~\ref{apx:construction}), 
we may assume that \(\Ccode^0\) and \(\Cstack^0\) have equal ranks, and likewise \(\Ccode^1\) and \(\Cstack^1\).

We claim that the circuit decouples the code if, for some invertible \(V\), the square
\begin{equation}
\begin{tikzpicture}[baseline=(current bounding box.center),>=To,
 every node/.style={inner sep=2pt},x=1.9cm,y=1.2cm]
\node (c0) at (0,0) {\(\Ccode^0\)};
\node (c1) at (1,0) {\(\Ccode^1\)};
\node (t0) at (0,-1) {\(\Cstack^0\)};
\node (t1) at (1,-1) {\(\Cstack^1\)};
\draw[->] (c0)--node[above]{\(\dcode^0\)}(c1);
\draw[->] (t0)--node[below]{\(\dstack^0\)}(t1);
\draw[->] (c0)--node[left]{\(U\)}(t0);
\draw[->] (c1)--node[right]{\(V\)}(t1);
\end{tikzpicture}
\label{eq:decoupling-square}
\end{equation}
commutes, \(\dstack^0U=V\dcode^0\). 
Indeed, by the topological condition~\eqref{eq:exactness}, the \(Z\)-type stabilizers of the code form the submodule \(\im H_Z^\dagger=\ker\dcode^0\), and the \(X\)-type stabilizers form \(\im(\dcode^0)^\dagger\); 
for the stack, these submodules are \(\ker\dstack^0\) and \(\im(\dstack^0)^\dagger\). 
Since \(U\) and \(V\) are invertible, the commuting square gives \(U\ker\dcode^0=\ker\dstack^0\), so the circuit maps the \(Z\)-type stabilizers of the code onto those of the stack. 
Taking the dagger of \(\dstack^0U=V\dcode^0\) gives \((U^\dagger)^{-1}(\dcode^0)^\dagger=(\dstack^0)^\dagger(V^\dagger)^{-1}\), 
so \((U^\dagger)^{-1}\im(\dcode^0)^\dagger=\im(\dstack^0)^\dagger\). 
The same thus holds for the \(X\)-type stabilizers. 
Theorem~\ref{thm:main} in Sec.~\ref{sec:main} states when such \(U\) and \(V\) exist, 
with \(U\) realized by a circuit of SUM and SWAP gates.

\begin{figure*}[t]
\centering
\begin{tikzpicture}[x={(0.62cm,0cm)},y={(0cm,0.62cm)},z={(0.31cm,-0.217cm)},
 line cap=round,line join=round,
 lab/.style={font=\scriptsize,align=center,text width=4.4cm},
 bg/.style={gray!28,line width=0.35pt}]
\newcommand{\lattice}{%
 \foreach \a in {-1,0,1}{\foreach \b in {-1,0,1}{%
  \draw[bg] (-1,\a,\b)--(1,\a,\b);
  \draw[bg] (\a,-1,\b)--(\a,1,\b);
  \draw[bg] (\a,\b,-1)--(\a,\b,1);}}}
\newcommand{\cubefaces}[1]{%
 \fill[#1!18] (0,0,1)--(1,0,1)--(1,1,1)--(0,1,1)--cycle;
 \fill[#1!18] (0,0,0)--(0,1,0)--(0,1,1)--(0,0,1)--cycle;
 \fill[#1!18] (0,1,0)--(1,1,0)--(1,1,1)--(0,1,1)--cycle;
 \draw[#1!55,dashed] (1,0,0)--(0,0,0) (1,0,0)--(1,1,0) (1,0,0)--(1,0,1);
 \draw[#1!80,thin] (0,0,0)--(0,1,0)--(1,1,0)--(1,1,1)--(1,0,1)--(0,0,1)--cycle
  (0,1,1)--(0,1,0) (0,1,1)--(1,1,1) (0,1,1)--(0,0,1);}
\node[font=\small,anchor=east] at (-2.2,0,0) {(a)};
\begin{scope}[shift={(-4.4,-0.6,0)},->,>=stealth,gray!70!black,thin]
 \draw (0,0,0)--(0.6,0,0) node[right,font=\tiny]{$x_1$};
 \draw (0,0,0)--(0,0.6,0) node[above,font=\tiny]{$x_2$};
 \draw (0,0,0)--(0,0,0.9) node[below,font=\tiny]{$x_3$};
\end{scope}
\begin{scope}
 \lattice
 \foreach \v in {(0,1,0),(-1,0,0),(0,0,-1),(0,0,1),(1,0,0),(0,-1,0)}
  \draw[qone] (0,0,0)--\v;
 \node[xone] at (0,0,0) {};
 \node[lab] at (0,-2.0,0) {$X$-type check on a vertex\\(six edges)};
\end{scope}
\begin{scope}[shift={(7,0,0)}]
 \lattice
 \draw[qone] (0,0,0)--(1,0,0)--(1,1,0)--(0,1,0)--cycle;
 \node[zone] at (0.5,0.5,0) {};
 \node[lab] at (0,-2.0,0) {$Z$-type check on a plaquette\\(four edges)};
\end{scope}
\begin{scope}[shift={(14,0,0)}]
 \lattice
 \draw[tcone!55,dashed] (1,0,0)--(0,0,0) (1,0,0)--(1,1,0) (1,0,0)--(1,0,1);
 \draw[qone] (0,0,0)--(0,1,0)--(1,1,0)--(1,1,1)--(1,0,1)--(0,0,1)--cycle
  (0,1,1)--(0,1,0) (0,1,1)--(1,1,1) (0,1,1)--(0,0,1);
 \foreach \m in {(0.5,0.5,1),(0.5,0.5,0),(0,0.5,0.5),(1,0.5,0.5),(0.5,0,0.5),(0.5,1,0.5)}
  \node[zone] at \m {};
 \node[lab] at (0,-2.0,0) {\shortstack{Oriented product of the\\$Z$-type checks on the six plaquettes\\of a cube equals $I$}};
\end{scope}
\begin{scope}[shift={(0,-4.9,0)}]
\node[font=\small,anchor=east] at (-2.2,0,0) {(b)};
\begin{scope}
 \lattice
 \foreach \q in {{(0,1,0)--(-1,1,0)},{(0,0,-1)--(-1,0,-1)},{(0,0,1)--(-1,0,1)},{(0,-1,0)--(-1,-1,0)}}
  \filldraw[tctwo!70,thin,fill=tctwo!30,fill opacity=0.5] (-1,0,0)--(0,0,0)--\q--cycle;
 \node[xtwo] at (-0.5,0,0) {};
 \node[lab] at (0,-2.0,0) {$X$-type check on an edge\\(four plaquettes)};
\end{scope}
\begin{scope}[shift={(7,0,0)}]
 \lattice
 \cubefaces{tctwo}
 \node[ztwo] at (0.5,0.5,0.5) {};
 \node[lab] at (0,-2.0,0) {$Z$-type check on a cube\\(six plaquettes)};
\end{scope}
\begin{scope}[shift={(14,0,0)}]
 \lattice
 \foreach \s in {-1,1}{\foreach \t in {-1,1}{
  \fill[tctwo!30,fill opacity=0.45] (0,0,0)--(\s,0,0)--(\s,\t,0)--(0,\t,0)--cycle;
  \fill[tctwo!30,fill opacity=0.45] (0,0,0)--(\s,0,0)--(\s,0,\t)--(0,0,\t)--cycle;
  \fill[tctwo!30,fill opacity=0.45] (0,0,0)--(0,\s,0)--(0,\s,\t)--(0,0,\t)--cycle;}}
 \foreach \m in {(0,0.5,0),(-0.5,0,0),(0,0,-0.5),(0,0,0.5),(0.5,0,0),(0,-0.5,0)}
  \node[xtwo] at \m {};
 \node[lab] at (0,-2.0,0) {\shortstack{Oriented product of the\\$X$-type checks on the six edges\\at a vertex equals $I$}};
\end{scope}
\end{scope}
\end{tikzpicture}
\caption{Three-dimensional toric codes. 
(a)~\(\TC_1^{(3)}(\F)\), with qudits on edges. 
(b)~\(\TC_2^{(3)}(\F)\), with qudits on plaquettes. 
The right panels show local relations among the checks; 
in each oriented product, every qudit involved is acted on by exactly two of the checks, with opposite exponents. 
In all figures, blue refers to \(\TC_1^{(3)}(\F)\) and orange to \(\TC_2^{(3)}(\F)\); 
stars mark \(X\)-type checks and diamonds \(Z\)-type checks, and colored edges and plaquettes mark qudits.}
\label{fig:toric-3d}
\end{figure*}
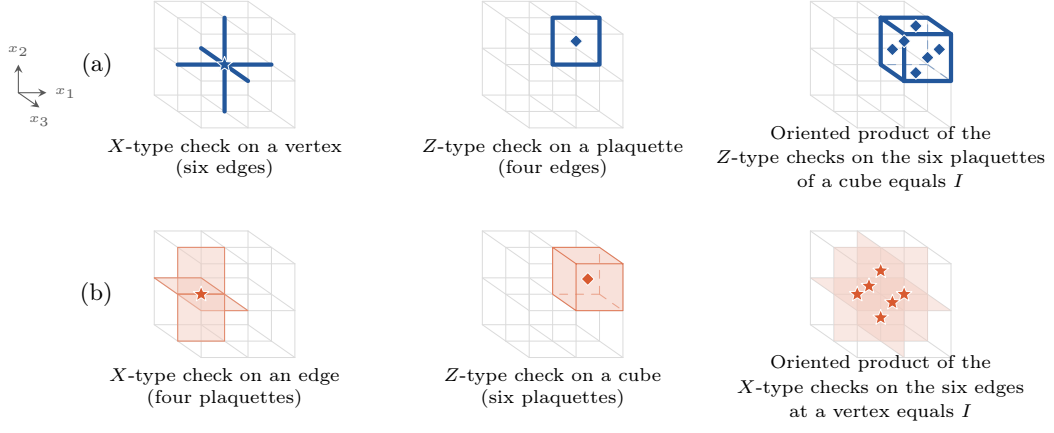

\subsection{Higher-form toric codes}
\label{subsec:toric}
The targets of a decoupling are stacks of higher-form toric codes. 
For \(1\leq j\leq D-1\), the \(j\)-form toric code \(\TC_j^{(D)}(\F)\) on the \(D\)-dimensional cubic lattice places one qudit on each \(j\)-cell \cite{Dennis2002,BombinMartinDelgado2007b,Kubica2015,Song2026}. 
Each \((j-1)\)-cell carries an \(X\)-type check acting on the \(j\)-cells that contain it, 
and each \((j+1)\)-cell carries a \(Z\)-type check acting on the \(j\)-cells in its boundary.

To fix the polynomial convention, let \(w_i=x_i-1\), and let \(e_1,\ldots,e_D\) be the edges of the unit cell at the origin, one along each direction. 
A wedge product \(e_{i_1}\wedge\cdots\wedge e_{i_k}\) is the oriented \(k\)-cell that these edges span, 
and multiplying it by \(x^\alpha\) translates it by \(\alpha\). 
The cells and their boundaries form the cellular chain complex of the cubic lattice, 
which in this polynomial form is the Koszul complex \(K_\bullet\): 
its terms are \(K_k=\textstyle\bigwedge^kR^D\), and its differential
\begin{equation}
 \begin{aligned}
 &\partial_k\,(e_{i_1}\wedge\cdots\wedge e_{i_k})\\
 &\quad=\sum_{a=1}^{k}(-1)^{a-1}w_{i_a}\,
 e_{i_1}\wedge\cdots\wedge\widehat{e}_{i_a}\wedge\cdots\wedge e_{i_k}
 \end{aligned}
 \label{eq:koszul}
\end{equation}
sends a \(k\)-cell to its oriented boundary. 
For \(I=\{i_1<\cdots<i_k\}\), the cells \(e_I=e_{i_1}\wedge\cdots\wedge e_{i_k}\) form a basis of \(K_k\), with \(e_\emptyset=1\). 
The augmentation \(\varepsilon:K_0=R\to\F\), \(x^\alpha\mapsto1\), adds up the coefficients of a \(0\)-chain; 
its kernel \(\mathfrak m=(w_1,\ldots,w_D)\) is the augmentation ideal. 
Since \(w_1,\ldots,w_D\) is a regular sequence, the sequence \(\cdots\to K_1\to K_0\xrightarrow{\varepsilon}\F\to0\) is exact \cite[Cor.~4.5.5]{Weibel1994},
so \(K_\bullet\) is a free resolution of the \(R\)-module \(\F\), on which \(R\) acts through \(\varepsilon\).

Following the convention of Sec.~\ref{subsec:polynomial}, 
we write \(T_j\) for \(K_\bullet\) regarded as a cochain complex with \(K_k\) in degree \(j-k\), so that the qudit module \(K_j\) lies in degree zero. 
The code \(\TC_j^{(D)}(\F)\) is represented by the part of \(T_j\) in degrees \(-1\) to \(1\),
\[
\begin{tikzpicture}[baseline=(b.base),>=To,
 x=1.35cm,every node/.style={inner sep=2pt}]
\foreach \n in {-1,0,1} \node[cxdeg] at (\n,0.6) {\(\n\)};
\node[cxdeg,anchor=east] at (-1.3,0.6) {degree};
\node (a) at (-1,0) {\(K_{j+1}\)};
\node (b) at (0,0) {\(K_j\)};
\node (c) at (1,0) {\(K_{j-1}\)};
\draw[->] (a)--node[above,cxlab]{\(\partial_{j+1}\)}(b);
\draw[->] (b)--node[above,cxlab]{\(\partial_j\)}(c);
\end{tikzpicture},
\]
that is, \(H_X=\partial_j\) and \(H_Z^\dagger=\partial_{j+1}\).

Figure~\ref{fig:toric-3d} shows the two toric codes in three dimensions.
In \(\TC_1^{(3)}(\F)\) \cite{Hamma2005}, qudits lie on edges,
\(X\)-type checks on vertices, and \(Z\)-type checks on plaquettes;
\(\TC_2^{(3)}(\F)\) is obtained by passing to the dual lattice and swapping \(X\) and \(Z\).
Both codes have local relations among their checks. 
In \(\TC_1^{(3)}(\F)\), the oriented product of the \(Z\)-type checks on the six plaquettes of a cube is the identity, 
and in \(\TC_2^{(3)}(\F)\), so is that of the \(X\)-type checks on the six edges at a vertex. 
In each product, every qudit involved is acted on by two of the checks with opposite exponents. 
The two codes differ in the dimension of their \(X\)-check excitations: a \(Z\)-type operator \(Z(v)\) with \(v\in K_j\) violates the \(X\)-type checks on the boundary \(\partial_jv\), so these excitations are pointlike in \(\TC_1^{(3)}(\F)\) and looplike in \(\TC_2^{(3)}(\F)\). 
Relations among the \(X\)-type checks and excitations of different dimensions are what the superselection profile of Sec.~\ref{subsec:profile} keeps track of.

For nonnegative integers \(t_1,\ldots,t_{D-1}\), the toric-code stack \(\bigoplus_{j=1}^{D-1}\TC_j^{(D)}(\F)^{\oplus t_j}\) consists of \(t_j\) copies of \(\TC_j^{(D)}(\F)\) for each \(j\). 
Its complex \(\Cstack=\bigoplus_jT_j^{\oplus t_j}\) serves as the target in Eq.~\eqref{eq:decoupling-square}.

\subsection{Superselection profiles}
\label{subsec:profile}
A finitely supported \(Z\)-type operator \(Z(v)\) creates the syndrome \(H_Xv\in R^s\) of violated \(X\)-type checks. 
Two syndromes represent the same excitation type if their difference is created by a finitely supported \(Z\)-type operator, that is, if it lies in \(\im H_X\). 
In two dimensions, the topological condition allows one to choose \(X\)-type checks with no local relations \cite[Lemma~7.1]{Haah2013}.
Every finitely supported syndrome can then occur, possibly created by an operator of infinite support. 
The pointlike excitation types therefore form the module \(\coker H_X=R^s/\im H_X\), in which addition describes fusion and the action of \(R\) describes translations.

In higher dimensions, local relations among the \(X\)-type checks (metachecks) restrict which syndromes can occur. 
We call the syndromes compatible with all of them \emph{admissible}. 
In \(\TC_2^{(3)}(\F)\), for example, the relation among the \(X\)-type checks at each vertex (Fig.~\ref{fig:toric-3d}) implies that every syndrome created by a \(Z\)-type operator consists of closed loops. 
In general, choose a relation map \(\gamma^\dagger:R^t\to R^s\) with \(\im\gamma^\dagger=\ker H_X^\dagger\), whose columns represent the local relations. 
The admissible syndromes form \(\ker\gamma\), and the pointlike excitation types form \(\ker\gamma/\im H_X\). 
This quotient misses the looplike and higher-dimensional excitations: for \(\TC_2^{(3)}(\F)\) it vanishes, although the code has looplike excitations. 
Capturing them requires the complete hierarchy of relations.

To describe all excitations independently of the chosen generators, consider the module \(M_X=\coker H_X^\dagger=R^q/\im H_X^\dagger\) of finitely supported \(X\)-type Pauli operators modulo \(X\)-type stabilizers. 
Choose a free resolution \(P_\bullet\) of \(M_X\) that begins with \(H_X^\dagger\) and \(\gamma^\dagger\),
\begin{equation}
 \cdots\longrightarrow R^t\xrightarrow{\gamma^\dagger}R^s
 \xrightarrow{H_X^\dagger}R^q\longrightarrow M_X\longrightarrow0.
 \label{eq:resolution}
\end{equation}
Its successive terms record the qudits, the \(X\)-type checks, the relations among checks, the relations among relations, and so on. 
Replacing each map of \(P_\bullet\) by its dagger gives the cochain complex
\begin{equation}
\begin{tikzpicture}[baseline=(a.base),>=To,
 x=1.35cm,every node/.style={inner sep=2pt}]
\foreach \n in {0,1,2} \node[cxdeg] at (\n,0.6) {\(\n\)};
\node[cxdeg,anchor=east] at (-0.3,0.6) {degree};
\node (z) at (-0.75,0) {\(0\)};
\node (a) at (0,0) {\(R^q\)};
\node (b) at (1,0) {\(R^s\)};
\node (c) at (2,0) {\(R^t\)};
\node (d) at (2.75,0) {\(\cdots\)};
\draw[->] (z)--(a);
\draw[->] (a)--node[above,cxlab]{\(H_X\)}(b);
\draw[->] (b)--node[above,cxlab]{\(\gamma\)}(c);
\draw[->] (c)--(d);
\end{tikzpicture}.
\label{eq:dual-complex}
\end{equation}
Its first two maps are the syndrome map and the relation map above.
Formally, Eq.~\eqref{eq:dual-complex} is obtained by applying \(\Hom_R(\overline{\,\cdot\,},R)\) to \(P_\bullet\), where \(\overline M\) denotes an \(R\)-module \(M\) with \(R\) acting through the antipode. 
It computes \(\mathbf R\Hom_R(\overline{M_X},R)\) \cite[Sec.~10.7]{Weibel1994}.

Its degree-zero cohomology, \(\ker H_X=\im H_Z^\dagger\), consists of \(Z\)-type stabilizers rather than excitations. 
Following Song \cite{Song2026}, we discard it and define the \emph{superselection profile} as the positive-degree truncation
\begin{equation}
 \mathscr E_X:=\tau_{\geq1}\mathbf R\Hom_R(\overline{M_X},R).
 \label{eq:superselection-profile}
\end{equation}
Here the canonical truncations \(\tau_{\geq k}\) and \(\tau_{\leq k}\) retain the cohomology in degrees \(n\geq k\) and \(n\leq k\), respectively. 
The cohomology modules
\[
 E_X^j:=H^j(\mathscr E_X)\cong\Ext_R^j(\overline{M_X},R),\qquad j\geq1,
\]
are the \emph{superselection layers} (called charge modules in Ref.~\cite{RubaYang2024}). 
The elements of \(E_X^j\) are types of \((j-1)\)-dimensional excitations \cite{Song2026}, with fusion and translations described as for \(\coker H_X\).
In particular, \(E_X^1\) is the quotient \(\ker\gamma/\im H_X\) found above. 
Only the degrees \(1\leq j\leq D-1\) can occur (Lemma~\ref{lem:degree-bound}, Appendix~\ref{apx:degree-bound}; see also Refs.~\cite{RubaYang2024,Song2026}), 
so a topological code in \(D\) dimensions has excitations of dimension at most \(D-2\).

The profile does not depend on the chosen generators. 
Indeed, \(M_X\) depends only on the \(X\)-type stabilizer submodule, and different resolutions give quasi-isomorphic complexes, that is, complexes related by cochain maps that induce isomorphisms on cohomology. 
We write \(\simeq\) for isomorphism in the derived category \(\mathrm{D}(R)\) of \(R\)-modules, in which quasi-isomorphic complexes become isomorphic.

For \(\TC_j^{(D)}(\F)\), applying the dagger to the part of \(T_j\) in nonnegative degrees gives a free resolution of \(M_X\), because the dual Koszul complex is exact below its top degree \cite{Song2026}. 
The complex~\eqref{eq:dual-complex} is then \(K_j\to K_{j-1}\to\cdots\to K_0\). 
Since \(K_\bullet\) resolves \(\F\), its only cohomology in positive degrees is \(K_0/\im\partial_1=\F\), in degree \(j\). 
Hence \(\mathscr E_X\simeq\F[-j]\), and \(\TC_j^{(D)}(\F)\) has a single nonzero layer \(E_X^j\cong\F\), generated by its \((j-1)\)-dimensional \(X\)-check excitation (Sec.~\ref{subsec:toric}).

The profile is unchanged by translation-invariant CSS-preserving Clifford circuits and by adjoining ancillas (Lemma~\ref{lem:invariance}, Appendix~\ref{apx:invariance}). 
Since the profile of \(\TC_j^{(D)}(\F)\) is \(\F[-j]\), a toric-code stack has the profile \(\bigoplus_j\F[-j]^{\oplus t_j}\). 
Therefore, a decoupling requires the profile of the code to have the same form.

\subsection{Coarse-graining}\label{subsec:coarse}\label{subsec:translation-action}
Coarse-graining to a superlattice \(\Lambda\subseteq\Z^D\) of index \(N:=[\Z^D:\Lambda]\) groups \(N\) unit cells into one, with \(Nq\) qudits. 
Only the translations in \(\Lambda\) remain; they span the subring \(\F[\Lambda]\) of Laurent polynomials whose exponents lie in \(\Lambda\) \cite{Haah2013}. 
Regarding \(R\)-modules as \(\F[\Lambda]\)-modules is called \emph{restriction of scalars} \cite[Ch.~2]{AtiyahMacdonald1969},
and we write \((\cdot)|_{\F[\Lambda]}\) for the restriction of a module, complex, or derived object. 
Since \(R\) is a free \(\F[\Lambda]\)-module with a basis of monomials representing \(\Z^D/\Lambda\), 
expanding \(H_X\) and \(H_Z\) in this basis gives the check matrices of the coarse-grained code. 
We say that a code can be decoupled on \(\Lambda\) if the coarse-grained code can be decoupled. 
By Lemma~\ref{lem:coarse-profile} (Appendix~\ref{apx:coarse-profile}), the coarse-grained code is again topological, and its profile and layers are the restrictions \(\mathscr E_X|_{\F[\Lambda]}\) and \(E_X^j|_{\F[\Lambda]}\) (see also Ref.~\cite{RubaYang2024} for the layers). 
In particular, every layer keeps its \(\F\)-dimension.

Since the layers of a toric-code stack are finite dimensional over \(\F\), a code with an infinite-dimensional layer, such as Haah's cubic code \cite{Haah2011}, cannot be decoupled on any superlattice. 
We therefore let the layers be finite dimensional, with \(t_j=\dim_{\F}E_X^j\). 
Regarding \(E_X^j\) as an \(\F\)-vector space of dimension \(t_j\), every translation acts on it by an invertible linear map, 
that is, by a matrix in \(\mathrm{GL}(t_j,\F)\) once a basis is chosen. 
The translations that act trivially on every layer form the kernel \(\Ltriv\) of the resulting homomorphism \(\Z^D\to\prod_j\mathrm{GL}(t_j,\F)\), a superlattice because this group is finite. 
From here through Sec.~\ref{subsec:inter-layer-extensions}, we coarse-grain to \(\Ltriv\), write \(R\) for \(\F[\Ltriv]\) with variables \(x_i=x^{\beta_i}\) for a basis \(\beta_1,\ldots,\beta_D\) of \(\Ltriv\), and keep the notation \(\mathscr E_X\) and \(E_X^j\) for the restricted profile and layers.
Every \(x_i\) then acts trivially on every layer, so
\begin{equation}
 \mathfrak m E_X^j=0,\qquad E_X^j\cong(R/\mathfrak m)^{\oplus t_j}\cong\F^{t_j},
 \label{eq:normalized}
\end{equation}
where \(\mathfrak m\) is the augmentation ideal of Sec.~\ref{subsec:toric}, now in the new variables. 
Hence every layer agrees with the corresponding layer of \(\bigoplus_j\TC_j^{(D)}(\F)^{\oplus t_j}\).

\section{Obstruction: inter-layer gluing}\label{sec:gluing}
After coarse-graining to \(\Ltriv\), every layer agrees with that of a toric-code stack, but the profile need not: it also records how the layers are glued together. 
Song showed that this gluing can be nonzero \cite{Song2026}. 
We now make the gluing precise and show, with an example, that it obstructs decoupling.

\subsection{Inter-layer gluing and its description}\label{subsec:gluing-classes}
Successive layers are related by the canonical truncation triangles \cite[Secs.~1.2 and 10.4]{Weibel1994},
\begin{equation}
 \tau_{\leq j-1}\mathscr E_X\longrightarrow
 \tau_{\leq j}\mathscr E_X\longrightarrow E_X^j[-j]
 \xrightarrow{\delta_j}(\tau_{\leq j-1}\mathscr E_X)[1].
 \label{eq:triangle}
\end{equation}
Here \(A[n]\) denotes a complex \(A\) shifted by \(n\), with \(A[n]^k=A^{k+n}\) and differential \((-1)^nd_A\), and an \(R\)-module is regarded as a complex in degree zero. 
The connecting morphism \(\delta_j\) records how the \(j\)th layer is attached to the lower ones. 
We call these morphisms the \emph{inter-layer gluing}. 
The profile \emph{splits} when all \(\delta_j\) vanish, equivalently when
\begin{equation}
 \mathscr E_X\simeq\bigoplus_{j\geq1}E_X^j[-j]
 \label{eq:profile-split}
\end{equation}
in \(\mathrm{D}(R)\) (Lemma~\ref{lem:gluing}, Appendix~\ref{apx:truncation}).

Since the layers already agree with those of \(\bigoplus_j\TC_j^{(D)}(\F)^{\oplus t_j}\), the profile has the form \(\bigoplus_j\F[-j]^{\oplus t_j}\) exactly when it splits. 
More generally, over a superlattice \(\Lambda\) of the original lattice, the profile has this form exactly when \(\Lambda\subseteq\Ltriv\) and every \(\delta_j\) restricts to zero over \(\F[\Lambda]\). 
By Sec.~\ref{subsec:profile}, a nonzero gluing therefore obstructs decoupling.

The gluing \(\delta_j\) maps the \(j\)th layer into the part \(\tau_{\leq j-1}\mathscr E_X\) of the profile below it, shifted by one.
Once this part has split as
\begin{equation}
 \tau_{\leq j-1}\mathscr E_X
 \simeq\bigoplus_{i=1}^{j-1}E_X^i[-i],
 \label{eq:lower-splitting}
\end{equation}
composing with this splitting turns \(\delta_j\) into a morphism from \(E_X^j[-j]\) to \(\bigoplus_{i<j}E_X^i[-i+1]\). 
Its component for the \(i\)th layer is a class
\begin{equation}
 \begin{aligned}
 \kappa_{ji}&\in\Hom_{\mathrm D(R)}\bigl(E_X^j[-j],E_X^i[-i+1]\bigr)\\
 &=\Ext_R^{\,j-i+1}(E_X^j,E_X^i),
 \end{aligned}
 \label{eq:attachments}
\end{equation}
where we shifted both arguments by \(j\) and used \(\Hom_{\mathrm D(R)}(A,B[n])=\Ext_R^n(A,B)\) for modules \(A\) and \(B\) \cite[Def.~10.7.1 and Cor.~10.7.5]{Weibel1994}. 
Thus \(\delta_j\) vanishes if and only if every \(\kappa_{ji}\) does. 
In that case, the triangle~\eqref{eq:triangle} splits and Eq.~\eqref{eq:lower-splitting} extends to degree \(j\) (Lemma~\ref{lem:split-triangle}, Appendix~\ref{apx:truncation}).
Physically, \(\kappa_{ji}\) describes how \((j-1)\)-dimensional excitations carry \((i-1)\)-dimensional ones.

These classes can be described by their values on unit cells.

\begin{lemma}\label{lem:values}
Let \(A\) and \(B\) be \(R\)-modules that are finite dimensional over \(\F\), with \(\mathfrak m A=\mathfrak m B=0\).
Every class \(\kappa\in\Ext_R^n(A,B)\) is represented, on the free resolution \(K_\bullet\otimes_{\F}A\) of \(A\), by a unique \(R\)-linear map \(f:K_n\otimes_{\F}A\to B\), and
\[
 \Ext_R^n(A,B)\cong\Hom_R\bigl(K_n\otimes_{\F}A,B\bigr)
 \cong\bigoplus_{|I|=n}\Hom_{\F}(A,B),
\]
where the second isomorphism sends \(f\) to the maps \(f_I:a\mapsto f(e_I\otimes a)\). 
Since translations act trivially on \(B\), the class \(\kappa\) thus assigns the value \(f_I\) to every unit \(n\)-cell of orientation \(I\).
\end{lemma}
By Eq.~\eqref{eq:normalized}, the gluing class \(\kappa_{ji}\) is the case \(A=E_X^j\) and \(B=E_X^i\).

\subsection{Concrete example: the glued code}\label{subsec:example-glued}
\begin{figure}[t]
\centering
\begin{tikzpicture}[x={(0.6cm,0cm)},y={(0cm,0.6cm)},z={(0.3cm,-0.21cm)},
 line cap=round,line join=round,
 lab/.style={font=\scriptsize,align=center,text width=3.6cm},
 bg/.style={gray!28,line width=0.35pt}]
\newcommand{\lattice}{%
 \foreach \a in {-1,0,1}{\foreach \b in {-1,0,1}{%
  \draw[bg] (-1,\a,\b)--(1,\a,\b);
  \draw[bg] (\a,-1,\b)--(\a,1,\b);
  \draw[bg] (\a,\b,-1)--(\a,\b,1);}}}
\begin{scope}
 \lattice
 \filldraw[tctwo!80,thin,fill=tctwo!30] (0,0,0)--(1,0,0)--(1,1,0)--(0,1,0)--cycle;
 \node[font=\scriptsize,text=tctwo!70!black] at (0.5,0.5,0) {$Z$};
 \foreach \m in {(0.5,0,0),(0.5,1,0),(0,0.5,0),(1,0.5,0)} \node[xtwo] at \m {};
 \node[xone] at (0,0,0) {};
 \node[lab] at (0,-1.95,0) {(a) $Z$ on a plaquette\\and the checks it violates};
\end{scope}
\begin{scope}[shift={(5.4,0,0)}]
 \lattice
 \fill[tctwo!18] (0,0,1)--(1,0,1)--(1,1,1)--(0,1,1)--cycle;
 \fill[tctwo!18] (0,0,0)--(0,1,0)--(0,1,1)--(0,0,1)--cycle;
 \fill[tctwo!18] (0,1,0)--(1,1,0)--(1,1,1)--(0,1,1)--cycle;
 \draw[tctwo!55,dashed] (1,0,0)--(0,0,0) (1,0,0)--(1,1,0) (1,0,0)--(1,0,1);
 \draw[tctwo!80,thin] (0,0,0)--(0,1,0)--(1,1,0)--(1,1,1)--(1,0,1)--(0,0,1)--cycle
  (0,1,1)--(0,1,0) (0,1,1)--(1,1,1) (0,1,1)--(0,0,1);
 \draw[qone] (0,0,0)--(0,0,1);
 \node[ztwo] at (0.5,0.5,0.5) {};
 \node[lab] at (0,-1.95,0) {(b) $Z$-type check on a cube\\and the qudits it acts on};
\end{scope}
\begin{scope}[shift={(-2.75,-0.7,0)},->,>=stealth,gray!70!black,thin]
 \draw (0,0,0)--(0.6,0,0) node[right,font=\tiny]{$x_1$};
 \draw (0,0,0)--(0,0.6,0) node[above,font=\tiny]{$x_2$};
 \draw (0,0,0)--(0,0,0.9) node[below,font=\tiny]{$x_3$};
\end{scope}
\end{tikzpicture}
\caption{Couplings in the code of Eq.~\eqref{eq:glued-example}. 
Panels (a) and (b) show the fourth columns of \(H_X(\GC)\) and \(H_Z(\GC)^\dagger\), respectively; 
the blue star in (a) and the blue edge in (b) come from the boxed entries.}
\label{fig:glued-example}
\end{figure}
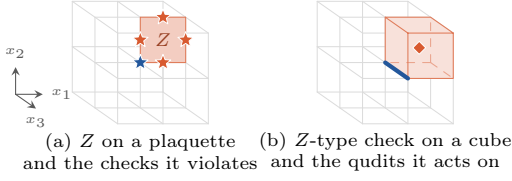

We illustrate the gluing with a topological code from Ref.~\cite{Song2026}, 
which we call the \emph{glued code} (GC). 
It is obtained from \(\TC_1^{(3)}(\F)\oplus\TC_2^{(3)}(\F)\) by adding the two coupling terms boxed in Eq.~\eqref{eq:glued-example} below. 
Its layers coincide with those of \(\TC_1^{(3)}(\F)\oplus\TC_2^{(3)}(\F)\), so translations act trivially on them and \(\Ltriv(\GC)=\Z^3\), but its gluing \(\delta_2(\GC)\) is nonzero. 
Physically, a looplike excitation of \(\TC_2^{(3)}(\F)\) carries a pointlike excitation of \(\TC_1^{(3)}(\F)\) (Fig.~\ref{fig:glued-example}(a)), 
a charge whose value is the area enclosed by the projection of the loop onto the \(x_1x_2\)-plane, up to sign. 
By Sec.~\ref{subsec:gluing-classes}, the GC therefore admits no decoupling on the original lattice.

\paragraph{The code.}
Take \(D=3\). 
The qudits and checks sit where they do in \(\TC_1^{(3)}(\F)\oplus\TC_2^{(3)}(\F)\) (Fig.~\ref{fig:toric-3d}).
Within a unit cell, we order them as \((e_1,e_2,e_3;e_{12},e_{13},e_{23})\) for the qudits, \((e_\emptyset;e_1,e_2,e_3)\) for the \(X\)-type checks, and \((e_{12},e_{13},e_{23};e_{123})\) for the \(Z\)-type checks, with those of \(\TC_1^{(3)}(\F)\) before the semicolons. 
In these bases, let
\begin{equation}
 \begin{aligned}
 H_X(\GC)&=\left(\begin{array}{ccc|ccc}
 w_1&w_2&w_3&\boxed{-1}&0&0\\ \hline
 0&0&0&-w_2&-w_3&0\\
 0&0&0&w_1&0&-w_3\\
 0&0&0&0&w_1&w_2
 \end{array}\right),\\
 H_Z(\GC)^\dagger&=\left(\begin{array}{ccc|c}
 -w_2&-w_3&0&0\\
 w_1&0&-w_3&0\\
 0&w_1&w_2&\boxed{1}\\ \hline
 0&0&0&w_3\\
 0&0&0&-w_2\\
 0&0&0&w_1
 \end{array}\right).
 \end{aligned}
 \label{eq:glued-example}
\end{equation}
The lines separate the blocks of the two toric codes.

\paragraph{The couplings.}
Each column of \(H_X(\GC)\) is the syndrome of \(Z\) on one qudit at the origin (Sec.~\ref{subsec:polynomial}). 
The fourth column (Fig.~\ref{fig:glued-example}(a)) is the syndrome of \(Z\) on the plaquette \(e_{12}\). 
As in \(\TC_2^{(3)}(\F)\), its entries \(-w_2=1-x_2\) and \(w_1=x_1-1\) state that \(Z\) on \(e_{12}\) violates the \(X\)-type checks on the bottom, top, left, and right edges of the plaquette with the exponents \(1,-1,-1,1\). 
The boxed entry adds a violation, with exponent \(-1\), of the \(X\)-type check of \(\TC_1^{(3)}(\F)\) at its corner \(e_\emptyset\). 
The fourth column of \(H_Z(\GC)^\dagger\) (Fig.~\ref{fig:glued-example}(b)) is the \(Z\)-type check of \(\TC_2^{(3)}(\F)\) on a cube, which acts on its six plaquettes and, through the boxed entry, on the \(x_3\)-edge \(e_3\) at its corner \(e_\emptyset\).

The boxed entries thus define maps \(c_X:K_2\to K_0\) and \(c_Z:K_3\to K_1\), 
with \(c_Xe_{12}=-e_\emptyset\) and \(c_Ze_{123}=e_3\), 
so that
\begin{equation}
 H_X(\GC)=\begin{pmatrix}\partial_1&c_X\\0&\partial_2\end{pmatrix},\qquad
 H_Z(\GC)^\dagger=\begin{pmatrix}\partial_2&c_Z\\0&\partial_3\end{pmatrix}.
 \label{eq:glued-blocks}
\end{equation}
Appendix~\ref{apx:glued} writes the GC as a mapping cone of the two toric codes, 
confirms that it is topological with layers \(E_X^1(\GC)\cong E_X^2(\GC)\cong\F\), 
and shows that \(\delta_2(\GC)\) is represented, up to sign, by \(f=\varepsilon c_X:K_2\to\F\), 
whose plaquette values \((f_{12},f_{13},f_{23})\) are \((-1,0,0)\). 
By Lemma~\ref{lem:values}, \(\delta_2(\GC)\neq0\), so the profile does not split.

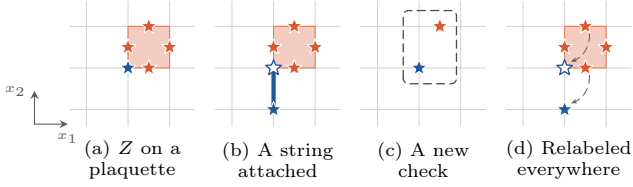
\begin{figure}[t]
\centering
\begin{tikzpicture}[x=0.55cm,y=0.55cm,line cap=round,line join=round,
 lab/.style={font=\scriptsize,align=center,text width=1.85cm},
 bg/.style={gray!35,line width=0.35pt},
 arr/.style={-{Stealth[length=2.4pt]},black!60,densely dashed,line width=0.4pt}]
\newcommand{\grid}{\foreach \i in {0,...,2}{\draw[bg] (\i,-0.4)--(\i,2.6);}%
 \foreach \j in {0,...,2}{\draw[bg] (-0.4,\j)--(2.6,\j);}}
\newcommand{\plaq}{\filldraw[tctwo!30,draw=tctwo!80,thin] (1,1) rectangle (2,2);
 \foreach \m in {(1.5,1),(1.5,2),(1,1.5),(2,1.5)} \node[xtwo] at \m {};}
\begin{scope}
 \grid \plaq
 \node[xone] at (1,1) {};
 \node[lab] at (1.1,-1.2) {(a) $Z$ on a\\plaquette};
\end{scope}
\begin{scope}[shift={(3.5,0)}]
 \grid \plaq
 \draw[qone] (1,0)--(1,1);
 \node[xoneopen] at (1,1) {}; \node[xone] at (1,0) {};
 \node[lab] at (1.1,-1.2) {(b) A string\\attached};
\end{scope}
\begin{scope}[shift={(7,0)}]
 \grid
 \node[xone] at (1,1) {}; \node[xtwo] at (1.5,2) {};
 \draw[regroup] (0.62,0.62) rectangle (1.9,2.38);
 \node[lab] at (1.1,-1.2) {(c) A new\\check};
\end{scope}
\begin{scope}[shift={(10.5,0)}]
 \grid \plaq
 \node[xoneopen] at (1,1) {}; \node[xone] at (1,0) {};
 \draw[arr] (1.6,1.82) to[bend left=35] (1.15,1.15);
 \draw[arr] (1.6,0.82) to[bend left=35] (1.15,0.15);
 \node[lab] at (1.1,-1.2) {(d) Relabeled\\everywhere};
\end{scope}
\begin{scope}[shift={(-1.25,-0.35)},->,>=stealth,gray!70!black,thin]
 \draw (0,0)--(0.8,0) node[below,font=\tiny]{$x_1$};
 \draw (0,0)--(0,0.8) node[left,font=\tiny]{$x_2$};
\end{scope}
\end{tikzpicture}
\caption{Why the charge cannot be detached from the loop on the original lattice (\(x_1x_2\)-plane, both codes superposed; open stars mark checks no longer violated). 
(a)~\(Z\) on a plaquette creates a loop and the charge \(-1\) at its corner. 
(b)~An attached string moves the charge but preserves the total. 
(c)~A new check: the \(X\)-type check of \(\TC_1^{(3)}(\F)\) at a vertex times a power of that of \(\TC_2^{(3)}(\F)\) on the \(x_1\)-edge above it. 
(d)~With every vertex relabeled, the top edge of a plaquette removes the charge at its corner, but the bottom edge creates it at the vertex below.}
\label{fig:glued-obstruction}
\end{figure}

\paragraph{Why the gluing obstructs decoupling.}
We call a violated \(X\)-type check of \(\TC_1^{(3)}(\F)\) a charge, with the exponent as its value, and \(Z\) on a path of edges a string. 
A string violates the checks of \(\TC_1^{(3)}(\F)\) at its two endpoints with opposite exponents, so strings move charges but preserve their total (Fig.~\ref{fig:glued-obstruction}(b)). 
\(Z\) on an \(x_1x_2\)-plaquette creates a small loop of violated checks of \(\TC_2^{(3)}(\F)\) together with the charge \(-1\) at its corner (Figs.~\ref{fig:glued-example}(a) and~\ref{fig:glued-obstruction}(a)); \(Z\) on the other plaquettes creates no charge. 
Hence \(Z\) on an oriented surface of plaquettes creates the loop \(\ell\) along its boundary with a total charge equal to minus the signed area enclosed by the projection of \(\ell\) onto the \(x_1x_2\)-plane.

Decoupling must detach the charge from the loop, as in \(\TC_1^{(3)}(\F)\oplus\TC_2^{(3)}(\F)\), with translation-invariant \(U\) and \(V\) in the square~\eqref{eq:decoupling-square}. 
A circuit with \(U=\bigl(\begin{smallmatrix}I&h^0\\0&I\end{smallmatrix}\bigr)\), where \(h^0:K_2\to K_1\), attaches a string to \(Z\) on each plaquette (Fig.~\ref{fig:glued-obstruction}(b)). 
Relabeling the checks, by multiplying each \(X\)-type check of \(\TC_1^{(3)}(\F)\) by powers of nearby \(X\)-type checks of \(\TC_2^{(3)}(\F)\) as recorded by \(h^1:K_1\to K_0\), 
acts on syndromes by \(V=\bigl(\begin{smallmatrix}I&h^1\\0&I\end{smallmatrix}\bigr)\) (Fig.~\ref{fig:glued-obstruction}(c)). 
By Eq.~\eqref{eq:glued-blocks}, the square commutes exactly when \(c_X=\partial_1h^0-h^1\partial_2\), 
that is, when the strings and the relabeling cancel the coupling. 
On the original lattice, this equation has no solution: 
evaluated on the plaquette \(e_{12}\), its left side has total charge \(-1\), 
whereas strings preserve the total charge and a translation-invariant relabeling contributes opposite charges from opposite edges (Fig.~\ref{fig:glued-obstruction}(d)).

\section{Main result: decoupling criterion}\label{sec:main}
Section~\ref{sec:gluing} showed that the gluing can survive coarse-graining to \(\Ltriv\) and then obstructs decoupling.
We now show that further coarse-graining always removes it, and that the resulting condition, that the profile is that of a toric-code stack,
also suffices for decoupling (Theorem~\ref{thm:main}). 
Together, these results yield two explicit algorithms, the second of polynomial cost.

\subsection{Removing the inter-layer gluing}\label{subsec:inter-layer-extensions}
We remove the gluing one layer at a time, in increasing degree.
For \(n>0\), coarse-graining by a factor of \(p\) in every direction turns \(p^n\) unit \(n\)-cells of each orientation \(I\) into one, which therefore carries the value \(p^nf_I=0\). 
The following lemma, proved together with Lemma~\ref{lem:values} in Appendix~\ref{apx:superlattice},
makes this precise.
\begin{lemma}\label{lem:ext-zero}
Let \(A\) and \(B\) be \(R\)-modules that are finite dimensional over \(\F\), with \(\mathfrak m A=\mathfrak m B=0\). 
For every \(n>0\), every class in \(\Ext_R^n(A,B)\) restricts to zero over \(\F[p\Ltriv]\).
\end{lemma}

Once the lower layers have split as in Eq.~\eqref{eq:lower-splitting},
every gluing class \(\kappa_{ji}\) of the next layer lies in \(\Ext\) of positive degree \(j-i+1\). 
Lemma~\ref{lem:ext-zero} therefore removes them after coarse-graining by a factor of \(p\) in every direction, and the splitting extends to that layer (Sec.~\ref{subsec:gluing-classes}). 
Restriction preserves the lower splitting and the trivial translation action, so the same argument applies on the new lattice.

\subsection{The decoupling criterion}\label{subsec:criterion}
We now return to the original lattice \(\Z^D\) and state the main result.
\begin{theorem}[Decoupling criterion]\label{thm:main}
Consider a translation-invariant topological \(p\)-ary CSS code on \(\Z^D\), \(D\geq2\), and let \(\Lambda\subseteq\Z^D\) be a superlattice. 
The code can be decoupled on \(\Lambda\) if and only if its profile over \(\F[\Lambda]\) is that of a toric-code stack,
\begin{equation}
 \mathscr E_X|_{\F[\Lambda]}\simeq\bigoplus_{j=1}^{D-1}\F[-j]^{\oplus t_j}
 \label{eq:criterion}
\end{equation}
in \(\mathrm{D}(\F[\Lambda])\), for some nonnegative integers \(t_j\). 
In that case, a decoupling onto \(\bigoplus_j\TC_j^{(D)}(\F)^{\oplus t_j}\) can be realized by a finite-depth local circuit of SUM and SWAP gates, possibly after adding ancillas.
\end{theorem}

Necessity follows from Sec.~\ref{subsec:profile}, applied to the code coarse-grained to \(\Lambda\): circuits and ancillas leave the profile unchanged, and a toric-code stack has the profile above.
Sufficiency is proved constructively in Appendix~\ref{apx:construction},
and Sec.~\ref{subsec:algorithm} turns the proof into an algorithm.

The profile of the \(Z\) sector is defined with \(X\) and \(Z\) exchanged. 
Since on-site qudit Fourier transforms exchange the two sectors, map SUM and SWAP circuits to SUM and SWAP circuits, and map \(\TC_j^{(D)}(\F)\) to \(\TC_{D-j}^{(D)}(\F)\), 
the criterion holds equally with this profile.

\subsection{The decoupling algorithms}\label{subsec:algorithm}
Lemma~\ref{lem:ext-zero} and Theorem~\ref{thm:main} turn into two algorithms. 
Algorithm~\ref{alg:coarse} coarse-grains the code so that its profile is that of a toric-code stack, 
or reports that no superlattice achieves this. 
Algorithm~\ref{alg:decoupling} then constructs the decoupling circuit on the coarse-grained lattice. 

\begin{algorithm}[t]
\caption{Coarse-graining}\label{alg:coarse}
\Input{check matrices \(H_X\), \(H_Z\) over \(R\) of a translation-invariant topological \(p\)-ary CSS code}
\Output{\(\Lsuff\) and the complex \(\Ccode\) restricted to \(\F[\Lsuff]\), or the statement that the code cannot be decoupled}
extend the complex~\eqref{eq:physical-window} to a bounded complex \(\Ccode\) of finite free \(R\)-modules (Lemma~\ref{lem:model}, Appendix~\ref{apx:models})\;\label{line:resolution}
compute the layers \(E_X^j=H^j(\Ccode)\)\; 
\If{some \(E_X^j\) is infinite dimensional over \(\F\)}{\Return{the code cannot be decoupled}\;\label{line:stop}}
compute the action of \(x_1,\ldots,x_D\) on the layers and its kernel \(\Ltriv\)\;\label{line:triv}
\(\Lsuff\gets p^{\max(m-1,0)}\Ltriv\), with \(m\) the number of nonzero layers\;\label{line:suff}
\Return{\(\Lsuff\) and \(\Ccode|_{\F[\Lsuff]}\)}\;
\end{algorithm}

When line~\ref{line:stop} stops, no superlattice works (Sec.~\ref{subsec:coarse}). 
Otherwise, the translations in \(\Ltriv\) act trivially on every layer, 
and line~\ref{line:suff} removes the gluing by sufficient coarse-graining (Sec.~\ref{subsec:inter-layer-extensions}), 
so the profile over \(\F[\Lsuff]\) has the form~\eqref{eq:criterion}. 
The restricted complex is that of the coarse-grained code (Lemma~\ref{lem:coarse-profile}, Appendix~\ref{apx:coarse-profile}), 
so it need not be recomputed. 

Line~\ref{line:suff} may coarse-grain less, at the cost of computing the gluing classes. 
For each layer, it then suffices to coarse-grain to a superlattice over which the gluing classes of that layer restrict to zero, and Appendix~\ref{apx:minimal} determines the smallest such superlattices. 
For the glued code, coarse-graining by a factor of \(p\) along a single direction suffices (Sec.~\ref{sec:example}). 

\begin{algorithm}[t]
\caption{Decoupling circuit}\label{alg:decoupling}
\Input{a complex \(\Ccode\) as in Lemma~\ref{lem:model} whose profile is that of a toric-code stack}
\Output{the multiplicities \(t_j\) and a finite-depth circuit of SUM and SWAP gates that decouples the code, with ancillas}
compute a basis of each cohomology module of \(\Ccode\), which gives \(t_j\) and representing cocycles\;\label{line:cohomology}
compute a cochain map \(\Gamma:\Cstack\to\Ccode\) that sends the generators of \(H^j(\Cstack)\) to these cocycles, and a contraction of its mapping cone, which gives a homotopy inverse \(\Delta\) (Appendix~\ref{apx:beta})\;\label{line:gamma}
construct a stable isomorphism of the check cokernels, adjoining identity \(X\)-type checks (Lemma~\ref{lem:stable-cokernel}, Appendix~\ref{apx:cokernels})\;\label{line:stable}
lift it to invertible \(U\) and \(V\), adjoining ancillas (Lemma~\ref{lem:invertible-lift}, Appendix~\ref{apx:lifting})\;\label{line:lift}
factor \(U\) into layers of SUM gates and on-site SWAP gates (Lemma~\ref{lem:gates}, Appendix~\ref{apx:circuit})\;\label{line:gates}
\Return{\(t_j\) and the circuit}\;
\end{algorithm}

Algorithm~\ref{alg:decoupling} follows the proof of sufficiency in Appendix~\ref{apx:construction}, 
applied to the output of Algorithm~\ref{alg:coarse} with \(\F[\Lsuff]\) in place of \(R\). 
Each of its lines solves linear equations over a Laurent polynomial ring, or evaluates explicit formulas in their solutions. 
In a fixed number of variables, such equations have solutions of polynomially bounded degree \cite[Thms.~3.2 and 3.4]{Aschenbrenner2004}, 
so the coefficients of the unknowns satisfy a linear system over \(\F\) of polynomial size. 

\begin{proposition}[Polynomial cost]\label{prop:complexity}
The number of arithmetic operations over \(\F\) used by Algorithm~\ref{alg:decoupling} is bounded by a polynomial in the number of coefficients of the differentials of its input, counted with all exponents in a common box. 
The degree of this polynomial depends only on the spatial dimension \(D\). 
The same polynomial bounds the depth of the output circuit and its number of ancillas per unit cell. 
\end{proposition}

The proof, in Appendix~\ref{apx:complexity}, rests on worst-case degree bounds, so this polynomial is far from practical. 
Algorithm~\ref{alg:decoupling} receives the code coarse-grained to \(\Lsuff\), roughly \([\Z^D:\Lsuff]\) times larger than the original code, 
and Proposition~\ref{prop:complexity} bounds neither this index nor the cost of Algorithm~\ref{alg:coarse}. 

Both algorithms are implemented, with exact arithmetic over \(\F\), in the open-source Python package \texttt{css-decoupling}~\cite{CSSDecoupling2026}, 
which verifies its outputs, 
including the commuting square~\eqref{eq:decoupling-square} and the action of the returned circuit. 
It solves the linear equations with Gröbner bases and by default coarse-grains less than line~\ref{line:suff} of Algorithm~\ref{alg:coarse}, using the gluing classes. 
For \(D=2\), it also implements the algorithm of Ref.~\cite{Wang2026}. 

\subsection{Logical qudits on a torus}\label{subsec:torus}
The decoupling criterion also governs finite systems. 
In two dimensions, a torus encodes the maximal number of logical qudits only when its periods are compatible with those of the anyons. 
Otherwise, it is said to be topologically frustrated \cite{Chen2025}. 
In any dimension, if the code can be decoupled on the period lattice of a torus, the decoupling circuit is compatible with the periodic boundary conditions. 
It maps the code on the torus to a toric-code stack, which encodes \(\sum_j\binom{D}{j}t_j\) logical qudits.

\begin{corollary}[Logical qudits on a torus]\label{cor:torus}
Let the layers \(E_X^j\) be finite dimensional over \(\F\), with \(t_j=\dim_{\F}E_X^j\). 
A torus with period lattice \(L\subseteq\Z^D\) encodes at most \(\sum_j\binom{D}{j}t_j\) logical qudits, with equality if and only if the code can be decoupled on \(L\).
\end{corollary}
The proof, in Appendix~\ref{apx:torus}, shows that each layer \(E_X^j\) contributes at most \(\binom{D}{j}t_j\) logical qudits and that a nontrivial translation action or a nonzero gluing on \(L\) lowers this count. 
By Theorem~\ref{thm:main} and Sec.~\ref{subsec:gluing-classes}, these are exactly the obstructions to decoupling on \(L\).

\section{Example: removing the gluing}\label{sec:example}
We return to the GC of Sec.~\ref{subsec:example-glued} and remove its gluing. 
Coarse-graining by a factor of \(p\) along a single direction already suffices, whereas Ref.~\cite{Song2026} coarse-grained in every direction for \(p=2\). 
On an \(L_1\times L_2\times L_3\) torus, the gluing also lowers the number of logical qudits unless \(p\) divides \(L_1L_2\).

\paragraph{The superlattice.}
There are \(m=2\) nonzero layers, so Algorithm~\ref{alg:coarse} coarse-grains to \(\Lsuff(\GC)=p\Z^3\). 
Coarse-graining along \(x_2\) alone, to \(\Z\times p\Z\times\Z\), already suffices; each new unit cell contains \(p\) original ones, stacked along \(x_2\).

\paragraph{The restricted gluing class.}
On this superlattice, the unit \(x_1x_2\)-plaquette consists of \(p\) original plaquettes, so its total charge is \(-p=0\) in \(\F\). 
The other two unit plaquettes project onto segments in the \(x_1x_2\)-plane and carry no charge. 
These charges are the values of the restricted gluing class (Lemma~\ref{lem:superlattice-values},
Appendix~\ref{apx:superlattice}), so by Lemma~\ref{lem:values},
\(\delta_2(\GC)\) restricts to zero and the coarse-grained profile splits as \(\F[-1]\oplus\F[-2]\). 
Coarse-graining along \(x_1\) works equally well, and index \(p\) is the smallest possible (Appendix~\ref{apx:minimal}).

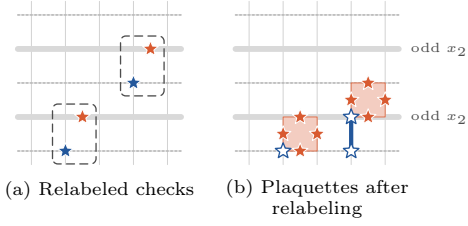
\begin{figure}[t]
\centering
\begin{tikzpicture}[x=0.45cm,y=0.45cm,line cap=round,line join=round,
 lab/.style={font=\scriptsize,align=center,text width=3cm},
 bg/.style={gray!35,line width=0.35pt}]
\newcommand{\grid}{\foreach \i in {0,...,4}{\draw[bg] (\i,-0.4)--(\i,4.4);}%
 \foreach \j in {0,...,4}{\draw[bg] (-0.4,\j)--(4.4,\j);}%
 \draw[gray!30,line width=2.2pt] (-0.4,1)--(4.4,1) (-0.4,3)--(4.4,3);%
 \draw[gray!55,densely dotted,line width=0.6pt] (-0.4,0)--(4.4,0)
  (-0.4,2)--(4.4,2) (-0.4,4)--(4.4,4);}
\begin{scope}
 \grid
 \node[xone] at (1,0) {}; \node[xtwo] at (1.5,1) {};
 \draw[regroup] (0.62,-0.38) rectangle (1.9,1.38);
 \node[xone] at (3,2) {}; \node[xtwo] at (3.5,3) {};
 \draw[regroup] (2.62,1.62) rectangle (3.9,3.38);
 \node[lab] at (2,-1.4) {(a) Relabeled checks\\\phantom{x}};
\end{scope}
\begin{scope}[shift={(6.4,0)}]
 \grid
 \filldraw[tctwo!30,draw=tctwo!80,thin] (1,0) rectangle (2,1);
 \foreach \m in {(1.5,0),(1.5,1),(1,0.5),(2,0.5)} \node[xtwo] at \m {};
 \node[xoneopen] at (1,0) {};
 \filldraw[tctwo!30,draw=tctwo!80,thin] (3,1) rectangle (4,2);
 \foreach \m in {(3.5,1),(3.5,2),(3,1.5),(4,1.5)} \node[xtwo] at \m {};
 \draw[qone] (3,0)--(3,1);
 \node[xoneopen] at (3,1) {}; \node[xoneopen] at (3,0) {};
 \node[font=\tiny,text=gray!60!black,anchor=west] at (4.5,1) {odd \(x_2\)};
 \node[font=\tiny,text=gray!60!black,anchor=west] at (4.5,3) {odd \(x_2\)};
 \node[lab] at (2,-1.4) {(b) Plaquettes after\\relabeling};
\end{scope}
\end{tikzpicture}
\caption{Removing the gluing for \(p=2\) after coarse-graining to
\(\Z\times2\Z\times\Z\) (drawn as in
Fig.~\ref{fig:glued-obstruction}; the new unit cells, one plaquette wide and two high, are bounded above and below
by dotted lines, and rows at odd \(x_2\) are shaded). (a)~Only the checks at vertices with even \(x_2\) are relabeled,
each with the check on the \(x_1\)-edge directly above it. (b)~\(Z\) on a
plaquette at even \(x_2\) (left) creates a loop without charge; \(Z\) on a
plaquette at odd \(x_2\) (right) creates two charges, which an attached
string removes.}
\label{fig:example-coarse}
\end{figure}

\paragraph{The decoupling procedure.}
Coarse-graining lifts the obstruction of Sec.~\ref{subsec:example-glued}: 
the relabeling and the strings now need to commute only with the translations in \(\Z\times p\Z\times\Z\), so they may treat the rows of a unit cell differently. 
For \(p=2\), we relabel only the checks of \(\TC_1^{(3)}(\F)\) at vertices with even \(x_2\), multiplying each by the check of \(\TC_2^{(3)}(\F)\) on the \(x_1\)-edge directly above it (Fig.~\ref{fig:example-coarse}(a)). 
Applying \(Z\) to an \(x_1x_2\)-plaquette at even \(x_2\) then creates no charge: its corner check is violated both directly and through the plaquette's top edge, and the two violations cancel since \(p=2\). 
Applying \(Z\) to an \(x_1x_2\)-plaquette at odd \(x_2\) creates charges at its corner and,
through its bottom edge, at the vertex below. 
The string on the \(x_2\)-edge joining them removes both (Fig.~\ref{fig:example-coarse}(b)).

For general \(p\), Green's theorem writes the charge of a loop \(\ell\),
which is minus the signed area it encloses, as
\[
 \oint_\ell x_2\,dx_1=\sum_\alpha\alpha_2\ell_\alpha
 \equiv\sum_\alpha(\alpha_2\bmod p)\,\ell_\alpha\ (\mathrm{mod}\ p),
\]
where the sum is the discrete version of the line integral, with \(\ell_\alpha\) the exponent of \(\ell\) on the \(x_1\)-edge at \(\alpha\).
We cancel the contribution of each \(x_1\)-edge by multiplying the check of \(\TC_1^{(3)}(\F)\) at the vertex below it by the check of \(\TC_2^{(3)}(\F)\) on the edge raised to the power \(-(\alpha_2\bmod p)\),
which repeats every \(p\) rows. 
Strings then remove the remaining charges, whose total vanishes.

\paragraph{The torus.}
Corollary~\ref{cor:torus} can be checked on this code. 
Here \(\Ltriv(\GC)=\Z^3\), yet an \(L_1\times L_2\times L_3\) torus encodes the maximal number of logical qudits, 6, if \(p\) divides \(L_1L_2\), and 4 otherwise, as found for \(p=2\) in Ref.~\cite{Song2026}. 
When \(p\nmid L_1L_2\), applying \(Z\) to an \(x_1x_2\)-plane gives a logical operator of \(\TC_2^{(3)}(\F)\), but here it also creates the total charge \(-L_1L_2\), which strings cannot remove. 
Applying \(Z\) to a closed string along \(x_3\) gives a logical operator of \(\TC_1^{(3)}(\F)\),
but here it is a stabilizer: the product of all \(Z\)-type checks on cubes is \(Z\) on all \(L_1L_2\) parallel strings along \(x_3\). 
These strings are equivalent up to stabilizers, so their product is the \(L_1L_2\)th power of one of them; since \(L_1L_2\) is invertible modulo \(p\), that string is itself a stabilizer. 
As above, the unit \(x_1x_2\)-plaquette of the period lattice consists of \(L_1L_2\) original plaquettes and carries the charge \(-L_1L_2\), while the other two carry none. 
The gluing therefore restricts to zero over the period lattice exactly when \(p\) divides \(L_1L_2\), in agreement with Corollary~\ref{cor:torus}.

\section{Discussion}\label{sec:discussion}
For fault-tolerant quantum computation, a decoupling serves as a dictionary between a code and its toric-code stack. 
The circuit, its inverse, and the map between syndromes are all local, 
so any property that such maps change only by constant factors is shared by the code and the stack. 
Single-shot decoding is one such property: it rests on the confinement of syndromes \cite{Bombin2015,Quintavalle2021}, which local maps distort only by bounded amounts. 
We therefore expect a code with finitely many excitation types in every layer to admit single-shot decoding of \(Z\)-type errors exactly when it has no pointlike \(X\)-check excitations, \(E_X^1=0\), and likewise with \(X\) and \(Z\) exchanged. 
Single-shot decoding of both error types then requires \(D\geq4\), as in the four-dimensional toric code \cite{Dennis2002}. 
The same dictionary turns toric-code decoders into decoders for the code, as for the color code \cite{KubicaDelfosse2023}: 
map the syndrome to that of the stack, decode the stack, and pull the correction back through the circuit. 
This would extend matching decoders for two-dimensional translation-invariant codes \cite{Tan2026} to higher-dimensional codes \cite{Jacob2025,Mian2026}, 
although the circuit correlates errors across the toric codes of the stack, so decoding them independently may cost performance. 

The dictionary extends to logical gates. 
On tori where a code can be decoupled, conjugation by the circuit maps its constant-depth logical gates to those of the toric-code stack, and back. 
The constant-depth gates of all such codes, non-Clifford ones included, therefore correspond to those of toric-code stacks, 
for which cup products give systematic constructions \cite{Barkeshli2023,Barkeshli2024a,Barkeshli2024b,WangYF2024,Hsin2024,Breuckmann2026}. 
For qubits, for instance, a constant-depth circuit of C\(^{D-1}Z\) gates implements a logical C\(^{D-1}Z\) gate on \(D\) copies of the \(D\)-dimensional toric code with qubits on edges \cite{WangYF2024}, 
and the transversal non-Clifford gate of the color code becomes a constant-depth gate on three toric codes \cite{Kubica2015}. 
Pulled back through a decoupling, such gates may supply non-Clifford gates for codes in which none is yet known \cite{VasmerKubica2022}. 

Up to constant factors, then, every code to which our results apply behaves like a stack of toric codes, 
and its advantages must lie in those constants and in finite geometry. 
The gross code illustrates this: 
it cannot be decoupled on the period lattice of its torus \cite{Chen2025}, 
so it encodes fewer logical qubits than a stack with the same layers (Corollary~\ref{cor:torus}), 
yet it is a leading candidate for low-overhead quantum memory \cite{Bravyi2024}. 
On such a torus, the circuit no longer fits the periodic boundary conditions, 
but it can still be applied away from a cut, leaving a toric-code stack with a domain wall along the cut, 
much as an anyon-permuting translation leaves an anyon-permuting domain wall in two dimensions. 
Decoding and logical operations on such tori may then reduce to those of toric codes with domain walls. 
Alternatively, one can choose period lattices, possibly twisted \cite{Liang2025}, on which the code can be decoupled. 

Our criterion decides whether a code can be decoupled, but not how cheaply. 
One cost is the translation symmetry that must be given up. 
When at most two layers are nonzero, as in every three-dimensional code, a single gluing class sets this cost beyond what is needed for translations to act trivially on every layer. 
In three dimensions, its values on the three unit plaquettes decide whether the gluing is absent, 
or disappears after coarse-graining by a factor of \(p\) along one direction or along two (Corollary~\ref{cor:minimal}). 
With three or more layers, possible only for \(D\geq4\), the gluing classes are no longer intrinsic: 
those of a higher layer are defined only once the lower layers have been split, and they depend on how this was done. 
Minimizing the coarse-graining layer by layer then gives only an upper bound, 
and the smallest superlattice on which the code can be decoupled is not known. 

The other cost is the circuit itself. 
Our construction aims at explicitness rather than economy: 
it does not minimize the depth or the range of the gates, 
and it adjoins ancillas (Appendix~\ref{apx:construction}). 
Yet our example can be decoupled without any (Sec.~\ref{sec:example}). 
Whether ancillas can always be avoided is open, 
as is whether retaining more translation symmetry forces a deeper or longer-range circuit. 

Our notion of equivalence is deliberately fine. 
Up to gapped interfaces and coarse-graining, toric codes are trivial, 
and in this coarser sense Yang and Yu have classified the Pauli stabilizer codes with finitely many excitation types in every layer, which they call mobile \cite{YangYu2026,RubaYang2025}. 
A decoupling retains more: it is an explicit circuit, it commutes with a prescribed superlattice of translations, and it counts the copies of each toric code. 
If all translations are kept, codes with the same layers can still differ in their gluing and then realize distinct translation-symmetry-enriched phases \cite{Song2026,Barkeshli2019}; 
the glued code, whose loops carry a charge set by the projected area they enclose, is the simplest example. 

Each of our assumptions marks a frontier. 
For composite qudit dimensions, the layers and their gluing are no longer enough: such qudits admit twisted quantum doubles \cite{Ellison2022} and Clifford obstructions that these cannot see \cite{Song2026}. 
Beyond CSS codes, a translation-invariant Clifford equivalence need not be a circuit at all, since nontrivial Clifford quantum cellular automata exist in three dimensions \cite{HaahFidkowskiHastings2023}. 
For foliated fracton codes such as the X-cube model, which have infinitely many excitation types, coarse-graining splits off layers of two-dimensional toric codes rather than removing an obstruction \cite{Shirley2018}; 
a criterion with such layers as additional building blocks might characterize foliated fracton phases \cite{Shirley2019}. 
\vspace{-\baselineskip}
\begin{acknowledgments}
This work was supported by the National Natural Science Foundation of China under Grants No.~12575022, No.~12522502, No.~12474145, and No.~12447101, the Scientific Research Innovation Capability Support Project for Young Faculty under Grant No.~ZY2025157, the Shanghai Committee of Science and Technology under Grant No.~25LZ2600800, the Tsinghua Dushi Program, and the Strategic Priority Research Program of the Chinese Academy of Sciences under Grant No.~XDB1680200.
\end{acknowledgments}
\vspace{-\baselineskip}
\section*{Data availability}
The Python package \texttt{css-decoupling}~\cite{CSSDecoupling2026} implements Algorithms~\ref{alg:coarse} and~\ref{alg:decoupling}. 
Version 0.1.0 is openly available at \url{https://github.com/hmx21/css-decoupling} and archived at \url{https://doi.org/10.5281/zenodo.23752878}. 
The Lean~4 formalization, version 0.1.0, is available at \url{https://github.com/hmx21/css-decoupling-lean} and archived at \url{https://doi.org/10.5281/zenodo.23752886}; it uses Lean v4.34.0 and Mathlib commit \texttt{5ed2965}. 
Using only the standard axioms of Lean, it proves Theorem~\ref{thm:main}, Propositions~\ref{prop:complexity} and~\ref{prop:superlattice}, Corollaries~\ref{cor:torus} and~\ref{cor:minimal}, Lemmas~\ref{lem:values}--\ref{lem:torus-layer}, the correctness of Algorithm~\ref{alg:coarse}, and the results for the glued code in Secs.~\ref{subsec:example-glued} and~\ref{sec:example}. 
It implements Algorithm~\ref{alg:decoupling} as a Lean program, for which it proves Proposition~\ref{prop:complexity}; it does not verify the Python package.

\appendix
\section{Complexes and the superselection profile}\label{apx:profile}
Section~\ref{subsec:profile} introduced the profile in the language of derived categories and stated several of its properties without proof.
This appendix fixes the conventions behind that language and proves these properties, together with a finite free model of the profile on which the other appendixes build.

\subsection{Conventions}\label{apx:conventions}
As in the main text, \(\simeq\) denotes isomorphism in the derived category \(\mathrm{D}(R)\), and \(A[n]\) is the shifted complex with \(A[n]^k=A^{k+n}\) and differential \((-1)^nd_A\). 
Thus \(A[1]\) is the translation \(TA\) of Ref.~\cite[Prop.~10.2.4]{Weibel1994}, written \(A[-1]\) there. 
The mapping cone of a cochain map \(c:A\to B\) has the terms \(B^n\oplus A^{n+1}\) and the differential \(\bigl(\begin{smallmatrix}d_B&c\\0&-d_A\end{smallmatrix}\bigr)\).
With \(\iota\) the inclusion of \(B\) and \(\delta\) the projection onto \(A[1]\), the sequence \(A\xrightarrow{c}B\xrightarrow{\iota}\operatorname{Cone}(c)\xrightarrow{\delta}A[1]\) is a distinguished triangle in \(\mathrm{D}(R)\) \cite[Def.~10.1.3 and Prop.~10.4.1]{Weibel1994}. 
Up to reordering the two summands and changing the sign of the \(A\)-component, this is the convention of Ref.~\cite{Weibel1994}.

For a cochain complex \(A\), the canonical truncations are
\[
 \begin{aligned}
 (\tau_{\geq k}A)^n&=
 \begin{cases}
 0,&n<k,\\
 A^k/\im d_A^{k-1},&n=k,\\
 A^n,&n>k,
 \end{cases}\\
 (\tau_{\leq k}A)^n&=
 \begin{cases}
 A^n,&n<k,\\
 \ker d_A^k,&n=k,\\
 0,&n>k,
 \end{cases}
 \end{aligned}
\]
with the induced differentials. 
They retain the cohomology of \(A\) in degrees \(n\geq k\) and \(n\leq k\), respectively.

For \(R\)-modules \(A\) and \(B\), regarded as complexes in degree zero,
\(\Hom_{\mathrm{D}(R)}(A,B[n])=\Ext_R^n(A,B)\) \cite[Def.~10.7.1 and Cor.~10.7.5]{Weibel1994}. 
Concretely, let \(P_\bullet\to A\) be a projective resolution with differentials \(\partial_n:P_n\to P_{n-1}\), placed with \(P_n\) in degree \(-n\). 
A cochain map \(P_\bullet\to B[n]\) has a single component \(f:P_n\to B\),
\[
\begin{tikzpicture}[baseline=(current bounding box.center),>=To,
 every node/.style={inner sep=2pt},x=1.6cm,y=1.25cm]
\foreach \x/\d in {0.2/-n-1,1.45/-n,2.7/-n+1}
 \node[cxdeg] at (\x,0.48) {\(\d\)};
\node[cxdeg,anchor=east] at (-0.95,0.48) {degree};
\node[anchor=east] at (-0.95,0) {\(P_\bullet:\)};
\node[anchor=east] at (-0.95,-1) {\(B[n]:\)};
\node (pl) at (-0.6,0) {\(\cdots\)};
\node (p1) at (0.2,0) {\(P_{n+1}\)};
\node (p0) at (1.45,0) {\(P_n\)};
\node (pm) at (2.7,0) {\(P_{n-1}\)};
\node (pr) at (3.45,0) {\(\cdots\)};
\node (al) at (-0.6,-1) {\(\cdots\)};
\node (a1) at (0.2,-1) {\(0\)};
\node (a0) at (1.45,-1) {\(B\)};
\node (am) at (2.7,-1) {\(0\)};
\node (ar) at (3.45,-1) {\(\cdots\)};
\draw[->] (pl)--(p1);
\draw[->] (p1)--node[above]{\(\partial_{n+1}\)}(p0);
\draw[->] (p0)--node[above]{\(\partial_n\)}(pm);
\draw[->] (pm)--(pr);
\draw[->] (al)--(a1);
\draw[->] (a1)--(a0);
\draw[->] (a0)--(am);
\draw[->] (am)--(ar);
\draw[->] (p1)--(a1);
\draw[->] (p0)--node[left]{\(f\)}(a0);
\draw[->] (pm)--(am);
\draw[->,dashed] (pm)--node[below right,pos=0.4,inner sep=1pt]{\(h\)}(a0);
\end{tikzpicture}
\]
and it commutes with the differentials exactly when \(f\,\partial_{n+1}=0\). 
In other words, \(f\) is an \(n\)-cocycle of \(\Hom_R(P_\bullet,B)\). 
Every class in \(\Ext_R^n(A,B)\) is represented by such a cocycle. 
Two cocycles represent the same class if and only if they differ by a coboundary \(h\,\partial_n\), with \(h:P_{n-1}\to B\) the dashed map, that is, exactly when the two cochain maps are homotopic.

\subsection{Degree bound and a finite free model}
\label{apx:models}\label{apx:degree-bound}
By Hilbert's syzygy theorem and localization, the ring \(R\) has global dimension at most \(D\) \cite[Thm.~3.2.2 and Cor.~4.3.8]{Weibel1994}.
Both results of this subsection rest on this bound: it limits the degrees of the layers, and it lets us represent the profile by a bounded complex of finite free modules, which Appendixes~\ref{apx:construction} and~\ref{apx:torus} use in explicit computations.

\begin{lemma}\label{lem:degree-bound}
For a translation-invariant topological CSS code in \(D\) dimensions,
\(E_X^j=0\) for \(j\geq D\).
\end{lemma}
\begin{proof}
By the topological condition~\eqref{eq:exactness},
\(\ker H_Z=\im H_X^\dagger\), so \(H_Z\) induces an isomorphism \(M_X=R^q/\im H_X^\dagger\cong\im H_Z\subseteq R^r\). 
Since \(R^r\) is free, the exact sequence \(0\to M_X\to R^r\to\coker H_Z\to0\) gives
\[
 \Ext_R^n(M_X,N)\cong\Ext_R^{n+1}(\coker H_Z,N),\qquad n\geq1,
\]
for every \(R\)-module \(N\). 
Since \(R\) has global dimension at most \(D\), the right-hand side vanishes for \(n\geq D\). 
Hence \(M_X\) has projective dimension at most \(D-1\). 
The antipode is a ring automorphism of \(R\), so \(\overline{M_X}\) has the same projective dimension, and \(E_X^j=\Ext_R^j(\overline{M_X},R)=0\) for \(j\geq D\).
\end{proof}
Together with the truncation in Eq.~\eqref{eq:superselection-profile},
this leaves only the degrees \(1\leq j\leq D-1\).

Since \(R\) is also Noetherian, every submodule of a finite free \(R\)-module is finitely generated and has a resolution by finite free modules of length at most \(D-1\).
Schreyer's algorithm computes such a resolution \cite[Thm.~15.10 and Cor.~15.11]{Eisenbud1995} without finding a basis of a projective module:
after its generators are multiplied by monomials, the submodule is defined over the polynomial ring \(\F[x_1,\ldots,x_D]\), where the algorithm applies,
and localizing to \(R\) preserves exactness and freeness.
This gives the following bounded model of the profile.

\begin{lemma}\label{lem:model}
The complex~\eqref{eq:physical-window} of a translation-invariant topological CSS code extends to a bounded complex of finite free \(R\)-modules, again denoted \(\Ccode\), that vanishes outside degrees \(|n|\leq D+2\), is exact in degrees \(n\leq0\), and whose part in nonnegative degrees has the form~\eqref{eq:dual-complex}. 
Then \(H^n(\Ccode)\cong E_X^n\) for \(n\geq1\) and \(\Ccode\simeq\mathscr E_X\). 
The complex \(\Cstack\) of a toric-code stack satisfies \(\Cstack\simeq\bigoplus_j\F[-j]^{\oplus t_j}\).
\end{lemma}
\begin{proof}
We extend Eq.~\eqref{eq:resolution} to a finite free resolution of \(M_X\) by resolving \(\ker\gamma^\dagger\), and we resolve \(\ker H_Z^\dagger\), both as above.
Since these resolutions have length at most \(D-1\), the complex below vanishes outside degrees \(|n|\leq D+2\).
In nonnegative degrees, we attach the complex~\eqref{eq:dual-complex} obtained from the first; in degrees below \(-1\), we attach the second. 
Then \(\Ccode\) is a bounded complex of finite free modules,
\[
\begin{tikzpicture}[baseline=(b.base),>=To,
 x=1.25cm,every node/.style={inner sep=2pt}]
\foreach \n in {-1,0,1,2} \node[cxdeg] at (\n,0.6) {\(\n\)};
\node[cxdeg,anchor=east] at (-1.6,0.6) {degree};
\node[anchor=east] at (-2.1,0) {\(\Ccode:\)};
\node (z) at (-1.85,0) {\(\cdots\)};
\node (a) at (-1,0) {\(R^r\)};
\node (b) at (0,0) {\(R^q\)};
\node (c) at (1,0) {\(R^s\)};
\node (d) at (2,0) {\(R^t\)};
\node (e) at (2.85,0) {\(\cdots\)};
\draw[->] (z)--(a);
\draw[->] (a)--node[above,cxlab]{\(H_Z^\dagger\)}(b);
\draw[->] (b)--node[above,cxlab]{\(H_X\)}(c);
\draw[->] (c)--node[above,cxlab]{\(\gamma\)}(d);
\draw[->] (d)--(e);
\end{tikzpicture}.
\]
The topological condition~\eqref{eq:exactness} gives \(\ker H_X=\im H_Z^\dagger\), and the attached resolution of \(\ker H_Z^\dagger\) makes \(\Ccode\) exact in all lower degrees. 
Hence \(H^n(\Ccode)=0\) for \(n\leq0\) and \(H^n(\Ccode)\cong E_X^n\) for \(n\geq1\).

The profile \(\mathscr E_X\) is represented by the truncation \(\tau_{\geq1}\) of the complex~\eqref{eq:dual-complex}. 
It has the same cohomology as \(\Ccode\), but its degree-one term is \(\coker\dcode^0\), which is not free in general. 
Consider the diagram
\[
\begin{tikzpicture}[baseline=(current bounding box.center),>=To,
 every node/.style={inner sep=2pt},x=1.18cm,y=1cm]
\foreach \x/\d in {1/-1,2.15/0,3.3/1,4.75/2}
 \node[cxdeg] at (\x,0.6) {\(\d\)};
\node[cxdeg,anchor=east] at (-0.35,0.6) {degree};
\node (c0) at (0,0) {\(\cdots\)};
\node (c1) at (1,0) {\(R^r\)};
\node (c2) at (2.15,0) {\(R^q\)};
\node (c3) at (3.3,0) {\(R^s\)};
\node (c4) at (4.75,0) {\(R^t\)};
\node (c5) at (5.75,0) {\(\cdots\)};
\node (s0) at (0,-1.2) {\(\cdots\)};
\node (s1) at (1,-1.2) {\(0\)};
\node (s2) at (2.15,-1.2) {\(0\)};
\node (s3) at (3.3,-1.2) {\(\coker \dcode^0\)};
\node (s4) at (4.75,-1.2) {\(R^t\)};
\node (s5) at (5.75,-1.2) {\(\cdots\)};
\node[anchor=east] at (-0.35,0) {\(\Ccode:\)};
\node[anchor=east] at (-0.35,-1.2) {\(\mathscr E_X:\)};
\draw[->] (c0)--(c1);
\draw[->] (c1)--node[above]{\(H_Z^\dagger\)}(c2);
\draw[->] (c2)--node[above]{\(H_X\)}(c3);
\draw[->] (c3)--node[above]{\(\gamma\)}(c4);
\draw[->] (c4)--(c5);
\draw[->] (s0)--(s1);
\draw[->] (s1)--(s2);
\draw[->] (s2)--(s3);
\draw[->] (s3)--node[below]{\(\tilde\gamma\)}(s4);
\draw[->] (s4)--(s5);
\draw[->] (c1)--node[right]{\(0\)}(s1);
\draw[->] (c2)--node[right]{\(0\)}(s2);
\draw[->] (c3)--node[right]{\(\pi_{\mathrm{code}}\)}(s3);
\draw[->] (c4)--node[right]{\(I\)}(s4);
\end{tikzpicture}
\]
where \(\pi_{\mathrm{code}}\) is the projection and \(\tilde\gamma[c]=\gamma c\) is well defined since \(\gamma H_X=0\).
Its vertical maps commute with the differentials because \(\pi_{\mathrm{code}}H_X=0\) and \(\tilde\gamma\pi_{\mathrm{code}}=\gamma\). 
The resulting cochain map \(\Ccode\to\mathscr E_X\) is a quasi-isomorphism, so \(\Ccode\simeq\mathscr E_X\).
Indeed, in degrees \(n\leq0\) both rows have zero cohomology. 
In degree one, both have the cohomology \(\ker\gamma/\im H_X\), on which \(\pi_{\mathrm{code}}\) induces the identity. 
From degree two on, the rows coincide, except that \(\gamma\) is replaced by \(\tilde\gamma\), which has the same image since \(\pi_{\mathrm{code}}\) is surjective.

For the toric-code stack, \(\Cstack=\bigoplus_jT_j^{\oplus t_j}\) is already a bounded complex of finite free modules. 
Applying the augmentation \(\varepsilon\) to the term \(K_0\), which lies in degree \(j\) of each copy of \(T_j\), gives a quasi-isomorphism \(\Cstack\to\bigoplus_j\F[-j]^{\oplus t_j}\), since \(K_\bullet\) is a free resolution of \(\F\) (Sec.~\ref{subsec:toric}).
\end{proof}

\subsection{Truncation triangles and gluing}\label{apx:truncation}
To decide when the profile splits, we identify each truncation triangle~\eqref{eq:triangle} with the distinguished triangle of a mapping cone, to which Lemma~\ref{lem:split-triangle} below applies.
For a cochain complex \(A\) and an integer \(j\), let \(c_j:A'=\tau_{\leq j-1}A\to A''=\tau_{\leq j}A\) be the inclusion.
The map \(\operatorname{Cone}(c_j)\to A''/A'\),
\((a'',a')\mapsto[a'']\), is a surjective cochain map whose kernel, the cone of the identity of \(A'\), is acyclic. 
The map is therefore a quasi-isomorphism. 
The quotient \(A''/A'\) is the two-term complex \(A^{j-1}/\ker d_A^{j-1}\to\ker d_A^j\) in degrees \(j-1\) and \(j\). 
Its differential is injective with cokernel \(H^j(A)\), so \(A''/A'\simeq H^j(A)[-j]\). 
Replacing the cone by \(H^j(A)[-j]\) gives the triangle
\[
 \tau_{\leq j-1}A\longrightarrow\tau_{\leq j}A\longrightarrow H^j(A)[-j]
 \xrightarrow{\delta_j}(\tau_{\leq j-1}A)[1],
\]
in which \(\delta_j\) is the inverse of the quasi-isomorphism \(\operatorname{Cone}(c_j)\to H^j(A)[-j]\) followed by the projection onto \((\tau_{\leq j-1}A)[1]\). 
For \(A=\mathscr E_X\),
this is Eq.~\eqref{eq:triangle} \cite[Example~10.4.9]{Weibel1994}.

\begin{lemma}\label{lem:split-triangle}
For a cochain map \(c:A\to B\) and its distinguished triangle \(A\xrightarrow{c}B\xrightarrow{\iota}\operatorname{Cone}(c)\xrightarrow{\delta}A[1]\),
the following are equivalent: (i)~\(\delta=0\); (ii)~\(\iota\) has a section, that is, a morphism \(\sigma:\operatorname{Cone}(c)\to B\) with \(\iota\sigma=I\); (iii)~there is an isomorphism \(A\oplus\operatorname{Cone}(c)\simeq B\) whose restriction to \(A\) is \(c\) and whose composite with \(\iota\) is the projection onto \(\operatorname{Cone}(c)\).
\end{lemma}
\begin{proof}
(i)\(\Rightarrow\)(ii). 
Applying \(\Hom_{\mathrm{D}(R)}(\operatorname{Cone}(c),-)\) to the triangle gives the exact sequence
\[
 \begin{aligned}
 \Hom(\operatorname{Cone}(c),B)&\xrightarrow{\iota\circ-}
 \Hom(\operatorname{Cone}(c),\operatorname{Cone}(c))\\
 &\xrightarrow{\delta\circ-}\Hom(\operatorname{Cone}(c),A[1])
 \end{aligned}
\]
of morphism groups in \(\mathrm{D}(R)\) \cite[Example~10.2.8]{Weibel1994}. 
Since \(\delta\circ I=\delta=0\),
exactness gives \(\sigma\) with \(\iota\sigma=I\).

(ii)\(\Rightarrow\)(iii). 
A morphism in \(\mathrm{D}(R)\) is an isomorphism if and only if it induces isomorphisms on all cohomology modules. 
The triangle gives the long exact sequence
\[
 H^n(B)\xrightarrow{H^n(\iota)}H^n(\operatorname{Cone}(c))
 \xrightarrow{H^n(\delta)}H^{n+1}(A)\to\cdots
\]
\cite[Cor.~10.1.4]{Weibel1994}. 
Since \(H^n(\iota)H^n(\sigma)=I\), every \(H^n(\iota)\) is surjective, so every connecting map \(H^n(\delta)\) vanishes. 
The sequence then breaks into short exact sequences \(0\to H^n(A)\to H^n(B)\to H^n(\operatorname{Cone}(c))\to0\), split by \(H^n(\sigma)\). 
Hence \((c\ \sigma):A\oplus\operatorname{Cone}(c)\to B\) induces isomorphisms on cohomology and is an isomorphism. 
Its restriction to \(A\) is \(c\), and its composite with \(\iota\) is \((\iota c\ \ \iota\sigma)=(0\ \ I)\), the projection onto \(\operatorname{Cone}(c)\), because consecutive maps of a distinguished triangle compose to zero.

(iii)\(\Rightarrow\)(i). 
Restricting the isomorphism to \(\operatorname{Cone}(c)\) gives a section \(\sigma\) of \(\iota\), and \(\delta=\delta\iota\sigma=0\), again because consecutive maps compose to zero.
\end{proof}

The next lemma justifies the description of a split profile in Sec.~\ref{subsec:gluing-classes}.

\begin{lemma}\label{lem:gluing}
The profile splits as in Eq.~\eqref{eq:profile-split} if and only if every \(\delta_j\) vanishes.
\end{lemma}
\begin{proof}
By the construction above,
the triangle~\eqref{eq:triangle} is isomorphic to that of the inclusion \(c_j\), with \(\operatorname{Cone}(c_j)\simeq E_X^j[-j]\) and \(\delta=\delta_j\), so Lemma~\ref{lem:split-triangle} applies to it. 
If all \(\delta_j\) vanish, the implication (i)\(\Rightarrow\)(iii) gives \(\tau_{\leq j}\mathscr E_X\simeq\tau_{\leq j-1}\mathscr E_X\oplus E_X^j[-j]\) for every \(j\), and induction on \(j\) gives Eq.~\eqref{eq:profile-split}.
Conversely, the truncations are functorial for cochain maps and preserve quasi-isomorphisms, so an isomorphism \(\mathscr E_X\simeq\bigoplus_iE_X^i[-i]\) in \(\mathrm{D}(R)\) identifies each triangle~\eqref{eq:triangle} with the corresponding triangle of \(\bigoplus_iE_X^i[-i]\), whose map onto \(E_X^j[-j]\) has a section. 
Every \(\delta_j\) then vanishes by the implication (ii)\(\Rightarrow\)(i).
\end{proof}

\subsection{Invariance of the profile}
\label{apx:invariance}
The necessity part of Theorem~\ref{thm:main} uses that a decoupling cannot change the profile. 
We check this for each of its ingredients,
together with the identity \(X\)-type checks adjoined in Appendix~\ref{apx:construction}.

\begin{lemma}\label{lem:invariance}
Up to isomorphism, the profile \(\mathscr E_X\) is unchanged by translation-invariant CSS-preserving Clifford circuits, by adjoining ancillas in the state \(\ket+\) or \(\ket0\), and by adjoining identity \(X\)-type checks.
\end{lemma}
\begin{proof}
The profile depends only on the module \(M_X=R^q/\im H_X^\dagger\) (Sec.~\ref{subsec:profile}). 
A translation-invariant CSS-preserving Clifford circuit acts on \(X\)-type representatives by an invertible matrix \((U^\dagger)^{-1}\) over \(R\) (Sec.~\ref{subsec:polynomial}).
If it maps one code onto another, it maps the \(X\)-type stabilizer submodule of the first onto that of the second and therefore induces an isomorphism of the modules \(M_X\). 
Adjoining \(a\) ancillas per unit cell in the state \(\ket+\) replaces \(H_X\) by \(\diag(H_X,I_a)\) and \(H_Z\) by \((H_Z\ 0)\). 
Then \(\im H_X^\dagger\) gains the summand \(R^a\), and \(M_X\) is unchanged. 
Adjoining them in the state \(\ket0\) replaces \(H_X\) by \((H_X\ 0)\) and \(H_Z\) by \(\diag(H_Z,I_a)\), which replaces \(M_X\) by \(M_X\oplus R^a\). 
Since \(R^a\) is free,
\(\mathbf R\Hom_R(\overline{R^a},R)\) is concentrated in degree zero and is removed by the truncation \(\tau_{\geq1}\). 
Finally, an identity \(X\)-type check adds a zero row to \(H_X\) and leaves \(\im H_X^\dagger\) unchanged.
\end{proof}

\subsection{Example: profile of the glued code}\label{apx:glued}
This subsection computes the profile and the gluing of the GC of Sec.~\ref{subsec:example-glued}, in the notation there.

\paragraph{The mapping cone.}
The block form~\eqref{eq:glued-blocks} describes the code as the two toric codes joined by the couplings. 
Algebraically, this is a mapping cone. 
Extended by zero, \(c_Z\) and \(c_X\) form a cochain map \(c:T_2[-1]\to T_1\) between the complexes of the two toric codes (Sec.~\ref{subsec:toric}):
\[
\begin{tikzpicture}[baseline=(current bounding box.center),>=To,
  x=1.12cm,y=1.2cm,every node/.style={inner sep=2pt}]
\foreach \n in {-2,-1,0,1,2,3} \node[cxdeg] at (\n,0.62) {\(\n\)};
\node[cxdeg,anchor=east] at (-2.4,0.62) {degree};
\node[anchor=east] at (-0.42,0) {\(T_2[-1]\)};
\node[anchor=east] at (-2.4,-1.2) {\(T_1\)};
\node (t3) at (0,0) {\(K_3\)};
\node (t2) at (1,0) {\(K_2\)};
\node (t1) at (2,0) {\(K_1\)};
\node (t0) at (3,0) {\(K_0\)};
\node (b3) at (-2,-1.2) {\(K_3\)};
\node (b2) at (-1,-1.2) {\(K_2\)};
\node (b1) at (0,-1.2) {\(K_1\)};
\node (b0) at (1,-1.2) {\(K_0\)};
\draw[->] (t3)--node[above,cxlab]{\(-\partial_3\)}(t2);
\draw[->] (t2)--node[above,cxlab]{\(-\partial_2\)}(t1);
\draw[->] (t1)--node[above,cxlab]{\(-\partial_1\)}(t0);
\draw[->] (b3)--node[below,cxlab]{\(\partial_3\)}(b2);
\draw[->] (b2)--node[below,cxlab]{\(\partial_2\)}(b1);
\draw[->] (b1)--node[below,cxlab]{\(\partial_1\)}(b0);
\draw[->] (t3)--node[left,cxlab]{\(c_Z\)}(b1);
\draw[->] (t2)--node[right,cxlab]{\(c_X\)}(b0);
\end{tikzpicture}.
\]
The shift by one degree reverses the sign of the differential in the top row, so the square containing \(c_Z\) and \(c_X\) commutes because \(\partial_1c_Z+c_X\partial_3=0\), 
as \(c_X\partial_3e_{123}=-w_3e_\emptyset=-\partial_1c_Ze_{123}\).

With the convention recalled in Appendix~\ref{apx:conventions}, the mapping cone \(\operatorname{Cone}(c)\) has the term \((T_1)^n\oplus(T_2)^n\) in degree \(n\) and the differential \(\bigl(\begin{smallmatrix}\partial&c\\0&\partial\end{smallmatrix}\bigr)\).
Written out, it is
\[
\begin{tikzpicture}[baseline=(current bounding box.center),>=To,
 every node/.style={inner sep=2pt}]
\foreach \n/\x in {-2/0,-1/1.75,0/3.85} \node[cxdeg] at (\x,0.6) {\(\n\)};
\foreach \n/\x in {1/1.75,2/3.85} \node[cxdeg] at (\x,-0.65) {\(\n\)};
\node[cxdeg,anchor=east] at (-0.35,0.6) {degree};
\node (a) at (0,0) {\(K_3\)};
\node (b) at (1.75,0) {\(K_2\oplus K_3\)};
\node (c) at (3.85,0) {\(K_1\oplus K_2\)};
\node (d) at (1.75,-1.25) {\(K_0\oplus K_1\)};
\node (e) at (3.85,-1.25) {\(K_0\)};
\draw[->] (a)--node[above,cxlab]{\(\left(\begin{smallmatrix}\partial_3\\0\end{smallmatrix}\right)\)}(b);
\draw[->] (b)--node[above,cxlab]{\(H_Z(\GC)^\dagger\)}(c);
\draw[->] (0.15,-1.25)--node[above,cxlab]{\(H_X(\GC)\)}(d);
\draw[->] (d)--node[above,cxlab]{\((0\ \partial_1)\)}(e);
\node[anchor=base west,inner sep=0pt] at (e.base east) {.};
\end{tikzpicture}
\]
By Eq.~\eqref{eq:glued-blocks}, its part in degrees \(-1\) to \(1\) is the complex~\eqref{eq:physical-window} of the code, and its outer maps are the local relations among the checks shown in Fig.~\ref{fig:toric-3d}.

\paragraph{The gluing class.}
Let \(\Ccode(\GC)=\operatorname{Cone}(c)\). 
The short exact sequence \(0\to T_1\to \Ccode(\GC)\to T_2\to0\) and its dual, together with Koszul exactness, show that the code is topological. 
They also show that \(\Ccode(\GC)\) is exact in degrees \(n\leq0\) and that its part in nonnegative degrees has the form~\eqref{eq:dual-complex}, so the quotient map \(\Ccode(\GC)\to\tau_{\geq1}\Ccode(\GC)\), whose target represents \(\mathscr E_X(\GC)\), is a quasi-isomorphism. 
Since \(T_j\simeq\F[-j]\), the only nonzero layers are \(E_X^1(\GC)\cong E_X^2(\GC)\cong\F\).

The inclusion \(T_1\hookrightarrow \Ccode(\GC)\) identifies \(T_1\) with \(\tau_{\leq1}\Ccode(\GC)\) in \(\mathrm D(R)\), so the rotated cone triangle is isomorphic to the truncation triangle:
\[
 \begin{array}{@{}c@{\;}c@{\;}c@{\;}c@{\;}c@{\;}c@{\;}c@{}}
  T_1&\longrightarrow&\Ccode(\GC)&\longrightarrow&
 T_2&\xrightarrow{\ -c[1]\ }&T_1[1]\\
 \big\downarrow{\scriptstyle\simeq}&&\big\downarrow{\scriptstyle\simeq}&&
 \big\downarrow{\scriptstyle\simeq}&&\big\downarrow{\scriptstyle\simeq}\\
 \F[-1]&\longrightarrow&\mathscr E_X(\GC)&\longrightarrow&
  \F[-2]&\xrightarrow{\ \delta_2(\GC)\ }&\F .
 \end{array}
\]
Thus \(\delta_2(\GC)\) corresponds to \(-c[1]\). 
Up to sign, it is represented by \(f=\varepsilon c_X:K_2\to\F\), whose plaquette values \((f_{12},f_{13},f_{23})\) are \((-1,0,0)\).

\section{Restriction to superlattices}\label{apx:coarse}
In Sec.~\ref{subsec:coarse}, we took the profile of the coarse-grained code to be the restriction of the profile of the original code; in other words, coarse-graining a code and taking its profile commute. 
This appendix justifies this and proves the results that rest on it.

\subsection{Profile of the coarse-grained code}
\label{apx:coarse-profile}
We first prove this commutation.

\begin{lemma}\label{lem:coarse-profile}
Let \(\Lambda\subseteq\Z^D\) be a superlattice. 
The coarse-grained code over \(\F[\Lambda]\) is again topological, and its profile and layers are \(\mathscr E_X|_{\F[\Lambda]}\) and \(E_X^j|_{\F[\Lambda]}\).
\end{lemma}
\begin{proof}
Let \(N=[\Z^D:\Lambda]\). 
Fix representatives \(\alpha_1,\ldots,\alpha_N\) of \(\Z^D/\Lambda\). 
They label the original unit cells within one unit cell of \(\Lambda\). 
Sorting the monomials of a Laurent polynomial by the coset of their exponents gives
\[
 R=\bigoplus_{k=1}^N\F[\Lambda]\,x^{\alpha_k},
\]
so \(R\) is a free \(\F[\Lambda]\)-module with basis \(x^{\alpha_1},\ldots,x^{\alpha_N}\). 
Restriction of scalars keeps a module and its maps but retains only the action of \(\F[\Lambda]\). 
It is exact and sends \(R^q\) to \(\F[\Lambda]^{Nq}\): the \(q\) qudits of each original unit cell become \(Nq\) qudits per unit cell of \(\Lambda\). 
Replacing each entry \(h\) of a matrix \(H\) over \(R\) by the \(N\times N\) matrix over \(\F[\Lambda]\) of multiplication by \(h\) in this basis gives the matrix \(H|_{\F[\Lambda]}\) of the restricted map. 
The check matrices of the coarse-grained code are \(H_X|_{\F[\Lambda]}\) and \(H_Z|_{\F[\Lambda]}\) \cite{Haah2013}. 
For example, for \(D=1\) and \(\Lambda=2\Z\) with representatives \(0\) and \(1\), multiplication by \(x\) sends \(1\mapsto x\) and \(x\mapsto x^2\cdot1\), so \(x|_{\F[\Lambda]}=\bigl(\begin{smallmatrix}0&x^2\\1&0\end{smallmatrix}\bigr)\): a unit translation moves the first original cell of a coarse-grained cell to the second, and the second to the first original cell of the next coarse-grained cell.

The profile is defined through the dagger and through duals into \(R\),
so we need to check that restriction is compatible with both. 
For the dagger, recall that the coefficient of \(x^\alpha\) in \(u^\dagger v\) is the commutation exponent of \(Z(v)\) with the translate of \(X(u)\) by \(\alpha\) (Sec.~\ref{subsec:polynomial}). 
The coarse-grained code sees only the translates by \(\alpha\in\Lambda\), so its symplectic form should keep only these coefficients. 
Accordingly, let \(\theta:R\to\F[\Lambda]\) keep the monomials with exponents in \(\Lambda\) and discard the others.
In the example above,
\(\theta(a_0+a_1x+a_2x^2+a_3x^3)=a_0+a_2x^2\). 
The map \(\theta\) is \(\F[\Lambda]\)-linear, commutes with the antipode because \(\Lambda=-\Lambda\), and satisfies \(\theta(x^{-\alpha_k}x^{\alpha_l})=1\) if \(k=l\) and \(0\) otherwise, since \(\alpha_l-\alpha_k\in\Lambda\) only for \(k=l\). 
Writing \(u=\sum_k x^{\alpha_k}u_k\) and \(v=\sum_l x^{\alpha_l}v_l\) with \(u_k,v_l\in\F[\Lambda]^q\), we obtain
\[
 \theta(u^\dagger v)=\sum_k u_k^\dagger v_k,
\]
the standard form in the expanded coordinates, as expected. 
Applying \(\theta\) to \((Hu)^\dagger v=u^\dagger(H^\dagger v)\) therefore shows that \((H|_{\F[\Lambda]})^\dagger=(H^\dagger)|_{\F[\Lambda]}\). 
Consequently, the coarse-grained check matrices satisfy the topological condition~\eqref{eq:exactness} over \(\F[\Lambda]\), and the coarse-grained code has the module \(M_X|_{\F[\Lambda]}\).

For duals, let \(P\) be a finite free \(R\)-module with basis \(e_i\).
Composition with \(\theta\) gives an \(\F[\Lambda]\)-linear map
\[
 \begin{aligned}
 \Hom_R(\overline P,R)|_{\F[\Lambda]}&\longrightarrow
 \Hom_{\F[\Lambda]}\bigl(\overline{P|_{\F[\Lambda]}},\F[\Lambda]\bigr),\\
 f&\longmapsto\theta\circ f,
 \end{aligned}
\]
natural in \(P\). 
It is bijective. 
An \(R\)-linear map \(f\) is determined by the elements \(f(e_i)\in R\), each of which has \(N\) coordinates in \(\F[\Lambda]\), \(f(e_i)=\sum_l x^{\alpha_l}f_{il}\). 
An \(\F[\Lambda]\)-linear map on \(\overline{P|_{\F[\Lambda]}}\) is determined by its values on the \(\F[\Lambda]\)-basis \(x^{\alpha_k}e_i\). 
Since \(R\) acts on \(\overline P\) through the antipode,
\((\theta\circ f)(x^{\alpha_k}e_i)=\theta(x^{-\alpha_k}f(e_i)) =f_{ik}\), so the map \(f\mapsto\theta\circ f\) matches the two sets of data.

Finally, restricting the resolution~\eqref{eq:resolution} term by term gives a free \(\F[\Lambda]\)-resolution of \(M_X|_{\F[\Lambda]}\), because restriction is exact and sends finite free modules to finite free modules. 
Since \((H_X^\dagger)|_{\F[\Lambda]}=(H_X|_{\F[\Lambda]})^\dagger\), it begins with the checks of the coarse-grained code, as Eq.~\eqref{eq:resolution} does for the original one. 
Applying the isomorphism above term by term identifies its dual with the restriction of the complex~\eqref{eq:dual-complex}. 
The identification respects the differentials because the isomorphism is natural. 
Since restriction is exact, it commutes with truncation and cohomology, so the profile and layers of the coarse-grained code are \(\mathscr E_X|_{\F[\Lambda]}\) and \(E_X^j|_{\F[\Lambda]}\).
\end{proof}

\subsection{Gluing classes on a superlattice}
\label{apx:superlattice}
Sections~\ref{subsec:gluing-classes} and~\ref{subsec:inter-layer-extensions} described a gluing class by its values on unit cells and stated that coarse-graining by a factor of \(p\) in every direction removes it. 
We now prove these statements and compute the values on a general superlattice, starting with Lemma~\ref{lem:values}, which we restate. 
Throughout this subsection,
\(R=\F[\Ltriv]\) as in Sec.~\ref{subsec:translation-action}, and \(A\) and \(B\) are \(R\)-modules that are finite dimensional over \(\F\),
with \(\mathfrak m A=\mathfrak m B=0\).

\begin{restated}{Lemma~\ref{lem:values}}
Every class \(\kappa\in\Ext_R^n(A,B)\) is represented, on the free resolution \(K_\bullet\otimes_{\F}A\) of \(A\), by a unique \(R\)-linear map \(f:K_n\otimes_{\F}A\to B\), and
\[
 \Ext_R^n(A,B)\cong\Hom_R\bigl(K_n\otimes_{\F}A,B\bigr)
 \cong\bigoplus_{|I|=n}\Hom_{\F}(A,B),
\]
where the second isomorphism sends \(f\) to the maps \(f_I:a\mapsto f(e_I\otimes a)\). 
Since translations act trivially on \(B\), the class \(\kappa\) thus assigns the value \(f_I\) to every unit \(n\)-cell of orientation \(I\).
\end{restated}
\begin{proof}
Since \(R\) acts on \(A\) through the augmentation, the Koszul resolution of \(\F\) gives the free resolution \(K_\bullet\otimes_{\F}A\) of \(A\), with \(R\) acting on the first factor. 
Its terms \(K_k\otimes_{\F}A\cong K_k^{\dim_{\F}A}\) have the basis \(e_I\otimes a\),
with \(a\) running over a basis of \(A\), and its augmentation \(r\otimes a\mapsto\varepsilon(r)a\) is \(R\)-linear. 
By Appendix~\ref{apx:conventions}, \(\Ext_R^n(A,B)\) consists of the \(R\)-linear maps \(f:K_n\otimes_{\F}A\to B\) with \(f\partial_{n+1}=0\), modulo the maps \(h\partial_n\). 
Each term of \(\partial\) in Eq.~\eqref{eq:koszul} multiplies by some \(w_i\), which acts as zero on \(B\). 
Hence \(f\partial_{n+1}=0\) for every \(f\) and \(h\partial_n=0\) for every \(h\), so \(\Ext_R^n(A,B)\) is the group of all \(R\)-linear maps \(f\). 
Such a map can take arbitrary values on the basis \(e_I\otimes a\). 
For each \(I\), its values on the \(e_I\otimes a\) form the linear map \(f_I\in\Hom_{\F}(A,B)\).
\end{proof}

For a superlattice \(\Lambda\subseteq\Ltriv\), translations in \(\Lambda\) still act trivially on \(A\) and \(B\). 
By Lemma~\ref{lem:values} over \(\F[\Lambda]\), a class restricted to \(\F[\Lambda]\) is therefore again a table of values, now on the unit cells of \(\Lambda\).

Let \(\kappa\in\Ext_R^n(A,B)\) with \(n>0\), and let \(f_I\) be its values from Lemma~\ref{lem:values}. 
For \(\nu_1,\ldots,\nu_n\in\Ltriv\), let \(\det_I(\nu_1,\ldots,\nu_n)\) be the \(n\times n\) minor of their coordinates, with respect to \(\beta_1,\ldots,\beta_D\), formed by the directions in \(I\).
Geometrically, it is the signed volume of the projection of the parallelotope spanned by the \(\nu_k\) onto these directions. 
We define \(\kappa(\nu_1,\ldots,\nu_n)\in\Hom_{\F}(A,B)\) by
\begin{equation}
 \kappa(\nu_1,\ldots,\nu_n)
 =\sum_{|I|=n}{\det}_I(\nu_1,\ldots,\nu_n)\,f_I.
 \label{eq:kappa-values}
\end{equation}

The next lemma computes the values of the restricted class in terms of this function.

\begin{lemma}\label{lem:superlattice-values}
Let \(\kappa\in\Ext_R^n(A,B)\) with \(n>0\), with values \(f_I\) as in Lemma~\ref{lem:values}, and let \(\Lambda\subseteq\Ltriv\) be a superlattice with basis \(\mu_1,\ldots,\mu_D\). 
Let \(K'_\bullet\) be the Koszul complex over \(\F[\Lambda]\) built from the \(x^{\mu_j}-1\) as in Sec.~\ref{subsec:toric}, with cells \(e'_J\). 
On the free resolution \(K'_\bullet\otimes_{\F}A\) of \(A|_{\F[\Lambda]}\), the restriction of \(\kappa\) to \(\F[\Lambda]\) takes the value
\[
 \kappa(\mu_{j_1},\ldots,\mu_{j_n})
 =\sum_{|I|=n}{\det}_I(\mu_{j_1},\ldots,\mu_{j_n})\,f_I
\]
on every cell \(e'_J\) with \(J=\{j_1<\cdots<j_n\}\). 
Here \({\det}_I\) is the \(n\times n\) minor of the coordinates of \(\mu_{j_1},\ldots,\mu_{j_n}\) with respect to \(\beta_1,\ldots,\beta_D\), formed by the directions in \(I\).
\end{lemma}
\begin{proof}
Let \(f\) represent \(\kappa\) as in Lemma~\ref{lem:values}. 
Since \(R\) is free over \(\F[\Lambda]\) (proof of Lemma~\ref{lem:coarse-profile}),
\(K_\bullet|_{\F[\Lambda]}\otimes_{\F}A\) is a free \(\F[\Lambda]\)-resolution of \(A|_{\F[\Lambda]}\), on which \(f\) still represents the restriction of \(\kappa\). 
To read off its values on the cells \(e'_J\), we move it to \(K'_\bullet\otimes_{\F}A\) along a chain map \(\chi:K'_\bullet\to K_\bullet|_{\F[\Lambda]}\) that lifts the identity of \(\F\).
Such a \(\chi\) writes each cell of \(\Lambda\) as a chain of unit cells of \(\Ltriv\).

We build \(\chi\) degree by degree. 
In degree zero, \(\chi\) is the inclusion \(\F[\Lambda]\subseteq R\). 
In degree one, \(\chi\) coarse-grains edges.
Let \(\ell_j\in K_1\) be the lattice path from the origin to \(\mu_j\) that takes \(\mu_{j,1}\) steps along \(\beta_1\), then \(\mu_{j,2}\) steps along \(\beta_2\), and so on; a step from \(\alpha\) to \(\alpha+\beta_i\) contributes \(x^\alpha e_i\), and a step from \(\alpha\) to \(\alpha-\beta_i\) contributes \(-x^{\alpha-\beta_i}e_i\).
We set \(\chi(e'_j)=\ell_j\), with \(e'_j=e'_{\{j\}}\). 
In higher degrees, \(\chi\) is the exterior product of these paths,
\[
 \chi(e'_{j_1}\wedge\cdots\wedge e'_{j_n})
 =\ell_{j_1}\wedge\cdots\wedge\ell_{j_n},\qquad j_1<\cdots<j_n,
\]
extended \(\F[\Lambda]\)-linearly; geometrically, the right-hand side is the prism spanned by the paths.

The intermediate vertices of \(\ell_j\) cancel in its boundary, so \(\partial\ell_j=x^{\mu_j}-1=\partial'e'_j\). 
The differential~\eqref{eq:koszul} is the derivation of the exterior algebra \(K_\bullet\) with \(\partial e_i=w_i\), so it obeys the Leibniz rule \(\partial(u\wedge v)=(\partial u)\wedge v+(-1)^ku\wedge\partial v\) for \(u\in K_k\), and likewise for \(\partial'\). 
Since each \(\partial\ell_j\) is a scalar in \(K_0=R\), it follows that \(\partial\chi(e'_J)=\sum_k(-1)^{k-1}(x^{\mu_{j_k}}-1)\,
\chi(e'_{J\setminus\{j_k\}})=\chi(\partial'e'_J)\). 
Thus \(\chi\) is a chain map, and in degree zero it lifts the identity of \(\F\). 
By the comparison theorem \cite[Thm.~2.2.6]{Weibel1994}, the restriction of \(\kappa\) is represented on \(K'_\bullet\otimes_{\F}A\) by \(f(\chi\otimes I_A)\). 
By Lemma~\ref{lem:values} over \(\F[\Lambda]\), it is the table of the values \(a\mapsto f(\chi(e'_J)\otimes a)\) on the cells \(e'_J\).

It remains to evaluate \(f\) on \(\chi(e'_J)\). 
Since translations act trivially on \(B\), \(f(x^\alpha e_I\otimes a)=f_I(a)\) for every \(\alpha\), so \(f\) evaluated on a chain in \(K_n\) depends only on the number of cells of each orientation \(I\) in it, counted with their coefficients. 
Setting every \(x_i\) to \(1\), that is, applying \(\varepsilon\) to the coefficients, computes these numbers, respects the wedge product, and turns each path \(\ell_j\) into its net displacement \(\sum_i\mu_{j,i}e_i\). 
The numbers for \(\chi(e'_J)\) are the coefficients of
\[
 \begin{aligned}
 &\Bigl(\sum_i\mu_{j_1,i}e_i\Bigr)\wedge\cdots\wedge
 \Bigl(\sum_i\mu_{j_n,i}e_i\Bigr)\\
 &\qquad=\sum_{|I|=n}{\det}_I(\mu_{j_1},\ldots,\mu_{j_n})\,e_I.
 \end{aligned}
\]
The restricted class therefore takes the value \(\sum_I\det_I(\mu_{j_1},\ldots,\mu_{j_n})\,f_I =\kappa(\mu_{j_1},\ldots,\mu_{j_n})\) on \(e'_J\).
\end{proof}

\begin{restated}{Lemma~\ref{lem:ext-zero}}
Let \(A\) and \(B\) be \(R\)-modules that are finite dimensional over \(\F\), with \(\mathfrak m A=\mathfrak m B=0\). 
For every \(n>0\), every class in \(\Ext_R^n(A,B)\) restricts to zero over \(\F[p\Ltriv]\).
\end{restated}
\begin{proof}
Apply Lemma~\ref{lem:superlattice-values} with \(\Lambda=p\Ltriv\) and the basis \(\mu_j=p\beta_j\). 
For \(n>0\), every \(\det_I(\mu_{j_1},\ldots,\mu_{j_n}) =p^n\det_I(\beta_{j_1},\ldots,\beta_{j_n})\) vanishes in \(\F\), so every value of the restricted class is zero. 
By Lemma~\ref{lem:values} over \(\F[p\Ltriv]\), the restriction is zero.
\end{proof}

\subsection{Minimizing the coarse-graining}\label{apx:minimal}
Line~\ref{line:suff} of Algorithm~\ref{alg:coarse} coarse-grains by a factor of \(p\) in every direction once for each layer it detaches, which may be more than necessary. 
We now determine the superlattices \(\Lambda\subseteq\Ltriv\) over which a given gluing class restricts to zero.

Let \(\pi:\Ltriv\to\F^D\) send a lattice vector to its coordinates,
with respect to \(\beta_1,\ldots,\beta_D\), modulo \(p\); its kernel is \(p\Ltriv\). 
The value \(\kappa(\nu_1,\ldots,\nu_n)\) depends only on \(\pi(\nu_1),\ldots,\pi(\nu_n)\). 
We say that \(\kappa\) \emph{vanishes on a subspace} \(W\subseteq\F^D\) if \(\kappa(\nu_1,\ldots,\nu_n)=0\) whenever \(\pi(\nu_k)\in W\) for all \(k\).

\begin{proposition}[Vanishing criterion for gluing]
\label{prop:superlattice}
Let \(A\) and \(B\) be as above, let \(\kappa\in\Ext_R^n(A,B)\) with \(n>0\), and let \(\Lambda\subseteq\Ltriv\) be a superlattice. 
The class \(\kappa\) restricts to zero over \(\F[\Lambda]\) if and only if \(\kappa\) vanishes on \(\pi(\Lambda)\). 
Moreover, let finitely many classes \(\kappa_a\in\Ext_R^{n_a}(A_a,B_a)\) be given, with \(n_a>0\) and \(R\)-modules \(A_a\) and \(B_a\) that are finite dimensional over \(\F\) and annihilated by \(\mathfrak m\).
Then
\begin{equation}
 \min_\Lambda\,[\Ltriv:\Lambda]=p^{c},\qquad c=\min_W\,(D-\dim_{\F}W),
 \label{eq:min-index}
\end{equation}
where \(\Lambda\) runs over the superlattices over which every \(\kappa_a\) restricts to zero, and \(W\) over the subspaces of \(\F^D\) on which every \(\kappa_a\) vanishes.
\end{proposition}

\begin{proof}
Let \(\mu_1,\ldots,\mu_D\) be a basis of \(\Lambda\). 
By Lemma~\ref{lem:superlattice-values}, the restriction of \(\kappa\) to \(\F[\Lambda]\) takes the values \(\kappa(\mu_{j_1},\ldots,\mu_{j_n})\) on the cells \(e'_J\), and by Lemma~\ref{lem:values} over \(\F[\Lambda]\), it vanishes exactly when all these values vanish. 
The value \(\kappa(\nu_1,\ldots,\nu_n)\) depends only on \(\pi(\nu_1),\ldots,\pi(\nu_n)\) and is multilinear and alternating in them, because every \(\det_I\) is. 
Every vector of \(\pi(\Lambda)\) is a linear combination of \(\pi(\mu_1),\ldots,\pi(\mu_D)\). 
Expanding by multilinearity and alternation, every value \(\kappa(\nu_1,\ldots,\nu_n)\) with all \(\pi(\nu_k)\in\pi(\Lambda)\) is a linear combination of the values \(\kappa(\mu_{j_1},\ldots,\mu_{j_n})\) with \(j_1<\cdots<j_n\). 
Hence \(\kappa\) vanishes on \(\pi(\Lambda)\) if and only if \(\kappa(\mu_{j_1},\ldots,\mu_{j_n})=0\) for all \(j_1<\cdots<j_n\).

For the second part, let \(c\) be as in Eq.~\eqref{eq:min-index}.
First, let \(\Lambda\) be a superlattice over which every \(\kappa_a\) restricts to zero. 
Its image \(\pi(\Lambda)\) is a subgroup, hence a subspace, of \(\F^D\). 
Since \(\pi\) is surjective,
composing it with the quotient map \(\F^D\to\F^D/\pi(\Lambda)\) gives a surjection \(\Ltriv\to\F^D/\pi(\Lambda)\) whose kernel contains \(\Lambda\). 
It induces a surjection \(\Ltriv/\Lambda\to\F^D/\pi(\Lambda)\), so \([\Ltriv:\Lambda]\geq|\F^D/\pi(\Lambda)|=p^{D-\dim_{\F}\pi(\Lambda)}\). 
By the first part, every \(\kappa_a\) vanishes on \(\pi(\Lambda)\), so \(\pi(\Lambda)\) is one of the subspaces \(W\) in Eq.~\eqref{eq:min-index}, and \(D-\dim_{\F}\pi(\Lambda)\geq c\). 
Hence \([\Ltriv:\Lambda]\geq p^c\).

Conversely, let \(W\) be a subspace on which every \(\kappa_a\) vanishes,
with \(D-\dim_{\F}W=c\), and let \(\Lambda=\pi^{-1}(W)\). 
It contains \(\ker\pi=p\Ltriv\), so it is a superlattice, and \(\pi\) induces an isomorphism \(\Ltriv/\Lambda\cong\F^D/W\), so \([\Ltriv:\Lambda]=p^{D-\dim_{\F}W}=p^c\). 
Since \(\pi\) is surjective,
\(\pi(\Lambda)=W\), so every \(\kappa_a\) vanishes on \(\pi(\Lambda)\) and, by the first part, restricts to zero over \(\F[\Lambda]\). 
The minimum in Eq.~\eqref{eq:min-index} is therefore \(p^c\).
\end{proof}

For \(\Lambda=p\Ltriv\),
\(\pi(\Lambda)=0\), and the proposition recovers Lemma~\ref{lem:ext-zero}.
In contrast, a superlattice of index coprime to \(p\) has \(\pi(\Lambda)=\F^D\), so coarse-graining to it cannot remove a nonzero gluing class. 
At each step of the layer-by-layer removal of Sec.~\ref{subsec:inter-layer-extensions}, one may therefore coarse-grain to any superlattice over which all gluing classes \(\kappa_{ji}\) of that step restrict to zero, instead of by \(p\) in every direction, with the current lattice in the role of \(\Ltriv\).

\begin{corollary}[Minimal coarse-graining]\label{cor:minimal}
Let the layers \(E_X^j\) be finite dimensional over \(\F\), at most two of them nonzero, as in every three-dimensional code, and let \(\kappa\) be the gluing class~\eqref{eq:attachments} between them, with \(\kappa=0\) if fewer than two are nonzero. 
Then the code can be decoupled on a superlattice \(\Lambda\) if and only if \(\Lambda\subseteq\Ltriv\) and \(\kappa\) vanishes on \(\pi(\Lambda)\). 
In three dimensions, the smallest such index \([\Ltriv:\Lambda]\) is \(1\), \(p\), or \(p^2\), depending on whether the plaquette values \(f_{12}\), \(f_{13}\),
\(f_{23}\in\Hom_{\F}(E_X^2,E_X^1)\) are all zero, linearly dependent but not all zero, or linearly independent.
\end{corollary}
\begin{proof}
With two nonzero layers, \(\kappa\) is the only gluing class, so the first statement follows from Theorem~\ref{thm:main},
Sec.~\ref{subsec:gluing-classes}, and Proposition~\ref{prop:superlattice}. 
For \(D=3\), \(\kappa\) vanishes on every line and, by Eq.~\eqref{eq:kappa-values}, on the plane spanned by \(\pi(\nu_1)\) and \(\pi(\nu_2)\) exactly when \(\sum_I\det_I(\nu_1,\nu_2)\,f_I=0\). 
As \(\nu_1\) and \(\nu_2\) vary,
\(\bigl(\det_I(\nu_1,\nu_2)\bigr)_I\) takes every nonzero value in \(\F^3\),
so \(c\) in Eq.~\eqref{eq:min-index} is \(0\), \(1\), or \(2\) in the three cases.
\end{proof}

\section{Construction of the decoupling circuit}\label{apx:construction}
Section~\ref{subsec:algorithm} sketched how to construct a decoupling circuit when, after coarse-graining, the profile of the code is isomorphic to that of a toric-code stack. 
This appendix carries out that construction and thereby proves Theorem~\ref{thm:main}, which we restate.

\begin{restated}{Theorem~\ref{thm:main}}[Decoupling criterion]
Consider a translation-invariant topological \(p\)-ary CSS code on \(\Z^D\), \(D\geq2\), and let \(\Lambda\subseteq\Z^D\) be a superlattice. 
The code can be decoupled on \(\Lambda\) if and only if its profile over \(\F[\Lambda]\) is that of a toric-code stack,
\begin{equation}
 \mathscr E_X|_{\F[\Lambda]}\simeq\bigoplus_{j=1}^{D-1}\F[-j]^{\oplus t_j}
 \tag{\ref{eq:criterion}}
\end{equation}
in \(\mathrm{D}(\F[\Lambda])\), for some nonnegative integers \(t_j\). 
In that case, a decoupling onto \(\bigoplus_j\TC_j^{(D)}(\F)^{\oplus t_j}\) can be realized by a finite-depth local circuit of SUM and SWAP gates, possibly after adding ancillas.
\end{restated}

Necessity was shown in Sec.~\ref{subsec:criterion}, so we prove sufficiency. 
After coarse-graining to \(\Lambda\), the code is again topological, with profile \(\mathscr E_X|_{\F[\Lambda]}\) (Lemma~\ref{lem:coarse-profile}), and a basis of \(\Lambda\) identifies \(\F[\Lambda]\) with \(R\). 
We may therefore assume that \(\Lambda=\Z^D\) and that
\begin{equation}
 \mathscr E_X\simeq\bigoplus_{j=1}^{D-1}\F[-j]^{\oplus t_j}
 \label{eq:toric-profile-hypothesis}
\end{equation}
in \(\mathrm{D}(R)\). 
By the claim in Sec.~\ref{subsec:polynomial}, it then suffices to construct invertible \(U\) and \(V\) that make the square~\eqref{eq:decoupling-square} commute, with \(U\) realized by a circuit. 
Ancillas and identity \(X\)-type checks may be adjoined to the code and to the stack. 
Such \(U\) and \(V\) induce an isomorphism between the check cokernels of \(\dcode^0=H_X\) and \(\dstack^0=\bigoplus_j\partial_j^{\oplus t_j}\), since \(V(\im\dcode^0)=\im(\dstack^0 U)=\im\dstack^0\). 
Our construction reverses this, in three steps. 
First, we match the check cokernels up to free summands (Lemma~\ref{lem:stable-cokernel}); this is the only step that uses hypothesis~\eqref{eq:toric-profile-hypothesis}. 
Second, we lift this match to invertible \(U\) and \(V\) (Lemma~\ref{lem:invertible-lift}).
Third, we factor \(U\) into layers of SUM and SWAP gates (Lemma~\ref{lem:gates}). 
The proof of Lemma~\ref{lem:stable-cokernel},
the most technical step, is deferred to Appendix~\ref{apx:beta}. 
Appendix~\ref{apx:complexity} bounds the cost of the construction.

\subsection{Matching the check cokernels}\label{apx:cokernels}
The check cokernel of the code is \(\coker\dcode^0=R^s/\im H_X\), and that of the stack is the direct sum of those of the toric codes,
\[
 \begin{aligned}
 \coker\dstack^0
 &=\bigoplus_{j=1}^{D-1}\bigl(K_{j-1}/\im\partial_j\bigr)^{\oplus t_j}\\
 &\cong\bigoplus_{j=1}^{D-1}\bigl(\im\partial_{j-1}\bigr)^{\oplus t_j},
 \end{aligned}
\]
where \(\partial_0=\varepsilon\) and the isomorphism follows from the exactness of the augmented Koszul complex (Sec.~\ref{subsec:toric}). 
The module \(\coker\dcode^0\) is the degree-one term of the truncation \(\tau_{\geq1}\Ccode\), which represents the profile \(\mathscr E_X\) (Lemma~\ref{lem:model}). 
We write \(\pi_{\mathrm{code}}:\Ccode^1\to\coker\dcode^0\) and \(\pi_{\mathrm{stack}}:\Cstack^1\to\coker\dstack^0\) for the quotient maps.

\begin{lemma}\label{lem:stable-cokernel}
Under hypothesis~\eqref{eq:toric-profile-hypothesis}, there exist nonnegative integers \(a,b\) and an isomorphism
\[
  \phi^2:\coker \dstack^0\oplus R^b\xrightarrow{\sim}\coker \dcode^0\oplus R^a.
\]
\end{lemma}

We outline the proof, which is given in Appendix~\ref{apx:beta}.
By Lemma~\ref{lem:model} and hypothesis~\eqref{eq:toric-profile-hypothesis},
\(\Ccode\) and \(\Cstack\) are quasi-isomorphic bounded complexes of finite free modules. 
The quasi-isomorphism \(\Cstack\to\Ccode\) can be represented by cochain maps \(\Gamma:\Cstack\to\Ccode\) and \(\Delta:\Ccode\to\Cstack\) that are mutually inverse up to homotopy \cite[Cor.~10.4.7]{Weibel1994}.
As cochain maps, \(\Gamma\) and \(\Delta\) induce maps \(\Gamma_*\) and \(\Delta_*\) between the check cokernels. 
The composites \(\Gamma_*\Delta_*\) and \(\Delta_*\Gamma_*\) differ from the identity because of the homotopy, but only by maps that pass through the free modules \(\Ccode^2\) and \(\Cstack^2\), respectively.
Adjoining \(\Ccode^2\) to \(\coker\dstack^0\) absorbs the first of these corrections: \(\coker\dcode^0\) becomes a direct summand of \(\coker\dstack^0\oplus\Ccode^2\). 
The second correction shows that the complementary summand is stably free, which suffices. 
A similar argument appears in Ref.~\cite{LiWilliamson2026} for codes with a single nonzero layer.

Note that the free summands in Lemma~\ref{lem:stable-cokernel} cannot be avoided: adjoining an identity \(X\)-type check adds a summand \(R\) to the check cokernel but leaves the profile unchanged (Lemma~\ref{lem:invariance}). 
We absorb them with disk complexes. 
For a finite free \(R\)-module \(P\), the disk complex \(D^j(P)\) has \(P\) in degrees \(j\) and \(j+1\), joined by the identity \(I_P\). 
Adjoining \(D^{-1}(P)\) or \(D^0(P)\) to a code adds ancillas in the state \(\ket0\) or \(\ket+\), respectively, one for each basis element of \(P\) per unit cell. 
Adjoining \(D^1(P)\) adds identity \(X\)-type checks.
We adjoin \(D^1(R^a)\) to \(\Ccode\) and \(D^1(R^b)\) to \(\Cstack\). 
This replaces \(\dcode^0\) by \(\bigl(\begin{smallmatrix}H_X\\0\end{smallmatrix}\bigr):R^q\to R^s\oplus R^a\) and \(\dstack^0\) likewise, so the two check cokernels gain the free summands \(R^a\) and \(R^b\), respectively. 
We keep the notation \(\Ccode\) and \(\Cstack\) for the enlarged complexes, so that \(\phi^2:\coker \dstack^0\to\coker \dcode^0\) is an isomorphism.

\subsection{Lifting to invertible maps}\label{apx:lifting}
The second step lifts the isomorphism of Lemma~\ref{lem:stable-cokernel} from the check cokernels to the terms in degrees zero and one, and adjoins ancillas to make the lifts invertible.

\begin{lemma}\label{lem:invertible-lift}
Let \(\phi^2:\coker \dstack^0\to\coker \dcode^0\) be an isomorphism, where \(\Ccode^0\), \(\Ccode^1\), \(\Cstack^0\), and \(\Cstack^1\) are finite free. 
After ancillas are adjoined to the code and to the stack, there are isomorphisms \(U\) and \(V\) between their terms in degrees zero and one that make the square~\eqref{eq:decoupling-square} commute.
\end{lemma}
\begin{proof}
We first lift \(\phi^2\) and its inverse \(\psi^2\) to the terms in degrees one and zero. 
Since \(\Ccode^0,\Ccode^1,\Cstack^0,\Cstack^1\) are finite free and the rows of the diagram below are exact at the degree-one terms and at the cokernels, the comparison theorem gives such lifts, unique up to homotopy \cite[Thm.~2.2.6 and Porism~2.2.7]{Weibel1994}. 
Writing the lifts as \(\phi^i:\Cstack^i\to \Ccode^i\) and \(\psi^i:\Ccode^i\to \Cstack^i\) for \(i=0,1\), we obtain the diagram
\begin{equation}
\begin{tikzpicture}[baseline=(current bounding box.center),>=To,
 every node/.style={inner sep=2pt},x=1cm,y=1cm]
\node (t0) at (0,0) {\(\Cstack^0\)};
\node (t1) at (2.15,0) {\(\Cstack^1\)};
\node (t2) at (4.55,0) {\(\coker \dstack^0\)};
\node (tz) at (6.2,0) {\(0\)};
\node (c0) at (0,-1.25) {\(\Ccode^0\)};
\node (c1) at (2.15,-1.25) {\(\Ccode^1\)};
\node (c2) at (4.55,-1.25) {\(\coker \dcode^0\)};
\node (cz) at (6.2,-1.25) {\(0\)};
\draw[->] (t0)--node[above]{\(\dstack^0\)}(t1);
\draw[->] (t1)--node[above]{\(\pi_{\mathrm{stack}}\)}(t2);
\draw[->] (t2)--(tz);
\draw[->] (c0)--node[below]{\(\dcode^0\)}(c1);
\draw[->] (c1)--node[below]{\(\pi_{\mathrm{code}}\)}(c2);
\draw[->] (c2)--(cz);
\draw[->] ([xshift=-4pt]t0.south)--node[left]{\(\phi^0\)}([xshift=-4pt]c0.north);
\draw[->] ([xshift=4pt]c0.north)--node[right]{\(\psi^0\)}([xshift=4pt]t0.south);
\draw[->] ([xshift=-4pt]t1.south)--node[left]{\(\phi^1\)}([xshift=-4pt]c1.north);
\draw[->] ([xshift=4pt]c1.north)--node[right]{\(\psi^1\)}([xshift=4pt]t1.south);
\draw[->] ([xshift=-4pt]t2.south)--node[left]{\(\phi^2\)}([xshift=-4pt]c2.north);
\draw[->] ([xshift=4pt]c2.north)--node[right]{\(\psi^2\)}([xshift=4pt]t2.south);
\end{tikzpicture}
\label{eq:physical-lifts}
\end{equation}
in which the downward maps and the upward maps each commute with the horizontal ones. 
Since \(\phi^1\psi^1\) and \(\psi^1\phi^1\) lift the identities of \(\coker\dcode^0\) and \(\coker\dstack^0\), there are homotopies \(\eta_{\mathrm{code}}:\Ccode^1\to\Ccode^0\) and \(\eta_{\mathrm{stack}}:\Cstack^1\to\Cstack^0\) with \(\dcode^0\eta_{\mathrm{code}}=I-\phi^1\psi^1\) and \(\dstack^0\eta_{\mathrm{stack}}=I-\psi^1\phi^1\).

Although \(\phi^2\) and \(\psi^2\) are mutually inverse, their lifts \(\phi^i\) and \(\psi^i\) need not be invertible. 
A pair of maps can nevertheless be assembled into an isomorphism. 
For maps \(f:A\to B\) and \(g:B\to A\), define
\[
  \Theta(f,g)=
 \begin{pmatrix}f&I_B-fg\\I_A&-g\end{pmatrix}:
 A\oplus B\longrightarrow B\oplus A.
\]
A direct computation shows that \(\Theta(g,f)\) is its inverse, even when \(f\) and \(g\) are not invertible. 
We will take \(V=\Theta(\psi^1,\phi^1)\), which needs both \(\Ccode^1\) and \(\Cstack^1\) on each side, and build \(U\) in the same way from maps between \(\Ccode^0\oplus\Cstack^1\) and \(\Cstack^0\oplus\Ccode^1\). 
We therefore adjoin ancillas,
\[
 \begin{aligned}
 \widehat C_{\mathrm{code}}&=\Ccode\oplus D^0(\Cstack^1)\oplus D^{-1}(\Cstack^0\oplus \Ccode^1),\\
 \widehat C_{\mathrm{stack}}&=\Cstack\oplus D^0(\Ccode^1)\oplus D^{-1}(\Ccode^0\oplus \Cstack^1),
 \end{aligned}
\]
whose terms in degrees zero and one are
\[
 \begin{aligned}
 \widehat C_{\mathrm{code}}^0&=(\Ccode^0\oplus \Cstack^1)\oplus(\Cstack^0\oplus \Ccode^1),\\
 \widehat C_{\mathrm{stack}}^0&=(\Cstack^0\oplus \Ccode^1)\oplus(\Ccode^0\oplus \Cstack^1),\\
 \widehat C_{\mathrm{code}}^1&=\Ccode^1\oplus \Cstack^1,\qquad
 \widehat C_{\mathrm{stack}}^1=\Cstack^1\oplus \Ccode^1.
 \end{aligned}
\]
In this order of summands, the degree-zero differentials are \(\widehat d^0_{\mathrm{code}}=(d'_{\mathrm{code}}\ 0)\) and \(\widehat d^0_{\mathrm{stack}}=(d'_{\mathrm{stack}}\ 0)\), with \(d'_{\mathrm{code}}=\diag(\dcode^0,I_{\Cstack^1})\) and \(d'_{\mathrm{stack}}=\diag(\dstack^0,I_{\Ccode^1})\).

In degree one, we set
\[
  V=\Theta(\psi^1,\phi^1):
 \widehat C_{\mathrm{code}}^1\xrightarrow{\sim}\widehat C_{\mathrm{stack}}^1.
\]
In degree zero, we take \(U=\Theta(\widehat\psi^0,\widehat\phi^0)\) for maps \(\widehat\psi^0:\Ccode^0\oplus\Cstack^1\to\Cstack^0\oplus\Ccode^1\) and \(\widehat\phi^0\) in the opposite direction. 
For the square to commute,
it suffices that these maps intertwine the degree-zero differentials with \(V\) and \(V^{-1}\),
\[
 d'_{\mathrm{stack}}\widehat\psi^0=Vd'_{\mathrm{code}},\qquad
 d'_{\mathrm{code}}\widehat\phi^0=V^{-1}d'_{\mathrm{stack}}.
\]
The lifts \(\psi^0\) and \(\phi^0\) satisfy \(\dstack^0\psi^0=\psi^1\dcode^0\) and \(\dcode^0\phi^0=\phi^1\dstack^0\) instead, with \(\psi^1\) and \(\phi^1\) in place of \(V\) and \(V^{-1}\).
The homotopies \(\eta\) correct this:
\[
 \widehat\phi^0=\begin{pmatrix}\phi^0&\eta_{\mathrm{code}}\\\dstack^0&-\psi^1\end{pmatrix},\qquad
 \widehat\psi^0=\begin{pmatrix}\psi^0&\eta_{\mathrm{stack}}\\\dcode^0&-\phi^1\end{pmatrix}
\]
satisfy both identities, by the commutativity of diagram~\eqref{eq:physical-lifts} and the defining identities of \(\eta_{\mathrm{code}}\) and \(\eta_{\mathrm{stack}}\). 
Then
\[
  U=\Theta(\widehat\psi^0,\widehat\phi^0)
   =\begin{pmatrix}\widehat\psi^0&I-\widehat\psi^0\widehat\phi^0\\
   I&-\widehat\phi^0\end{pmatrix}:
   \widehat C_{\mathrm{code}}^0\xrightarrow{\sim}\widehat C_{\mathrm{stack}}^0,
\]
and since \(d'_{\mathrm{stack}}\widehat\psi^0\widehat\phi^0=Vd'_{\mathrm{code}}\widehat\phi^0=d'_{\mathrm{stack}}\),
\[
 \begin{aligned}
 \widehat d^0_{\mathrm{stack}}U
 &=\bigl(d'_{\mathrm{stack}}\widehat\psi^0,\
 d'_{\mathrm{stack}}-d'_{\mathrm{stack}}\widehat\psi^0\widehat\phi^0\bigr)\\
 &=(Vd'_{\mathrm{code}},0)=V\widehat d^0_{\mathrm{code}}.
 \end{aligned}
\]
This is the commuting square~\eqref{eq:decoupling-square} for \(\widehat C_{\mathrm{code}}\) and \(\widehat C_{\mathrm{stack}}\), with \(U\) and \(V\) invertible.
\end{proof}

Disk complexes and their duals are exact, so \(\widehat C_{\mathrm{code}}\) and \(\widehat C_{\mathrm{stack}}\) still satisfy the topological condition~\eqref{eq:exactness}. 
By the claim in Sec.~\ref{subsec:polynomial}, a circuit that acts on \(Z\)-type representatives by \(U\) therefore maps the code, with its ancillas, onto the toric-code stack with its ancillas.

\subsection{Realizing the circuit}\label{apx:circuit}
Finally, we turn \(U\) into a circuit.

\begin{lemma}\label{lem:gates}
The isomorphism \(U\) of Lemma~\ref{lem:invertible-lift} is the action on \(Z\)-type representatives of a finite-depth local circuit of SUM and SWAP gates whose layers commute with all translations.
\end{lemma}
\begin{proof}
Like every \(\Theta(f,g)\), the map \(U\) factors into two unitriangular matrices and a block swap,
\[
 U=
  \begin{pmatrix}I&\widehat\psi^0\\0&I\end{pmatrix}
 \begin{pmatrix}I&0\\-\widehat\phi^0&I\end{pmatrix}
 \begin{pmatrix}0&I\\I&0\end{pmatrix}.
\]
Since the entries of \(\widehat\phi^0\) and \(\widehat\psi^0\) are Laurent polynomials, each unitriangular factor is a finite product of elementary matrices. 
Each of them differs from \(I\) by a single entry \(\lambda x^\alpha\), with \(\lambda\in\F\), in some row \(i\) and column \(j\ne i\). 
Each elementary matrix is realized by a single layer of SUM gates \cite{Haah2013}. 
Indeed, the SUM gate of weight \(\lambda\), \(\ket{a,b}\mapsto\ket{a,b+\lambda a}\), sends \(Z_1\mapsto Z_1\) and \(Z_2\mapsto Z_1^{-\lambda}Z_2\) \cite{Hostens2005}. 
A layer of SUM gates of weight \(-\lambda\), in which each qudit \(j\) is controlled by qudit \(i\) in the unit cell displaced by \(\alpha\), therefore acts on \(Z\)-type representatives by this elementary matrix. 
The block swap permutes qudits within each unit cell and is implemented by on-site SWAP gates.
The resulting circuit has finite depth and is local, in the sense that each gate acts on qudits within a bounded distance of one another.
\end{proof}

Together with Lemma~\ref{lem:stable-cokernel},
Lemmas~\ref{lem:invertible-lift} and~\ref{lem:gates} prove Theorem~\ref{thm:main}.

\subsection{Proof of the stable isomorphism}\label{apx:beta}
The proof is constructive and has the following steps. 
We first find a cochain map \(\Gamma:\Cstack\to\Ccode\) that represents the isomorphism of profiles,
and obtain a homotopy inverse \(\Delta\) from a contraction of its mapping cone. 
We then pass to the maps that \(\Gamma\) and \(\Delta\) induce on the check cokernels and absorb the errors by which these fail to be mutually inverse.

\paragraph{Constructing \(\Gamma\).}
Here \(\Ccode\) and \(\Cstack\) are the complexes before any disk complexes are adjoined. 
By Lemma~\ref{lem:model} and hypothesis~\eqref{eq:toric-profile-hypothesis},
\(\Cstack\simeq\bigoplus_j\F[-j]^{\oplus t_j}\simeq\mathscr E_X\simeq \Ccode\) in \(\mathrm{D}(R)\). 
Since \(\Cstack\) is a bounded complex of projective modules, morphisms from \(\Cstack\) in \(\mathrm{D}(R)\) are homotopy classes of cochain maps \cite[Cor.~10.4.7]{Weibel1994}, so this isomorphism is represented by a cochain map \(\Gamma:\Cstack\to\Ccode\), which is a quasi-isomorphism. 
The map \(\Gamma\) can be computed explicitly, one copy of \(T_j\) at a time. 
Let \(\Gamma_{j,k}:T_j\to\Ccode\) be its restriction to the \(k\)th copy of \(T_j\). 
Its components must make the diagram
\[
\begin{tikzpicture}[baseline=(current bounding box.center),>=To,
 every node/.style={inner sep=2pt},x=1.75cm,y=1.3cm]
\foreach \x/\d in {1/j-2,2/j-1,3/j,4/j+1}
 \node[cxdeg] at (\x,0.55) {\(\d\)};
\node[cxdeg,anchor=east] at (0.2,0.55) {degree};
\node[anchor=east] at (0.2,0) {\(T_j:\)};
\node[anchor=east] at (0.2,-1.3) {\(\Ccode:\)};
\node (t0) at (0.4,0) {\(\cdots\)};
\node (t1) at (1,0) {\(K_2\)};
\node (t2) at (2,0) {\(K_1\)};
\node (t3) at (3,0) {\(K_0\)};
\node (t4) at (4,0) {\(0\)};
\node (c0) at (0.4,-1.3) {\(\cdots\)};
\node (c1) at (1,-1.3) {\(\Ccode^{j-2}\)};
\node (c2) at (2,-1.3) {\(\Ccode^{j-1}\)};
\node (c3) at (3,-1.3) {\(\Ccode^{j}\)};
\node (c4) at (4,-1.3) {\(\Ccode^{j+1}\)};
\draw[->] (t0)--(t1);
\draw[->] (t1)--node[above,cxlab]{\(\partial_2\)}(t2);
\draw[->] (t2)--node[above,cxlab]{\(\partial_1\)}(t3);
\draw[->] (t3)--(t4);
\draw[->] (c0)--(c1);
\draw[->] (c1)--node[below,cxlab]{\(\dcode^{j-2}\)}(c2);
\draw[->] (c2)--node[below,cxlab]{\(\dcode^{j-1}\)}(c3);
\draw[->] (c3)--node[below,cxlab]{\(\dcode^{j}\)}(c4);
\draw[->] (t1)--node[left,cxlab]{\(\Gamma^{j-2}_{j,k}\)}(c1);
\draw[->] (t2)--node[left,cxlab]{\(\Gamma^{j-1}_{j,k}\)}(c2);
\draw[->] (t3)--node[left,cxlab]{\(\Gamma^{j}_{j,k}\)}(c3);
\draw[->] (t4)--(c4);
\end{tikzpicture}
\]
commute. 
The top component \(\Gamma^j_{j,k}\) is determined by its value on \(e_\emptyset\). 
For \(k=1,\ldots,t_j\), we choose these values to be cocycles of \(\Ccode\) representing a basis of \(E_X^j\); this makes \(\Gamma\) a quasi-isomorphism. 
The other components then solve the linear system \(\dcode\Gamma_{j,k}=\Gamma_{j,k}\partial\) over \(R\), with the value on \(e_\emptyset\) prescribed. 
A solution exists. 
A constant change of basis of the copies of \(T_j\) makes the quasi-isomorphism above send the generators \(e_\emptyset\) to the chosen cocycles up to coboundaries, 
and subtracting \(\dcode h+h\dstack\), for a map \(h\) that is nonzero only on these generators, removes the coboundaries. 
Appendix~\ref{apx:complexity} bounds the cost of solving this system. 
Combined together, the \(\Gamma_{j,k}\) give \(\Gamma\) with \(\dcode\Gamma=\Gamma\dstack\).

\paragraph{Inverting \(\Gamma\) up to homotopy.}
To find a homotopy inverse of \(\Gamma\), we contract its mapping cone. 
The mapping cone \(G=\operatorname{Cone}(\Gamma)\) of Appendix~\ref{apx:conventions} has terms \(G^n=\Ccode^n\oplus \Cstack^{n+1}\) and differential
\[
 \begin{aligned}
 d_G&=\begin{pmatrix}\dcode&\Gamma\\0&-\dstack\end{pmatrix},\\
 d_G(c,t)&=(\dcode c+\Gamma t,\,-\dstack t).
 \end{aligned}
\]
Like \(\Ccode\) and \(\Cstack\), it is bounded with finite free terms, and it is acyclic because \(\Gamma\) is a quasi-isomorphism \cite[Cor.~1.5.4]{Weibel1994}.
We construct maps \(h_G:G^n\to G^{n-1}\) with \(d_Gh_G+h_Gd_G=I\), that is, a contraction of \(G\). 
Since \(G\) is acyclic, \(d_G^n\) maps \(G^n\) onto \(\ker d_G^{n+1}\) for every \(n\). 
We first choose sections \(\sigma^n:\ker d_G^{n+1}\to G^n\) of these surjections, going down in \(n\). 
A section exists whenever \(\ker d_G^{n+1}\) is projective, that is, a direct summand of a free module. 
This holds when \(n\) is the highest degree of \(G\), since then \(\ker d_G^{n+1}=0\). 
If it holds for some \(n\), we choose \(\sigma^n\) and obtain
\[
 G^n=\ker d_G^n\oplus\sigma^n(\ker d_G^{n+1}),
\]
since \(g=(g-\sigma^nd_G^ng)+\sigma^nd_G^ng\) and \(d_G^n\sigma^n=I\). 
Thus \(\ker d_G^n\) is a direct summand of the free module \(G^n\), and the condition holds for \(n-1\). 
By induction, all the sections \(\sigma^n\) exist. 
The maps
\[
  h_G(g+\sigma^ng')=\sigma^{n-1}g,\qquad g\in\ker d_G^n,\
  g'\in\ker d_G^{n+1},
\]
then satisfy \(d_Gh_G+h_Gd_G=I\): on \(g\), the two terms give \(g\) and \(0\), and on \(\sigma^ng'\), they give \(0\) and \(\sigma^ng'\).
In an algorithm, we find the components \(h_G^n:G^n\to G^{n-1}\) going down in \(n\) by solving \(d_G^{n-1}h_G^n=I-h_G^{n+1}d_G^n\);
the right side maps \(G^n\) into \(\ker d_G^n=\im d_G^{n-1}\), so a solution exists because \(G^n\) is free.

We read off \(\Delta\) from the blocks of the contraction,
\[
 h_G=\begin{pmatrix}h&*\\\Delta&-h'\end{pmatrix}:
 \Ccode^n\oplus \Cstack^{n+1}\longrightarrow \Ccode^{n-1}\oplus \Cstack^n,
\]
where the entries marked \(*\) are not needed. 
With the degrees suppressed, \(d_Gh_G+h_Gd_G\) has the blocks
\[
 \begin{pmatrix}
 \dcode h+h\dcode+\Gamma\Delta&*\\
 \Delta \dcode-\dstack\Delta&\dstack h'+h'\dstack+\Delta\Gamma
 \end{pmatrix}.
\]
Comparing with \(I\) shows that \(\Delta:\Ccode\to \Cstack\) is a cochain map and that
\begin{equation}
 I-\Gamma\Delta=\dcode h+h\dcode,\qquad
 I-\Delta\Gamma=\dstack h'+h'\dstack;
 \label{eq:apx-homotopies}
\end{equation}
that is, \(\Delta\) is a homotopy inverse of \(\Gamma\).

\paragraph{Passing to the cokernels.}
As cochain maps, \(\Gamma\) and \(\Delta\) map \(\im\dstack^0\) into \(\im\dcode^0\) and \(\im\dcode^0\) into \(\im\dstack^0\), respectively, so they induce maps \(\Gamma_*:\coker \dstack^0\to\coker \dcode^0\) and \(\Delta_*:\coker \dcode^0\to\coker \dstack^0\):
\[
\begin{tikzpicture}[baseline=(current bounding box.center),>=To,
 every node/.style={inner sep=2pt},x=1cm,y=1cm]
\node (t0) at (0,0) {\(\Cstack^0\)};
\node (t1) at (2.15,0) {\(\Cstack^1\)};
\node (t2) at (4.55,0) {\(\coker \dstack^0\)};
\node (tz) at (6.2,0) {\(0\)};
\node (c0) at (0,-1.25) {\(\Ccode^0\)};
\node (c1) at (2.15,-1.25) {\(\Ccode^1\)};
\node (c2) at (4.55,-1.25) {\(\coker \dcode^0\)};
\node (cz) at (6.2,-1.25) {\(0\)};
\draw[->] (t0)--node[above]{\(\dstack^0\)}(t1);
\draw[->] (t1)--node[above]{\(\pi_{\mathrm{stack}}\)}(t2);
\draw[->] (t2)--(tz);
\draw[->] (c0)--node[below]{\(\dcode^0\)}(c1);
\draw[->] (c1)--node[below]{\(\pi_{\mathrm{code}}\)}(c2);
\draw[->] (c2)--(cz);
\draw[->] ([xshift=-4pt]t0.south)--node[left]{\(\Gamma\)}([xshift=-4pt]c0.north);
\draw[->] ([xshift=4pt]c0.north)--node[right]{\(\Delta\)}([xshift=4pt]t0.south);
\draw[->] ([xshift=-4pt]t1.south)--node[left]{\(\Gamma\)}([xshift=-4pt]c1.north);
\draw[->] ([xshift=4pt]c1.north)--node[right]{\(\Delta\)}([xshift=4pt]t1.south);
\draw[->] ([xshift=-4pt]t2.south)--node[left]{\(\Gamma_*\)}([xshift=-4pt]c2.north);
\draw[->] ([xshift=4pt]c2.north)--node[right]{\(\Delta_*\)}([xshift=4pt]t2.south);
\end{tikzpicture}
\]
with \(\Gamma_*[t]=[\Gamma t]\) and \(\Delta_*[c]=[\Delta c]\).

\paragraph{Absorbing the errors.}
If the homotopies vanished, \(\Gamma_*\) and \(\Delta_*\) would be mutually inverse. 
In general,
evaluate the first identity in Eq.~\eqref{eq:apx-homotopies} on \(c\in \Ccode^1\) and pass to classes modulo \(\im\dcode^0\). 
The term \(\dcode^0hc\) drops out, and we obtain \([c]-\Gamma_*\Delta_*[c]=[h\dcode^1c]\).
Likewise, the second identity gives \([t]-\Delta_*\Gamma_*[t]=[h'\dstack^1t]\) for \(t\in \Cstack^1\).
Thus
\begin{equation}
 I-\Gamma_*\Delta_*=\zeta_1\xi_1,\qquad I-\Delta_*\Gamma_*=\zeta_2\xi_2,
 \label{eq:apx-errors}
\end{equation}
where \(\xi_1:\coker \dcode^0\to \Ccode^2\) and \(\xi_2:\coker \dstack^0\to \Cstack^2\) are induced by \(\dcode^1\) and \(\dstack^1\), and \(\zeta_1=\pi_{\mathrm{code}}h:\Ccode^2\to\coker \dcode^0\) and \(\zeta_2=\pi_{\mathrm{stack}}h':\Cstack^2\to\coker \dstack^0\). 
The maps \(\xi_1\) and \(\xi_2\) are well defined because the differentials square to zero. 
The maps \(\Gamma_*\) and \(\Delta_*\) are mutually inverse up to maps that factor through the free modules \(\Ccode^2\) and \(\Cstack^2\).

We now absorb these errors by showing that \(\coker \dstack^0\oplus \Ccode^2\) is isomorphic to \(\coker \dcode^0\) plus a finite free module. 
Define \(\iota:\coker \dcode^0\to\coker \dstack^0\oplus \Ccode^2\) and \(\rho:\coker \dstack^0\oplus \Ccode^2\to\coker \dcode^0\) by \(\iota[c]=(\Delta_*[c],\xi_1[c])\) and \(\rho([t],c)=\Gamma_*[t]+\zeta_1c\). 
By the first identity in Eq.~\eqref{eq:apx-errors}, \(\rho\iota=I\), so \(\iota\) is a section of \(\rho\). 
With \(Q=\ker\rho\), we obtain
\[
 \coker \dstack^0\oplus \Ccode^2=\iota(\coker \dcode^0)\oplus Q\cong\coker \dcode^0\oplus Q.
\]
If \(([t],c)\in Q\), then \(\Gamma_*[t]=-\zeta_1c\), and the second identity in Eq.~\eqref{eq:apx-errors} gives \([t]=\Delta_*\Gamma_*[t]+\zeta_2\xi_2[t]=\zeta_2\xi_2[t]-\Delta_*\zeta_1c\).
Hence the map \(Q\to \Cstack^2\oplus \Ccode^2\), \(([t],c)\mapsto(\xi_2[t],c)\),
has the left inverse \((t',c)\mapsto(I-\iota\rho)(\zeta_2t'-\Delta_*\zeta_1c,\,c)\), where \(I-\iota\rho\) is the projection onto \(Q\). 
Thus \(Q\) is isomorphic to a direct summand of the finite free module \(\Cstack^2\oplus \Ccode^2\), so it is finitely generated and projective. 
It is even free \cite{Swan1978}, but we never compute a basis: an explicit isomorphism \(Q\oplus R^k\cong R^a\) for some \(k\) and \(a\) suffices, and the complexes already provide one.
Since \(\Ccode\) and \(\Cstack\) are exact in degrees \(n\leq0\), their parts in degrees at most one are finite free resolutions of \(\coker\dcode^0\) and \(\coker\dstack^0\). 
Lifting \(\rho\) to these resolutions, with \(\Ccode^2\) adjoined in degree one, and taking the mapping cone gives a bounded complex of finite free modules whose only cohomology is \(Q\), in degree zero. 
As \(Q\) is projective, this complex splits degree by degree, 
and each splitting map is a section of a surjection or a retraction of an injection, found by solving a linear system. 
Comparing the terms in even and odd degrees then gives the isomorphism \(Q\oplus R^k\cong R^a\). 
This is the argument of the generalized Schanuel lemma \cite[\S5A]{Lam1999}; 
as for the contraction above, it only uses sections of surjections onto projective modules. 
With \(b=\rank \Ccode^2+k\), adding \(R^k\) to both sides gives the isomorphism \(\phi^2:\coker \dstack^0\oplus R^b\to\coker \dcode^0\oplus R^a\) of Lemma~\ref{lem:stable-cokernel}. 

\subsection{Cost of the construction}\label{apx:complexity}
We now bound the cost of Algorithm~\ref{alg:decoupling} and prove Proposition~\ref{prop:complexity}, which we restate.

\begin{restated}{Proposition~\ref{prop:complexity}}[Polynomial cost]
The number of arithmetic operations over \(\F\) used by Algorithm~\ref{alg:decoupling} is bounded by a polynomial in the number of coefficients of the differentials of its input, counted with all exponents in a common box. 
The degree of this polynomial depends only on the spatial dimension \(D\). 
The same polynomial bounds the depth of the output circuit and its number of ancillas per unit cell.
\end{restated}

Throughout, \(R\) denotes the ring of the coarse-grained lattice. 
We write the differentials of the input \(\Ccode\) as matrices of Laurent polynomials whose exponents lie in a common box \(\{-h,\ldots,h\}^D\). 
The size in Proposition~\ref{prop:complexity} is the number of their coefficients in this box, so it bounds both the sizes of these matrices and their degrees. 
Like the complex of Lemma~\ref{lem:model}, \(\Ccode\) vanishes outside degrees \(|n|\leq D+2\).

\paragraph{Linear systems.}
Every step of the construction finds either generators of the solutions of a homogeneous system \(Ay=0\) over \(R\), 
or one solution of a system \(Ay=b\) that is known to be solvable. 
We replace each \(x_i^{-1}\) by a new variable \(y_i\), clear negative exponents by multiplying with monomials, and add unknown multiples of \(x_iy_i-1\); 
this gives an equivalent system over the polynomial ring in \(2D\) variables. 
If this system has \(m\) equations and degree \(d\), 
its solutions are generated by solutions of degree at most \((2md)^{2^{2D}}\), 
and if it is solvable, it has a solution of this degree \cite[Thms.~3.2 and 3.4]{Aschenbrenner2004}. 
Expanding the unknowns up to this degree and comparing coefficients gives a linear system over \(\F\) of polynomial size, which Gaussian elimination solves. 
Each step therefore costs a polynomial in the sizes and degrees of \(A\) and \(b\), and its output has polynomially bounded size and degree.

\paragraph{The lines of Algorithm~\ref{alg:decoupling}.}
In line~\ref{line:cohomology}, for each degree \(n\), we compute generators \(z_1,\ldots,z_g\) of \(\ker\dcode^n\), 
and generators \(u_\alpha\) of the module of vectors \(u\in R^g\) with \(\sum_iu_iz_i\in\im\dcode^{n-1}\). 
Since \(\mathfrak m\) annihilates \(H^n(\Ccode)\), this module contains \(\mathfrak mR^g\), 
so \(H^n(\Ccode)\cong\F^g/\operatorname{span}\{u_\alpha(1)\}\), where \(u_\alpha(1)\) is the value of \(u_\alpha\) at \(x_1=\cdots=x_D=1\). 
Gaussian elimination then gives \(t_n\) and cocycles representing a basis, as in the first step of the algorithm of Ref.~\cite{Wang2026} for \(D=2\). 
In line~\ref{line:gamma}, \(\Gamma\) solves the solvable system of Appendix~\ref{apx:beta}, 
and the components of \(h_G\) solve the solvable systems given there, one degree at a time.
In line~\ref{line:stable}, \(\Gamma_*\), \(\Delta_*\), \(\xi_i\), \(\zeta_i\), \(\iota\), and \(\rho\) are products of known matrices, 
and the isomorphism \(Q\oplus R^k\cong R^a\) needs a lift of \(\rho\) and the splitting maps, each a solution of a solvable system. 
In line~\ref{line:lift}, the lifts \(\phi^i\) and \(\psi^i\) and the homotopies \(\eta\) solve solvable systems, and \(U\) and \(V\) are block matrices built from them. 
In line~\ref{line:gates}, \(U\) is a block swap times elementary matrices, one for each monomial in the entries of \(\widehat\phi^0\) and \(\widehat\psi^0\).
The depth of the circuit is therefore at most the number of these monomials plus the number of qudits per unit cell, which bounds the SWAP gates of the block swap.
The number of ancillas per unit cell is the total rank of the adjoined free modules.

\paragraph{Total cost.}
The number of steps, and the number of times the output of one step enters another, are bounded in terms of \(D\). 
Composing the polynomial bounds a bounded number of times again gives a polynomial in the dense size of the input, which proves Proposition~\ref{prop:complexity}.

\section{Logical qudits on a torus}\label{apx:torus}
We prove Corollary~\ref{cor:torus}, which we restate. 
Here \(R=\F[\Z^D]\) is the ring of the original lattice and \(L\subseteq\Z^D\) is the period lattice. 
As in Ref.~\cite{Wang2026},
periodic boundary conditions are imposed by tensoring with \(R_L=R/(x^\alpha-1:\alpha\in L)\). 
For a complex \(A\) of \(R\)-modules, 
we write \(A_L\) for the derived tensor product \(A\otimes_RR_L\), 
computed after replacing \(A\) by a quasi-isomorphic bounded complex of free \(R\)-modules. 
Since \(R_L=R\otimes_{\F[L]}\F\) and \(R\) is free over \(\F[L]\), 
\(A_L\) is also \(A|_{\F[L]}\otimes_{\F[L]}\F\), computed in the same way, so it depends only on \(A|_{\F[L]}\); 
we use the same notation for complexes over \(\F[L]\).

\begin{restated}{Corollary~\ref{cor:torus}}[Logical qudits on a torus]
Let the layers \(E_X^j\) be finite dimensional over \(\F\), with \(t_j=\dim_{\F}E_X^j\). 
A torus with period lattice \(L\subseteq\Z^D\) encodes at most \(\sum_j\binom{D}{j}t_j\) logical qudits, with equality if and only if the code can be decoupled on \(L\).
\end{restated}

We first count the contribution of a single layer.

\begin{lemma}\label{lem:torus-layer}
Let \(E\) be an \(\F[L]\)-module that is finite dimensional over \(\F\), and let \(1\leq j\leq D-1\). 
Then \(\dim_{\F}H^0\bigl(E[-j]_L\bigr)\leq\binom{D}{j}\dim_{\F}E\), with equality if and only if \(L\) acts trivially on \(E\).
\end{lemma}
\begin{proof}
Let \(K'_\bullet\) be the Koszul complex over \(\F[L]\) built from the \(x^{\mu_a}-1\) for a basis \(\mu_1,\ldots,\mu_D\) of \(L\), a free resolution of \(\F\). 
Computing with it, \(H^0(E[-j]_L)\) is the homology of \(K'_\bullet\otimes_{\F[L]}E\) at the term \(K'_j\otimes E\), which has dimension \(\binom{D}{j}\dim_{\F}E\). 
The differentials multiply by the \(x^{\mu_a}-1\), so they vanish if \(L\) acts trivially on \(E\). 
Otherwise \((x^{\mu_a}-1)b\neq0\) for some \(a\) and \(b\in E\). 
For a \((j+1)\)-cell \(e'_J\) with \(a\in J\),
which exists since \(j\leq D-1\), the image of \(e'_J\otimes b\) in \(K'_j\otimes E\) has the nonzero component \(\pm e'_{J\setminus\{a\}}\otimes(x^{\mu_a}-1)b\), and the inequality is strict.
\end{proof}

\begin{proof}[Proof of Corollary~\ref{cor:torus}]
On the torus, the number of logical qudits is \(k=\dim_{\F}(\ker H_X/\im H_Z^\dagger)\) over \(R_L\), the cohomology in degree zero of \((\Ccode)_L\). 
Since \(\Ccode\simeq\mathscr E_X\) (Lemma~\ref{lem:model}), \(k=\dim_{\F}H^0\bigl((\mathscr E_X)_L\bigr)\).
Tensoring the truncation triangles~\eqref{eq:triangle} with \(R_L\) gives exact sequences
\[
 \begin{aligned}
 &H^0\bigl((\tau_{\leq j-1}\mathscr E_X)_L\bigr)
 \to H^0\bigl((\tau_{\leq j}\mathscr E_X)_L\bigr)\\
 &\qquad\to H^0\bigl(E_X^j[-j]_L\bigr)
 \xrightarrow{(\delta_j)_*}H^1\bigl((\tau_{\leq j-1}\mathscr E_X)_L\bigr),
 \end{aligned}
\]
so the dimension of the second term is at most the sum of those of its neighbors minus the rank of \((\delta_j)_*\). 
Summing over \(j\) and applying Lemma~\ref{lem:torus-layer} to every layer gives \(k\leq\sum_j\binom{D}{j}t_j\), with strict inequality unless \(L\subseteq\Ltriv\) and every \((\delta_j)_*\) vanishes.

Now let \(L\subseteq\Ltriv\). 
If the profile does not split over \(\F[L]\), then for the smallest \(j\) with \(\delta_j\neq0\), some gluing class \(\kappa_{ji}\) is nonzero, so it has a nonzero value on some cell of \(L\) (Lemma~\ref{lem:values}). 
Since \(L\) acts trivially on the layers, the differentials in the proof of Lemma~\ref{lem:torus-layer} vanish, 
so \(H^0(E_X^j[-j]_L)\) has one copy of \(E_X^j\) for each orientation \(J\) of the \(j\)-cells of \(L\). 
The map \((\delta_j)_*\) sends the copy at \(J\) to the copies of \(E_X^i\) at the faces \(J\setminus I\), each through the value of \(\kappa_{ji}\) on \(I\), so a nonzero value makes it nonzero. 
Equality therefore requires \(L\subseteq\Ltriv\) and a split profile over \(\F[L]\), so the code can be decoupled on \(L\) by Sec.~\ref{subsec:gluing-classes} and Theorem~\ref{thm:main}.

Conversely, if the code can be decoupled on \(L\), its profile over \(\F[L]\) is \(\bigoplus_j\F[-j]^{\oplus t_j}\) (Theorem~\ref{thm:main}), so Lemma~\ref{lem:torus-layer} with \(E=\F\) gives \(k=\sum_j\binom{D}{j}t_j\). 
\end{proof}

\clearpage
\bibliography{references}

\begin{thebibliography}{61}%
\makeatletter
\providecommand \@ifxundefined [1]{%
 \@ifx{#1\undefined}
}%
\providecommand \@ifnum [1]{%
 \ifnum #1\expandafter \@firstoftwo
 \else \expandafter \@secondoftwo
 \fi
}%
\providecommand \@ifx [1]{%
 \ifx #1\expandafter \@firstoftwo
 \else \expandafter \@secondoftwo
 \fi
}%
\providecommand \natexlab [1]{#1}%
\providecommand \enquote  [1]{``#1''}%
\providecommand \bibnamefont  [1]{#1}%
\providecommand \bibfnamefont [1]{#1}%
\providecommand \citenamefont [1]{#1}%
\providecommand \href@noop [0]{\@secondoftwo}%
\providecommand \href [0]{\begingroup \@sanitize@url \@href}%
\providecommand \@href[1]{\@@startlink{#1}\@@href}%
\providecommand \@@href[1]{\endgroup#1\@@endlink}%
\providecommand \@sanitize@url [0]{\catcode `\\12\catcode `\$12\catcode
  `\&12\catcode `\#12\catcode `\^12\catcode `\_12\catcode `\%12\relax}%
\providecommand \@@startlink[1]{}%
\providecommand \@@endlink[0]{}%
\providecommand \url  [0]{\begingroup\@sanitize@url \@url }%
\providecommand \@url [1]{\endgroup\@href {#1}{\urlprefix }}%
\providecommand \urlprefix  [0]{URL }%
\providecommand \Eprint [0]{\href }%
\providecommand \doibase [0]{https://doi.org/}%
\providecommand \selectlanguage [0]{\@gobble}%
\providecommand \bibinfo  [0]{\@secondoftwo}%
\providecommand \bibfield  [0]{\@secondoftwo}%
\providecommand \translation [1]{[#1]}%
\providecommand \BibitemOpen [0]{}%
\providecommand \bibitemStop [0]{}%
\providecommand \bibitemNoStop [0]{.\EOS\space}%
\providecommand \EOS [0]{\spacefactor3000\relax}%
\providecommand \BibitemShut  [1]{\csname bibitem#1\endcsname}%
\let\auto@bib@innerbib\@empty
%</preamble>
\bibitem [{\citenamefont {Kitaev}(2003)}]{Kitaev2003}%
  \BibitemOpen
  \bibfield  {author} {\bibinfo {author} {\bibfnamefont {A.~Y.}\ \bibnamefont
  {Kitaev}},\ }\bibfield  {title} {\bibinfo {title} {Fault-tolerant quantum
  computation by anyons},\ }\href
  {https://doi.org/10.1016/S0003-4916(02)00018-0} {\bibfield  {journal}
  {\bibinfo  {journal} {Ann. Phys.}\ }\textbf {\bibinfo {volume} {303}},\
  \bibinfo {pages} {2} (\bibinfo {year} {2003})},\ \Eprint
  {https://arxiv.org/abs/quant-ph/9707021} {arXiv:quant-ph/9707021}
  \BibitemShut {NoStop}%
\bibitem [{\citenamefont {Dennis}\ \emph {et~al.}(2002)\citenamefont {Dennis},
  \citenamefont {Kitaev}, \citenamefont {Landahl},\ and\ \citenamefont
  {Preskill}}]{Dennis2002}%
  \BibitemOpen
  \bibfield  {author} {\bibinfo {author} {\bibfnamefont {E.}~\bibnamefont
  {Dennis}}, \bibinfo {author} {\bibfnamefont {A.}~\bibnamefont {Kitaev}},
  \bibinfo {author} {\bibfnamefont {A.}~\bibnamefont {Landahl}},\ and\ \bibinfo
  {author} {\bibfnamefont {J.}~\bibnamefont {Preskill}},\ }\bibfield  {title}
  {\bibinfo {title} {Topological quantum memory},\ }\href
  {https://doi.org/10.1063/1.1499754} {\bibfield  {journal} {\bibinfo
  {journal} {J. Math. Phys.}\ }\textbf {\bibinfo {volume} {43}},\ \bibinfo
  {pages} {4452} (\bibinfo {year} {2002})},\ \Eprint
  {https://arxiv.org/abs/quant-ph/0110143} {arXiv:quant-ph/0110143}
  \BibitemShut {NoStop}%
\bibitem [{\citenamefont {Bomb\'in}\ and\ \citenamefont
  {Martin-Delgado}(2007)}]{BombinMartinDelgado2007c}%
  \BibitemOpen
  \bibfield  {author} {\bibinfo {author} {\bibfnamefont {H.}~\bibnamefont
  {Bomb\'in}}\ and\ \bibinfo {author} {\bibfnamefont {M.~A.}\ \bibnamefont
  {Martin-Delgado}},\ }\bibfield  {title} {\bibinfo {title} {Topological
  computation without braiding},\ }\href
  {https://doi.org/10.1103/PhysRevLett.98.160502} {\bibfield  {journal}
  {\bibinfo  {journal} {Phys. Rev. Lett.}\ }\textbf {\bibinfo {volume} {98}},\
  \bibinfo {pages} {160502} (\bibinfo {year} {2007})},\ \Eprint
  {https://arxiv.org/abs/quant-ph/0610024} {arXiv:quant-ph/0610024}
  \BibitemShut {NoStop}%
\bibitem [{\citenamefont {Kubica}\ and\ \citenamefont
  {Beverland}(2015)}]{KubicaBeverland2015}%
  \BibitemOpen
  \bibfield  {author} {\bibinfo {author} {\bibfnamefont {A.}~\bibnamefont
  {Kubica}}\ and\ \bibinfo {author} {\bibfnamefont {M.~E.}\ \bibnamefont
  {Beverland}},\ }\bibfield  {title} {\bibinfo {title} {Universal transversal
  gates with color codes: A simplified approach},\ }\href
  {https://doi.org/10.1103/PhysRevA.91.032330} {\bibfield  {journal} {\bibinfo
  {journal} {Phys. Rev. A}\ }\textbf {\bibinfo {volume} {91}},\ \bibinfo
  {pages} {032330} (\bibinfo {year} {2015})},\ \Eprint
  {https://arxiv.org/abs/1410.0069} {arXiv:1410.0069} \BibitemShut {NoStop}%
\bibitem [{\citenamefont {Kubica}\ \emph {et~al.}(2015)\citenamefont {Kubica},
  \citenamefont {Yoshida},\ and\ \citenamefont {Pastawski}}]{Kubica2015}%
  \BibitemOpen
  \bibfield  {author} {\bibinfo {author} {\bibfnamefont {A.}~\bibnamefont
  {Kubica}}, \bibinfo {author} {\bibfnamefont {B.}~\bibnamefont {Yoshida}},\
  and\ \bibinfo {author} {\bibfnamefont {F.}~\bibnamefont {Pastawski}},\
  }\bibfield  {title} {\bibinfo {title} {Unfolding the color code},\ }\href
  {https://doi.org/10.1088/1367-2630/17/8/083026} {\bibfield  {journal}
  {\bibinfo  {journal} {New J. Phys.}\ }\textbf {\bibinfo {volume} {17}},\
  \bibinfo {pages} {083026} (\bibinfo {year} {2015})},\ \Eprint
  {https://arxiv.org/abs/1503.02065} {arXiv:1503.02065} \BibitemShut {NoStop}%
\bibitem [{\citenamefont {Vasmer}\ and\ \citenamefont
  {Browne}(2019)}]{VasmerBrowne2019}%
  \BibitemOpen
  \bibfield  {author} {\bibinfo {author} {\bibfnamefont {M.}~\bibnamefont
  {Vasmer}}\ and\ \bibinfo {author} {\bibfnamefont {D.~E.}\ \bibnamefont
  {Browne}},\ }\bibfield  {title} {\bibinfo {title} {Three-dimensional surface
  codes: Transversal gates and fault-tolerant architectures},\ }\href
  {https://doi.org/10.1103/PhysRevA.100.012312} {\bibfield  {journal} {\bibinfo
   {journal} {Phys. Rev. A}\ }\textbf {\bibinfo {volume} {100}},\ \bibinfo
  {pages} {012312} (\bibinfo {year} {2019})},\ \Eprint
  {https://arxiv.org/abs/1801.04255} {arXiv:1801.04255} \BibitemShut {NoStop}%
\bibitem [{\citenamefont {Bomb\'in}(2015)}]{Bombin2015}%
  \BibitemOpen
  \bibfield  {author} {\bibinfo {author} {\bibfnamefont {H.}~\bibnamefont
  {Bomb\'in}},\ }\bibfield  {title} {\bibinfo {title} {Single-shot
  fault-tolerant quantum error correction},\ }\href
  {https://doi.org/10.1103/PhysRevX.5.031043} {\bibfield  {journal} {\bibinfo
  {journal} {Phys. Rev. X}\ }\textbf {\bibinfo {volume} {5}},\ \bibinfo {pages}
  {031043} (\bibinfo {year} {2015})},\ \Eprint
  {https://arxiv.org/abs/1404.5504} {arXiv:1404.5504} \BibitemShut {NoStop}%
\bibitem [{\citenamefont {Jacob}\ \emph {et~al.}(2025)\citenamefont {Jacob},
  \citenamefont {McLauchlan},\ and\ \citenamefont {Browne}}]{Jacob2025}%
  \BibitemOpen
  \bibfield  {author} {\bibinfo {author} {\bibfnamefont {A.}~\bibnamefont
  {Jacob}}, \bibinfo {author} {\bibfnamefont {C.}~\bibnamefont {McLauchlan}},\
  and\ \bibinfo {author} {\bibfnamefont {D.~E.}\ \bibnamefont {Browne}},\
  }\href {https://arxiv.org/abs/2508.08191v3} {\bibinfo {title} {Single-shot
  decoding and fault-tolerant gates with trivariate tricycle codes}} (\bibinfo
  {year} {2025}),\ \bibinfo {note} {version 3},\ \Eprint
  {https://arxiv.org/abs/2508.08191} {arXiv:2508.08191 [quant-ph]} \BibitemShut
  {NoStop}%
\bibitem [{\citenamefont {Menon}\ \emph {et~al.}(2026)\citenamefont {Menon},
  \citenamefont {Bonilla~Ataides}, \citenamefont {Mehta}, \citenamefont {Gu},
  \citenamefont {Tan},\ and\ \citenamefont {Lukin}}]{Menon2026MagicTricycles}%
  \BibitemOpen
  \bibfield  {author} {\bibinfo {author} {\bibfnamefont {V.}~\bibnamefont
  {Menon}}, \bibinfo {author} {\bibfnamefont {J.~P.}\ \bibnamefont
  {Bonilla~Ataides}}, \bibinfo {author} {\bibfnamefont {R.}~\bibnamefont
  {Mehta}}, \bibinfo {author} {\bibfnamefont {A.}~\bibnamefont {Gu}}, \bibinfo
  {author} {\bibfnamefont {D.~B.}\ \bibnamefont {Tan}},\ and\ \bibinfo {author}
  {\bibfnamefont {M.~D.}\ \bibnamefont {Lukin}},\ }\bibfield  {title} {\bibinfo
  {title} {Magic tricycles: Efficient magic-state generation with finite
  block-length quantum {LDPC} codes},\ }\href
  {https://doi.org/10.1103/ghhp-cytl} {\bibfield  {journal} {\bibinfo
  {journal} {Phys. Rev. X}\ }\textbf {\bibinfo {volume} {16}},\ \bibinfo
  {pages} {021014} (\bibinfo {year} {2026})}\BibitemShut {NoStop}%
\bibitem [{\citenamefont {Mian}\ \emph {et~al.}(2026)\citenamefont {Mian},
  \citenamefont {Gwilliam},\ and\ \citenamefont {Krastanov}}]{Mian2026}%
  \BibitemOpen
  \bibfield  {author} {\bibinfo {author} {\bibfnamefont {F.~A.}\ \bibnamefont
  {Mian}}, \bibinfo {author} {\bibfnamefont {O.}~\bibnamefont {Gwilliam}},\
  and\ \bibinfo {author} {\bibfnamefont {S.}~\bibnamefont {Krastanov}},\ }\href
  {https://arxiv.org/abs/2601.18879v2} {\bibinfo {title} {Multivariate
  multicycle codes for complete single-shot decoding}} (\bibinfo {year}
  {2026}),\ \bibinfo {note} {version 2},\ \Eprint
  {https://arxiv.org/abs/2601.18879} {arXiv:2601.18879 [quant-ph]} \BibitemShut
  {NoStop}%
\bibitem [{\citenamefont {Kubica}\ and\ \citenamefont
  {Delfosse}(2023)}]{KubicaDelfosse2023}%
  \BibitemOpen
  \bibfield  {author} {\bibinfo {author} {\bibfnamefont {A.}~\bibnamefont
  {Kubica}}\ and\ \bibinfo {author} {\bibfnamefont {N.}~\bibnamefont
  {Delfosse}},\ }\bibfield  {title} {\bibinfo {title} {Efficient color code
  decoders in {$d\geq 2$} dimensions from toric code decoders},\ }\href
  {https://doi.org/10.22331/q-2023-02-21-929} {\bibfield  {journal} {\bibinfo
  {journal} {Quantum}\ }\textbf {\bibinfo {volume} {7}},\ \bibinfo {pages}
  {929} (\bibinfo {year} {2023})},\ \Eprint {https://arxiv.org/abs/1905.07393}
  {arXiv:1905.07393} \BibitemShut {NoStop}%
\bibitem [{\citenamefont {Chen}\ \emph {et~al.}(2010)\citenamefont {Chen},
  \citenamefont {Gu},\ and\ \citenamefont {Wen}}]{ChenGuWen2010}%
  \BibitemOpen
  \bibfield  {author} {\bibinfo {author} {\bibfnamefont {X.}~\bibnamefont
  {Chen}}, \bibinfo {author} {\bibfnamefont {Z.-C.}\ \bibnamefont {Gu}},\ and\
  \bibinfo {author} {\bibfnamefont {X.-G.}\ \bibnamefont {Wen}},\ }\bibfield
  {title} {\bibinfo {title} {Local unitary transformation, long-range quantum
  entanglement, wave function renormalization, and topological order},\ }\href
  {https://doi.org/10.1103/PhysRevB.82.155138} {\bibfield  {journal} {\bibinfo
  {journal} {Phys. Rev. B}\ }\textbf {\bibinfo {volume} {82}},\ \bibinfo
  {pages} {155138} (\bibinfo {year} {2010})},\ \Eprint
  {https://arxiv.org/abs/1004.3835} {arXiv:1004.3835} \BibitemShut {NoStop}%
\bibitem [{\citenamefont {Calderbank}\ and\ \citenamefont
  {Shor}(1996)}]{CalderbankShor1996}%
  \BibitemOpen
  \bibfield  {author} {\bibinfo {author} {\bibfnamefont {A.~R.}\ \bibnamefont
  {Calderbank}}\ and\ \bibinfo {author} {\bibfnamefont {P.~W.}\ \bibnamefont
  {Shor}},\ }\bibfield  {title} {\bibinfo {title} {Good quantum
  error-correcting codes exist},\ }\href
  {https://doi.org/10.1103/PhysRevA.54.1098} {\bibfield  {journal} {\bibinfo
  {journal} {Phys. Rev. A}\ }\textbf {\bibinfo {volume} {54}},\ \bibinfo
  {pages} {1098} (\bibinfo {year} {1996})},\ \Eprint
  {https://arxiv.org/abs/quant-ph/9512032} {arXiv:quant-ph/9512032}
  \BibitemShut {NoStop}%
\bibitem [{\citenamefont {Steane}(1996)}]{Steane1996}%
  \BibitemOpen
  \bibfield  {author} {\bibinfo {author} {\bibfnamefont {A.~M.}\ \bibnamefont
  {Steane}},\ }\bibfield  {title} {\bibinfo {title} {Error correcting codes in
  quantum theory},\ }\href {https://doi.org/10.1103/PhysRevLett.77.793}
  {\bibfield  {journal} {\bibinfo  {journal} {Phys. Rev. Lett.}\ }\textbf
  {\bibinfo {volume} {77}},\ \bibinfo {pages} {793} (\bibinfo {year}
  {1996})}\BibitemShut {NoStop}%
\bibitem [{\citenamefont {Bravyi}\ \emph {et~al.}(2024)\citenamefont {Bravyi},
  \citenamefont {Cross}, \citenamefont {Gambetta}, \citenamefont {Maslov},
  \citenamefont {Rall},\ and\ \citenamefont {Yoder}}]{Bravyi2024}%
  \BibitemOpen
  \bibfield  {author} {\bibinfo {author} {\bibfnamefont {S.}~\bibnamefont
  {Bravyi}}, \bibinfo {author} {\bibfnamefont {A.~W.}\ \bibnamefont {Cross}},
  \bibinfo {author} {\bibfnamefont {J.~M.}\ \bibnamefont {Gambetta}}, \bibinfo
  {author} {\bibfnamefont {D.}~\bibnamefont {Maslov}}, \bibinfo {author}
  {\bibfnamefont {P.}~\bibnamefont {Rall}},\ and\ \bibinfo {author}
  {\bibfnamefont {T.~J.}\ \bibnamefont {Yoder}},\ }\bibfield  {title} {\bibinfo
  {title} {High-threshold and low-overhead fault-tolerant quantum memory},\
  }\href {https://doi.org/10.1038/s41586-024-07107-7} {\bibfield  {journal}
  {\bibinfo  {journal} {Nature}\ }\textbf {\bibinfo {volume} {627}},\ \bibinfo
  {pages} {778} (\bibinfo {year} {2024})},\ \Eprint
  {https://arxiv.org/abs/2308.07915} {arXiv:2308.07915} \BibitemShut {NoStop}%
\bibitem [{\citenamefont {Song}(2026)}]{Song2026}%
  \BibitemOpen
  \bibfield  {author} {\bibinfo {author} {\bibfnamefont {H.}~\bibnamefont
  {Song}},\ }\href {https://arxiv.org/abs/2608.09913v2} {\bibinfo {title}
  {Superselection rigidity of topological {Calderbank--Shor--Steane} codes:
  From {Koszul}-complex models to nonsplit extensions}} (\bibinfo {year}
  {2026}),\ \bibinfo {note} {version 2},\ \Eprint
  {https://arxiv.org/abs/2608.09913} {arXiv:2608.09913 [quant-ph]} \BibitemShut
  {NoStop}%
\bibitem [{\citenamefont {Wen}(2003)}]{Wen2003}%
  \BibitemOpen
  \bibfield  {author} {\bibinfo {author} {\bibfnamefont {X.-G.}\ \bibnamefont
  {Wen}},\ }\bibfield  {title} {\bibinfo {title} {Quantum orders in an exact
  soluble model},\ }\href {https://doi.org/10.1103/PhysRevLett.90.016803}
  {\bibfield  {journal} {\bibinfo  {journal} {Phys. Rev. Lett.}\ }\textbf
  {\bibinfo {volume} {90}},\ \bibinfo {pages} {016803} (\bibinfo {year}
  {2003})},\ \Eprint {https://arxiv.org/abs/quant-ph/0205004}
  {arXiv:quant-ph/0205004} \BibitemShut {NoStop}%
\bibitem [{\citenamefont {Kitaev}(2006)}]{Kitaev2006}%
  \BibitemOpen
  \bibfield  {author} {\bibinfo {author} {\bibfnamefont {A.}~\bibnamefont
  {Kitaev}},\ }\bibfield  {title} {\bibinfo {title} {Anyons in an exactly
  solved model and beyond},\ }\href {https://doi.org/10.1016/j.aop.2005.10.005}
  {\bibfield  {journal} {\bibinfo  {journal} {Ann. Phys.}\ }\textbf {\bibinfo
  {volume} {321}},\ \bibinfo {pages} {2} (\bibinfo {year} {2006})},\ \Eprint
  {https://arxiv.org/abs/cond-mat/0506438} {arXiv:cond-mat/0506438}
  \BibitemShut {NoStop}%
\bibitem [{\citenamefont {Rao}\ and\ \citenamefont
  {Sodemann}(2021)}]{RaoSodemann2021}%
  \BibitemOpen
  \bibfield  {author} {\bibinfo {author} {\bibfnamefont {P.}~\bibnamefont
  {Rao}}\ and\ \bibinfo {author} {\bibfnamefont {I.}~\bibnamefont {Sodemann}},\
  }\bibfield  {title} {\bibinfo {title} {Theory of weak symmetry breaking of
  translations in {$\mathbb{Z}_2$} topologically ordered states and its
  relation to topological superconductivity from an exact lattice
  {$\mathbb{Z}_2$} charge-flux attachment},\ }\href
  {https://doi.org/10.1103/PhysRevResearch.3.023120} {\bibfield  {journal}
  {\bibinfo  {journal} {Phys. Rev. Research}\ }\textbf {\bibinfo {volume}
  {3}},\ \bibinfo {pages} {023120} (\bibinfo {year} {2021})},\ \Eprint
  {https://arxiv.org/abs/2012.12280} {arXiv:2012.12280} \BibitemShut {NoStop}%
\bibitem [{\citenamefont {Bomb{\'i}n}\ \emph {et~al.}(2012)\citenamefont
  {Bomb{\'i}n}, \citenamefont {Duclos-Cianci},\ and\ \citenamefont
  {Poulin}}]{BombinDuclosPoulin2012}%
  \BibitemOpen
  \bibfield  {author} {\bibinfo {author} {\bibfnamefont {H.}~\bibnamefont
  {Bomb{\'i}n}}, \bibinfo {author} {\bibfnamefont {G.}~\bibnamefont
  {Duclos-Cianci}},\ and\ \bibinfo {author} {\bibfnamefont {D.}~\bibnamefont
  {Poulin}},\ }\bibfield  {title} {\bibinfo {title} {Universal topological
  phase of two-dimensional stabilizer codes},\ }\href
  {https://doi.org/10.1088/1367-2630/14/7/073048} {\bibfield  {journal}
  {\bibinfo  {journal} {New J. Phys.}\ }\textbf {\bibinfo {volume} {14}},\
  \bibinfo {pages} {073048} (\bibinfo {year} {2012})},\ \Eprint
  {https://arxiv.org/abs/1103.4606} {arXiv:1103.4606} \BibitemShut {NoStop}%
\bibitem [{\citenamefont {Bombin}(2014)}]{Bombin2014}%
  \BibitemOpen
  \bibfield  {author} {\bibinfo {author} {\bibfnamefont {H.}~\bibnamefont
  {Bombin}},\ }\bibfield  {title} {\bibinfo {title} {Structure of {2D}
  topological stabilizer codes},\ }\href
  {https://doi.org/10.1007/s00220-014-1893-4} {\bibfield  {journal} {\bibinfo
  {journal} {Commun. Math. Phys.}\ }\textbf {\bibinfo {volume} {327}},\
  \bibinfo {pages} {387} (\bibinfo {year} {2014})},\ \Eprint
  {https://arxiv.org/abs/1107.2707} {arXiv:1107.2707} \BibitemShut {NoStop}%
\bibitem [{\citenamefont {Haah}(2016)}]{Haah2016}%
  \BibitemOpen
  \bibfield  {author} {\bibinfo {author} {\bibfnamefont {J.}~\bibnamefont
  {Haah}},\ }\bibfield  {title} {\bibinfo {title} {Algebraic methods for
  quantum codes on lattices},\ }\href
  {https://doi.org/10.15446/recolma.v50n2.62214} {\bibfield  {journal}
  {\bibinfo  {journal} {Rev. Colomb. Mat.}\ }\textbf {\bibinfo {volume} {50}},\
  \bibinfo {pages} {299} (\bibinfo {year} {2016})},\ \Eprint
  {https://arxiv.org/abs/1607.01387} {arXiv:1607.01387} \BibitemShut {NoStop}%
\bibitem [{\citenamefont {Haah}(2021)}]{Haah2021}%
  \BibitemOpen
  \bibfield  {author} {\bibinfo {author} {\bibfnamefont {J.}~\bibnamefont
  {Haah}},\ }\bibfield  {title} {\bibinfo {title} {Classification of
  translation invariant topological {Pauli} stabilizer codes for prime
  dimensional qudits on two-dimensional lattices},\ }\href
  {https://doi.org/10.1063/5.0021068} {\bibfield  {journal} {\bibinfo
  {journal} {J. Math. Phys.}\ }\textbf {\bibinfo {volume} {62}},\ \bibinfo
  {pages} {012201} (\bibinfo {year} {2021})},\ \Eprint
  {https://arxiv.org/abs/1812.11193} {arXiv:1812.11193} \BibitemShut {NoStop}%
\bibitem [{\citenamefont {Wang}\ \emph {et~al.}(2026)\citenamefont {Wang},
  \citenamefont {Ni}, \citenamefont {He}, \citenamefont {Liu},\ and\
  \citenamefont {Gu}}]{Wang2026}%
  \BibitemOpen
  \bibfield  {author} {\bibinfo {author} {\bibfnamefont {Y.}~\bibnamefont
  {Wang}}, \bibinfo {author} {\bibfnamefont {Z.}~\bibnamefont {Ni}}, \bibinfo
  {author} {\bibfnamefont {M.}~\bibnamefont {He}}, \bibinfo {author}
  {\bibfnamefont {J.}~\bibnamefont {Liu}},\ and\ \bibinfo {author}
  {\bibfnamefont {Y.}~\bibnamefont {Gu}},\ }\href
  {https://arxiv.org/abs/2608.09915v1} {\bibinfo {title} {Decoupling {2D}
  translation-invariant topological {CSS} codes}} (\bibinfo {year} {2026}),\
  \bibinfo {note} {version 1},\ \Eprint {https://arxiv.org/abs/2608.09915}
  {arXiv:2608.09915 [quant-ph]} \BibitemShut {NoStop}%
\bibitem [{\citenamefont {Ruba}\ and\ \citenamefont
  {Yang}(2024)}]{RubaYang2024}%
  \BibitemOpen
  \bibfield  {author} {\bibinfo {author} {\bibfnamefont {B.}~\bibnamefont
  {Ruba}}\ and\ \bibinfo {author} {\bibfnamefont {B.}~\bibnamefont {Yang}},\
  }\bibfield  {title} {\bibinfo {title} {Homological invariants of {Pauli}
  stabilizer codes},\ }\href {https://doi.org/10.1007/s00220-024-04991-y}
  {\bibfield  {journal} {\bibinfo  {journal} {Commun. Math. Phys.}\ }\textbf
  {\bibinfo {volume} {405}},\ \bibinfo {pages} {126} (\bibinfo {year}
  {2024})},\ \Eprint {https://arxiv.org/abs/2204.06023} {arXiv:2204.06023}
  \BibitemShut {NoStop}%
\bibitem [{\citenamefont {Nandkishore}\ and\ \citenamefont
  {Hermele}(2019)}]{NandkishoreHermele2019}%
  \BibitemOpen
  \bibfield  {author} {\bibinfo {author} {\bibfnamefont {R.~M.}\ \bibnamefont
  {Nandkishore}}\ and\ \bibinfo {author} {\bibfnamefont {M.}~\bibnamefont
  {Hermele}},\ }\bibfield  {title} {\bibinfo {title} {Fractons},\ }\href
  {https://doi.org/10.1146/annurev-conmatphys-031218-013604} {\bibfield
  {journal} {\bibinfo  {journal} {Annu. Rev. Condens. Matter Phys.}\ }\textbf
  {\bibinfo {volume} {10}},\ \bibinfo {pages} {295} (\bibinfo {year} {2019})},\
  \Eprint {https://arxiv.org/abs/1803.11196} {arXiv:1803.11196} \BibitemShut
  {NoStop}%
\bibitem [{\citenamefont {Pretko}\ \emph {et~al.}(2020)\citenamefont {Pretko},
  \citenamefont {Chen},\ and\ \citenamefont {You}}]{PretkoChenYou2020}%
  \BibitemOpen
  \bibfield  {author} {\bibinfo {author} {\bibfnamefont {M.}~\bibnamefont
  {Pretko}}, \bibinfo {author} {\bibfnamefont {X.}~\bibnamefont {Chen}},\ and\
  \bibinfo {author} {\bibfnamefont {Y.}~\bibnamefont {You}},\ }\bibfield
  {title} {\bibinfo {title} {Fracton phases of matter},\ }\href
  {https://doi.org/10.1142/S0217751X20300033} {\bibfield  {journal} {\bibinfo
  {journal} {Int. J. Mod. Phys. A}\ }\textbf {\bibinfo {volume} {35}},\
  \bibinfo {pages} {2030003} (\bibinfo {year} {2020})},\ \Eprint
  {https://arxiv.org/abs/2001.01722} {arXiv:2001.01722} \BibitemShut {NoStop}%
\bibitem [{\citenamefont {Haah}(2011)}]{Haah2011}%
  \BibitemOpen
  \bibfield  {author} {\bibinfo {author} {\bibfnamefont {J.}~\bibnamefont
  {Haah}},\ }\bibfield  {title} {\bibinfo {title} {Local stabilizer codes in
  three dimensions without string logical operators},\ }\href
  {https://doi.org/10.1103/PhysRevA.83.042330} {\bibfield  {journal} {\bibinfo
  {journal} {Phys. Rev. A}\ }\textbf {\bibinfo {volume} {83}},\ \bibinfo
  {pages} {042330} (\bibinfo {year} {2011})},\ \Eprint
  {https://arxiv.org/abs/1101.1962} {arXiv:1101.1962} \BibitemShut {NoStop}%
\bibitem [{\citenamefont {Vijay}\ \emph {et~al.}(2016)\citenamefont {Vijay},
  \citenamefont {Haah},\ and\ \citenamefont {Fu}}]{Vijay2016}%
  \BibitemOpen
  \bibfield  {author} {\bibinfo {author} {\bibfnamefont {S.}~\bibnamefont
  {Vijay}}, \bibinfo {author} {\bibfnamefont {J.}~\bibnamefont {Haah}},\ and\
  \bibinfo {author} {\bibfnamefont {L.}~\bibnamefont {Fu}},\ }\bibfield
  {title} {\bibinfo {title} {Fracton topological order, generalized lattice
  gauge theory, and duality},\ }\href
  {https://doi.org/10.1103/PhysRevB.94.235157} {\bibfield  {journal} {\bibinfo
  {journal} {Phys. Rev. B}\ }\textbf {\bibinfo {volume} {94}},\ \bibinfo
  {pages} {235157} (\bibinfo {year} {2016})},\ \Eprint
  {https://arxiv.org/abs/1603.04442} {arXiv:1603.04442} \BibitemShut {NoStop}%
\bibitem [{\citenamefont {Li}\ and\ \citenamefont
  {Williamson}(2026)}]{LiWilliamson2026}%
  \BibitemOpen
  \bibfield  {author} {\bibinfo {author} {\bibfnamefont {A.}~\bibnamefont
  {Li}}\ and\ \bibinfo {author} {\bibfnamefont {D.~J.}\ \bibnamefont
  {Williamson}},\ }\href {https://arxiv.org/abs/2608.20981v1} {\bibinfo {title}
  {A classification of translation-invariant quantum codes in any dimension}}
  (\bibinfo {year} {2026}),\ \bibinfo {note} {version 1},\ \Eprint
  {https://arxiv.org/abs/2608.20981} {arXiv:2608.20981 [quant-ph]} \BibitemShut
  {NoStop}%
\bibitem [{\citenamefont {Bomb{\'i}n}\ and\ \citenamefont
  {Martin-Delgado}(2006)}]{BombinMartinDelgado2006}%
  \BibitemOpen
  \bibfield  {author} {\bibinfo {author} {\bibfnamefont {H.}~\bibnamefont
  {Bomb{\'i}n}}\ and\ \bibinfo {author} {\bibfnamefont {M.~A.}\ \bibnamefont
  {Martin-Delgado}},\ }\bibfield  {title} {\bibinfo {title} {Topological
  quantum distillation},\ }\href
  {https://doi.org/10.1103/PhysRevLett.97.180501} {\bibfield  {journal}
  {\bibinfo  {journal} {Phys. Rev. Lett.}\ }\textbf {\bibinfo {volume} {97}},\
  \bibinfo {pages} {180501} (\bibinfo {year} {2006})},\ \Eprint
  {https://arxiv.org/abs/quant-ph/0605138} {arXiv:quant-ph/0605138}
  \BibitemShut {NoStop}%
\bibitem [{\citenamefont {Bomb{\'i}n}\ and\ \citenamefont
  {Martin-Delgado}(2007{\natexlab{a}})}]{BombinMartinDelgado2007}%
  \BibitemOpen
  \bibfield  {author} {\bibinfo {author} {\bibfnamefont {H.}~\bibnamefont
  {Bomb{\'i}n}}\ and\ \bibinfo {author} {\bibfnamefont {M.~A.}\ \bibnamefont
  {Martin-Delgado}},\ }\bibfield  {title} {\bibinfo {title} {Exact topological
  quantum order in {$D=3$} and beyond: Branyons and brane-net condensates},\
  }\href {https://doi.org/10.1103/PhysRevB.75.075103} {\bibfield  {journal}
  {\bibinfo  {journal} {Phys. Rev. B}\ }\textbf {\bibinfo {volume} {75}},\
  \bibinfo {pages} {075103} (\bibinfo {year} {2007}{\natexlab{a}})},\ \Eprint
  {https://arxiv.org/abs/cond-mat/0607736} {arXiv:cond-mat/0607736}
  \BibitemShut {NoStop}%
\bibitem [{\citenamefont {Haah}(2013)}]{Haah2013}%
  \BibitemOpen
  \bibfield  {author} {\bibinfo {author} {\bibfnamefont {J.}~\bibnamefont
  {Haah}},\ }\bibfield  {title} {\bibinfo {title} {Commuting {Pauli}
  {Hamiltonians} as maps between free modules},\ }\href
  {https://doi.org/10.1007/s00220-013-1810-2} {\bibfield  {journal} {\bibinfo
  {journal} {Commun. Math. Phys.}\ }\textbf {\bibinfo {volume} {324}},\
  \bibinfo {pages} {351} (\bibinfo {year} {2013})},\ \Eprint
  {https://arxiv.org/abs/1204.1063} {arXiv:1204.1063} \BibitemShut {NoStop}%
\bibitem [{\citenamefont {Hostens}\ \emph {et~al.}(2005)\citenamefont
  {Hostens}, \citenamefont {Dehaene},\ and\ \citenamefont {{De
  Moor}}}]{Hostens2005}%
  \BibitemOpen
  \bibfield  {author} {\bibinfo {author} {\bibfnamefont {E.}~\bibnamefont
  {Hostens}}, \bibinfo {author} {\bibfnamefont {J.}~\bibnamefont {Dehaene}},\
  and\ \bibinfo {author} {\bibfnamefont {B.}~\bibnamefont {{De Moor}}},\
  }\bibfield  {title} {\bibinfo {title} {Stabilizer states and {Clifford}
  operations for systems of arbitrary dimensions and modular arithmetic},\
  }\href {https://doi.org/10.1103/PhysRevA.71.042315} {\bibfield  {journal}
  {\bibinfo  {journal} {Phys. Rev. A}\ }\textbf {\bibinfo {volume} {71}},\
  \bibinfo {pages} {042315} (\bibinfo {year} {2005})},\ \Eprint
  {https://arxiv.org/abs/quant-ph/0408190} {arXiv:quant-ph/0408190}
  \BibitemShut {NoStop}%
\bibitem [{\citenamefont {Bomb{\'i}n}\ and\ \citenamefont
  {Martin-Delgado}(2007{\natexlab{b}})}]{BombinMartinDelgado2007b}%
  \BibitemOpen
  \bibfield  {author} {\bibinfo {author} {\bibfnamefont {H.}~\bibnamefont
  {Bomb{\'i}n}}\ and\ \bibinfo {author} {\bibfnamefont {M.~A.}\ \bibnamefont
  {Martin-Delgado}},\ }\bibfield  {title} {\bibinfo {title} {Homological error
  correction: Classical and quantum codes},\ }\href
  {https://doi.org/10.1063/1.2731356} {\bibfield  {journal} {\bibinfo
  {journal} {J. Math. Phys.}\ }\textbf {\bibinfo {volume} {48}},\ \bibinfo
  {pages} {052105} (\bibinfo {year} {2007}{\natexlab{b}})},\ \Eprint
  {https://arxiv.org/abs/quant-ph/0605094} {arXiv:quant-ph/0605094}
  \BibitemShut {NoStop}%
\bibitem [{\citenamefont {Weibel}(1994)}]{Weibel1994}%
  \BibitemOpen
  \bibfield  {author} {\bibinfo {author} {\bibfnamefont {C.~A.}\ \bibnamefont
  {Weibel}},\ }\href {https://doi.org/10.1017/CBO9781139644136} {\emph
  {\bibinfo {title} {An Introduction to Homological Algebra}}},\ \bibinfo
  {series} {Cambridge Studies in Advanced Mathematics}, Vol.~\bibinfo {volume}
  {38}\ (\bibinfo  {publisher} {Cambridge University Press},\ \bibinfo {year}
  {1994})\BibitemShut {NoStop}%
\bibitem [{\citenamefont {Hamma}\ \emph {et~al.}(2005)\citenamefont {Hamma},
  \citenamefont {Zanardi},\ and\ \citenamefont {Wen}}]{Hamma2005}%
  \BibitemOpen
  \bibfield  {author} {\bibinfo {author} {\bibfnamefont {A.}~\bibnamefont
  {Hamma}}, \bibinfo {author} {\bibfnamefont {P.}~\bibnamefont {Zanardi}},\
  and\ \bibinfo {author} {\bibfnamefont {X.-G.}\ \bibnamefont {Wen}},\
  }\bibfield  {title} {\bibinfo {title} {String and membrane condensation on
  three-dimensional lattices},\ }\href
  {https://doi.org/10.1103/PhysRevB.72.035307} {\bibfield  {journal} {\bibinfo
  {journal} {Phys. Rev. B}\ }\textbf {\bibinfo {volume} {72}},\ \bibinfo
  {pages} {035307} (\bibinfo {year} {2005})},\ \Eprint
  {https://arxiv.org/abs/cond-mat/0411752} {arXiv:cond-mat/0411752}
  \BibitemShut {NoStop}%
\bibitem [{\citenamefont {Atiyah}\ and\ \citenamefont
  {Macdonald}(1969)}]{AtiyahMacdonald1969}%
  \BibitemOpen
  \bibfield  {author} {\bibinfo {author} {\bibfnamefont {M.~F.}\ \bibnamefont
  {Atiyah}}\ and\ \bibinfo {author} {\bibfnamefont {I.~G.}\ \bibnamefont
  {Macdonald}},\ }\href@noop {} {\emph {\bibinfo {title} {Introduction to
  Commutative Algebra}}}\ (\bibinfo  {publisher} {Addison-Wesley},\ \bibinfo
  {address} {Reading, MA},\ \bibinfo {year} {1969})\BibitemShut {NoStop}%
\bibitem [{\citenamefont {Aschenbrenner}(2004)}]{Aschenbrenner2004}%
  \BibitemOpen
  \bibfield  {author} {\bibinfo {author} {\bibfnamefont {M.}~\bibnamefont
  {Aschenbrenner}},\ }\bibfield  {title} {\bibinfo {title} {Ideal membership in
  polynomial rings over the integers},\ }\href
  {https://doi.org/10.1090/S0894-0347-04-00451-5} {\bibfield  {journal}
  {\bibinfo  {journal} {J. Amer. Math. Soc.}\ }\textbf {\bibinfo {volume}
  {17}},\ \bibinfo {pages} {407} (\bibinfo {year} {2004})}\BibitemShut
  {NoStop}%
\bibitem [{\citenamefont {He}(2026)}]{CSSDecoupling2026}%
  \BibitemOpen
  \bibfield  {author} {\bibinfo {author} {\bibfnamefont {M.}~\bibnamefont
  {He}},\ }\href {https://doi.org/10.5281/zenodo.23752878} {\bibinfo {title}
  {css-decoupling: decoupling translation-invariant topological {CSS} codes}}
  (\bibinfo {year} {2026}),\ \bibinfo {note} {version 0.1.0}\BibitemShut
  {NoStop}%
\bibitem [{\citenamefont {Chen}\ \emph {et~al.}(2025)\citenamefont {Chen},
  \citenamefont {Liu}, \citenamefont {Zhang}, \citenamefont {Liang},
  \citenamefont {Chen}, \citenamefont {Liu},\ and\ \citenamefont
  {Song}}]{Chen2025}%
  \BibitemOpen
  \bibfield  {author} {\bibinfo {author} {\bibfnamefont {K.}~\bibnamefont
  {Chen}}, \bibinfo {author} {\bibfnamefont {Y.}~\bibnamefont {Liu}}, \bibinfo
  {author} {\bibfnamefont {Y.}~\bibnamefont {Zhang}}, \bibinfo {author}
  {\bibfnamefont {Z.}~\bibnamefont {Liang}}, \bibinfo {author} {\bibfnamefont
  {Y.-A.}\ \bibnamefont {Chen}}, \bibinfo {author} {\bibfnamefont
  {K.}~\bibnamefont {Liu}},\ and\ \bibinfo {author} {\bibfnamefont
  {H.}~\bibnamefont {Song}},\ }\bibfield  {title} {\bibinfo {title} {Anyon
  theory and topological frustration of high-efficiency quantum low-density
  parity-check codes},\ }\href {https://doi.org/10.1103/86j7-cmsw} {\bibfield
  {journal} {\bibinfo  {journal} {Phys. Rev. Lett.}\ }\textbf {\bibinfo
  {volume} {135}},\ \bibinfo {pages} {076603} (\bibinfo {year} {2025})},\
  \Eprint {https://arxiv.org/abs/2503.04699} {arXiv:2503.04699} \BibitemShut
  {NoStop}%
\bibitem [{\citenamefont {Quintavalle}\ \emph {et~al.}(2021)\citenamefont
  {Quintavalle}, \citenamefont {Vasmer}, \citenamefont {Roffe},\ and\
  \citenamefont {Campbell}}]{Quintavalle2021}%
  \BibitemOpen
  \bibfield  {author} {\bibinfo {author} {\bibfnamefont {A.~O.}\ \bibnamefont
  {Quintavalle}}, \bibinfo {author} {\bibfnamefont {M.}~\bibnamefont {Vasmer}},
  \bibinfo {author} {\bibfnamefont {J.}~\bibnamefont {Roffe}},\ and\ \bibinfo
  {author} {\bibfnamefont {E.~T.}\ \bibnamefont {Campbell}},\ }\bibfield
  {title} {\bibinfo {title} {Single-shot error correction of three-dimensional
  homological product codes},\ }\href
  {https://doi.org/10.1103/PRXQuantum.2.020340} {\bibfield  {journal} {\bibinfo
   {journal} {PRX Quantum}\ }\textbf {\bibinfo {volume} {2}},\ \bibinfo {pages}
  {020340} (\bibinfo {year} {2021})},\ \Eprint
  {https://arxiv.org/abs/2009.11790} {arXiv:2009.11790} \BibitemShut {NoStop}%
\bibitem [{\citenamefont {Tan}\ \emph {et~al.}(2026)\citenamefont {Tan},
  \citenamefont {Gill}, \citenamefont {Huang}, \citenamefont {Liu},
  \citenamefont {Zhao}, \citenamefont {Dehghani}, \citenamefont {Kubica},
  \citenamefont {Zhou},\ and\ \citenamefont {Dua}}]{Tan2026}%
  \BibitemOpen
  \bibfield  {author} {\bibinfo {author} {\bibfnamefont {S.~J.~S.}\
  \bibnamefont {Tan}}, \bibinfo {author} {\bibfnamefont {I.}~\bibnamefont
  {Gill}}, \bibinfo {author} {\bibfnamefont {E.}~\bibnamefont {Huang}},
  \bibinfo {author} {\bibfnamefont {P.}~\bibnamefont {Liu}}, \bibinfo {author}
  {\bibfnamefont {C.}~\bibnamefont {Zhao}}, \bibinfo {author} {\bibfnamefont
  {H.}~\bibnamefont {Dehghani}}, \bibinfo {author} {\bibfnamefont
  {A.}~\bibnamefont {Kubica}}, \bibinfo {author} {\bibfnamefont
  {H.}~\bibnamefont {Zhou}},\ and\ \bibinfo {author} {\bibfnamefont
  {A.}~\bibnamefont {Dua}},\ }\href {https://arxiv.org/abs/2603.05402v2}
  {\bibinfo {title} {Generalized matching decoders for {2D} topological
  translationally-invariant codes}} (\bibinfo {year} {2026}),\ \bibinfo {note}
  {version 2},\ \Eprint {https://arxiv.org/abs/2603.05402} {arXiv:2603.05402
  [quant-ph]} \BibitemShut {NoStop}%
\bibitem [{\citenamefont {Barkeshli}\ \emph {et~al.}(2023)\citenamefont
  {Barkeshli}, \citenamefont {Chen}, \citenamefont {Huang}, \citenamefont
  {Kobayashi}, \citenamefont {Tantivasadakarn},\ and\ \citenamefont
  {Zhu}}]{Barkeshli2023}%
  \BibitemOpen
  \bibfield  {author} {\bibinfo {author} {\bibfnamefont {M.}~\bibnamefont
  {Barkeshli}}, \bibinfo {author} {\bibfnamefont {Y.-A.}\ \bibnamefont {Chen}},
  \bibinfo {author} {\bibfnamefont {S.-J.}\ \bibnamefont {Huang}}, \bibinfo
  {author} {\bibfnamefont {R.}~\bibnamefont {Kobayashi}}, \bibinfo {author}
  {\bibfnamefont {N.}~\bibnamefont {Tantivasadakarn}},\ and\ \bibinfo {author}
  {\bibfnamefont {G.}~\bibnamefont {Zhu}},\ }\bibfield  {title} {\bibinfo
  {title} {Codimension-2 defects and higher symmetries in (3+1){D} topological
  phases},\ }\href {https://doi.org/10.21468/SciPostPhys.14.4.065} {\bibfield
  {journal} {\bibinfo  {journal} {SciPost Phys.}\ }\textbf {\bibinfo {volume}
  {14}},\ \bibinfo {pages} {065} (\bibinfo {year} {2023})},\ \Eprint
  {https://arxiv.org/abs/2208.07367} {arXiv:2208.07367} \BibitemShut {NoStop}%
\bibitem [{\citenamefont {Barkeshli}\ \emph
  {et~al.}(2024{\natexlab{a}})\citenamefont {Barkeshli}, \citenamefont {Chen},
  \citenamefont {Hsin},\ and\ \citenamefont {Kobayashi}}]{Barkeshli2024a}%
  \BibitemOpen
  \bibfield  {author} {\bibinfo {author} {\bibfnamefont {M.}~\bibnamefont
  {Barkeshli}}, \bibinfo {author} {\bibfnamefont {Y.-A.}\ \bibnamefont {Chen}},
  \bibinfo {author} {\bibfnamefont {P.-S.}\ \bibnamefont {Hsin}},\ and\
  \bibinfo {author} {\bibfnamefont {R.}~\bibnamefont {Kobayashi}},\ }\bibfield
  {title} {\bibinfo {title} {Higher-group symmetry in finite gauge theory and
  stabilizer codes},\ }\href {https://doi.org/10.21468/SciPostPhys.16.4.089}
  {\bibfield  {journal} {\bibinfo  {journal} {SciPost Phys.}\ }\textbf
  {\bibinfo {volume} {16}},\ \bibinfo {pages} {089} (\bibinfo {year}
  {2024}{\natexlab{a}})},\ \Eprint {https://arxiv.org/abs/2211.11764}
  {arXiv:2211.11764} \BibitemShut {NoStop}%
\bibitem [{\citenamefont {Barkeshli}\ \emph
  {et~al.}(2024{\natexlab{b}})\citenamefont {Barkeshli}, \citenamefont {Hsin},\
  and\ \citenamefont {Kobayashi}}]{Barkeshli2024b}%
  \BibitemOpen
  \bibfield  {author} {\bibinfo {author} {\bibfnamefont {M.}~\bibnamefont
  {Barkeshli}}, \bibinfo {author} {\bibfnamefont {P.-S.}\ \bibnamefont
  {Hsin}},\ and\ \bibinfo {author} {\bibfnamefont {R.}~\bibnamefont
  {Kobayashi}},\ }\bibfield  {title} {\bibinfo {title} {Higher-group symmetry
  of (3+1){D} fermionic {$\mathbb{Z}_2$} gauge theory: Logical {CCZ}, {CS}, and
  {T} gates from higher symmetry},\ }\href
  {https://doi.org/10.21468/SciPostPhys.16.5.122} {\bibfield  {journal}
  {\bibinfo  {journal} {SciPost Phys.}\ }\textbf {\bibinfo {volume} {16}},\
  \bibinfo {pages} {122} (\bibinfo {year} {2024}{\natexlab{b}})},\ \Eprint
  {https://arxiv.org/abs/2311.05674} {arXiv:2311.05674} \BibitemShut {NoStop}%
\bibitem [{\citenamefont {Wang}\ \emph {et~al.}(2024)\citenamefont {Wang},
  \citenamefont {Wang}, \citenamefont {Chen}, \citenamefont {Zhang},
  \citenamefont {Zhang}, \citenamefont {Hu}, \citenamefont {Chen},
  \citenamefont {Gu},\ and\ \citenamefont {Liu}}]{WangYF2024}%
  \BibitemOpen
  \bibfield  {author} {\bibinfo {author} {\bibfnamefont {Y.-F.}\ \bibnamefont
  {Wang}}, \bibinfo {author} {\bibfnamefont {Y.}~\bibnamefont {Wang}}, \bibinfo
  {author} {\bibfnamefont {Y.-A.}\ \bibnamefont {Chen}}, \bibinfo {author}
  {\bibfnamefont {W.}~\bibnamefont {Zhang}}, \bibinfo {author} {\bibfnamefont
  {T.}~\bibnamefont {Zhang}}, \bibinfo {author} {\bibfnamefont
  {J.}~\bibnamefont {Hu}}, \bibinfo {author} {\bibfnamefont {W.}~\bibnamefont
  {Chen}}, \bibinfo {author} {\bibfnamefont {Y.}~\bibnamefont {Gu}},\ and\
  \bibinfo {author} {\bibfnamefont {Z.-W.}\ \bibnamefont {Liu}},\ }\bibfield
  {title} {\bibinfo {title} {Efficient fault-tolerant implementations of
  non-{{Clifford}} gates with reconfigurable atom arrays},\ }\href
  {https://doi.org/10.1038/s41534-024-00945-3} {\bibfield  {journal} {\bibinfo
  {journal} {npj Quantum Inf.}\ }\textbf {\bibinfo {volume} {10}},\ \bibinfo
  {pages} {136} (\bibinfo {year} {2024})},\ \Eprint
  {https://arxiv.org/abs/2312.09111} {arXiv:2312.09111} \BibitemShut {NoStop}%
\bibitem [{\citenamefont {Hsin}\ \emph {et~al.}(2024)\citenamefont {Hsin},
  \citenamefont {Kobayashi},\ and\ \citenamefont {Zhu}}]{Hsin2024}%
  \BibitemOpen
  \bibfield  {author} {\bibinfo {author} {\bibfnamefont {P.-S.}\ \bibnamefont
  {Hsin}}, \bibinfo {author} {\bibfnamefont {R.}~\bibnamefont {Kobayashi}},\
  and\ \bibinfo {author} {\bibfnamefont {G.}~\bibnamefont {Zhu}},\ }\href
  {https://arxiv.org/abs/2411.15848v3} {\bibinfo {title} {Classifying logical
  gates in quantum codes via cohomology operations and symmetry}} (\bibinfo
  {year} {2024}),\ \bibinfo {note} {version 3},\ \Eprint
  {https://arxiv.org/abs/2411.15848} {arXiv:2411.15848 [quant-ph]} \BibitemShut
  {NoStop}%
\bibitem [{\citenamefont {Breuckmann}\ \emph {et~al.}(2026)\citenamefont
  {Breuckmann}, \citenamefont {Davydova}, \citenamefont {Eberhardt},\ and\
  \citenamefont {Tantivasadakarn}}]{Breuckmann2026}%
  \BibitemOpen
  \bibfield  {author} {\bibinfo {author} {\bibfnamefont {N.~P.}\ \bibnamefont
  {Breuckmann}}, \bibinfo {author} {\bibfnamefont {M.}~\bibnamefont
  {Davydova}}, \bibinfo {author} {\bibfnamefont {J.~N.}\ \bibnamefont
  {Eberhardt}},\ and\ \bibinfo {author} {\bibfnamefont {N.}~\bibnamefont
  {Tantivasadakarn}},\ }\bibfield  {title} {\bibinfo {title} {Cups and gates
  {I}: Cohomology invariants and logical quantum operations},\ }\href
  {https://doi.org/10.1007/s00220-026-05570-z} {\bibfield  {journal} {\bibinfo
  {journal} {Commun. Math. Phys.}\ }\textbf {\bibinfo {volume} {407}},\
  \bibinfo {pages} {86} (\bibinfo {year} {2026})},\ \Eprint
  {https://arxiv.org/abs/2410.16250} {arXiv:2410.16250} \BibitemShut {NoStop}%
\bibitem [{\citenamefont {Vasmer}\ and\ \citenamefont
  {Kubica}(2022)}]{VasmerKubica2022}%
  \BibitemOpen
  \bibfield  {author} {\bibinfo {author} {\bibfnamefont {M.}~\bibnamefont
  {Vasmer}}\ and\ \bibinfo {author} {\bibfnamefont {A.}~\bibnamefont
  {Kubica}},\ }\bibfield  {title} {\bibinfo {title} {Morphing quantum codes},\
  }\href {https://doi.org/10.1103/PRXQuantum.3.030319} {\bibfield  {journal}
  {\bibinfo  {journal} {PRX Quantum}\ }\textbf {\bibinfo {volume} {3}},\
  \bibinfo {pages} {030319} (\bibinfo {year} {2022})},\ \Eprint
  {https://arxiv.org/abs/2112.01446} {arXiv:2112.01446} \BibitemShut {NoStop}%
\bibitem [{\citenamefont {Liang}\ \emph {et~al.}(2025)\citenamefont {Liang},
  \citenamefont {Liu}, \citenamefont {Song},\ and\ \citenamefont
  {Chen}}]{Liang2025}%
  \BibitemOpen
  \bibfield  {author} {\bibinfo {author} {\bibfnamefont {Z.}~\bibnamefont
  {Liang}}, \bibinfo {author} {\bibfnamefont {K.}~\bibnamefont {Liu}}, \bibinfo
  {author} {\bibfnamefont {H.}~\bibnamefont {Song}},\ and\ \bibinfo {author}
  {\bibfnamefont {Y.-A.}\ \bibnamefont {Chen}},\ }\bibfield  {title} {\bibinfo
  {title} {Generalized toric codes on twisted tori for quantum error
  correction},\ }\href {https://doi.org/10.1103/rmy6-9n89} {\bibfield
  {journal} {\bibinfo  {journal} {PRX Quantum}\ }\textbf {\bibinfo {volume}
  {6}},\ \bibinfo {pages} {020357} (\bibinfo {year} {2025})},\ \Eprint
  {https://arxiv.org/abs/2503.03827} {arXiv:2503.03827} \BibitemShut {NoStop}%
\bibitem [{\citenamefont {Yang}\ and\ \citenamefont {Yu}(2026)}]{YangYu2026}%
  \BibitemOpen
  \bibfield  {author} {\bibinfo {author} {\bibfnamefont {B.}~\bibnamefont
  {Yang}}\ and\ \bibinfo {author} {\bibfnamefont {M.}~\bibnamefont {Yu}},\
  }\href {https://arxiv.org/abs/2604.24847v1} {\bibinfo {title} {The
  classification of {Pauli} stabilizer codes: A lattice and continuum
  treatise}} (\bibinfo {year} {2026}),\ \bibinfo {note} {version 1},\ \Eprint
  {https://arxiv.org/abs/2604.24847} {arXiv:2604.24847 [math-ph]} \BibitemShut
  {NoStop}%
\bibitem [{\citenamefont {Ruba}\ and\ \citenamefont
  {Yang}(2025)}]{RubaYang2025}%
  \BibitemOpen
  \bibfield  {author} {\bibinfo {author} {\bibfnamefont {B.}~\bibnamefont
  {Ruba}}\ and\ \bibinfo {author} {\bibfnamefont {B.}~\bibnamefont {Yang}},\
  }\href {https://arxiv.org/abs/2509.10418v2} {\bibinfo {title} {Witt groups
  and bulk-boundary correspondence for stabilizer states}} (\bibinfo {year}
  {2025}),\ \bibinfo {note} {version 2},\ \Eprint
  {https://arxiv.org/abs/2509.10418} {arXiv:2509.10418 [math-ph]} \BibitemShut
  {NoStop}%
\bibitem [{\citenamefont {Barkeshli}\ \emph {et~al.}(2019)\citenamefont
  {Barkeshli}, \citenamefont {Bonderson}, \citenamefont {Cheng},\ and\
  \citenamefont {Wang}}]{Barkeshli2019}%
  \BibitemOpen
  \bibfield  {author} {\bibinfo {author} {\bibfnamefont {M.}~\bibnamefont
  {Barkeshli}}, \bibinfo {author} {\bibfnamefont {P.}~\bibnamefont
  {Bonderson}}, \bibinfo {author} {\bibfnamefont {M.}~\bibnamefont {Cheng}},\
  and\ \bibinfo {author} {\bibfnamefont {Z.}~\bibnamefont {Wang}},\ }\bibfield
  {title} {\bibinfo {title} {Symmetry fractionalization, defects, and gauging
  of topological phases},\ }\href {https://doi.org/10.1103/PhysRevB.100.115147}
  {\bibfield  {journal} {\bibinfo  {journal} {Phys. Rev. B}\ }\textbf {\bibinfo
  {volume} {100}},\ \bibinfo {pages} {115147} (\bibinfo {year} {2019})},\
  \Eprint {https://arxiv.org/abs/1410.4540} {arXiv:1410.4540} \BibitemShut
  {NoStop}%
\bibitem [{\citenamefont {Ellison}\ \emph {et~al.}(2022)\citenamefont
  {Ellison}, \citenamefont {Chen}, \citenamefont {Dua}, \citenamefont
  {Shirley}, \citenamefont {Tantivasadakarn},\ and\ \citenamefont
  {Williamson}}]{Ellison2022}%
  \BibitemOpen
  \bibfield  {author} {\bibinfo {author} {\bibfnamefont {T.~D.}\ \bibnamefont
  {Ellison}}, \bibinfo {author} {\bibfnamefont {Y.-A.}\ \bibnamefont {Chen}},
  \bibinfo {author} {\bibfnamefont {A.}~\bibnamefont {Dua}}, \bibinfo {author}
  {\bibfnamefont {W.}~\bibnamefont {Shirley}}, \bibinfo {author} {\bibfnamefont
  {N.}~\bibnamefont {Tantivasadakarn}},\ and\ \bibinfo {author} {\bibfnamefont
  {D.~J.}\ \bibnamefont {Williamson}},\ }\bibfield  {title} {\bibinfo {title}
  {{Pauli} stabilizer models of twisted quantum doubles},\ }\href
  {https://doi.org/10.1103/PRXQuantum.3.010353} {\bibfield  {journal} {\bibinfo
   {journal} {PRX Quantum}\ }\textbf {\bibinfo {volume} {3}},\ \bibinfo {pages}
  {010353} (\bibinfo {year} {2022})},\ \Eprint
  {https://arxiv.org/abs/2112.11394} {arXiv:2112.11394} \BibitemShut {NoStop}%
\bibitem [{\citenamefont {Haah}\ \emph {et~al.}(2023)\citenamefont {Haah},
  \citenamefont {Fidkowski},\ and\ \citenamefont
  {Hastings}}]{HaahFidkowskiHastings2023}%
  \BibitemOpen
  \bibfield  {author} {\bibinfo {author} {\bibfnamefont {J.}~\bibnamefont
  {Haah}}, \bibinfo {author} {\bibfnamefont {L.}~\bibnamefont {Fidkowski}},\
  and\ \bibinfo {author} {\bibfnamefont {M.~B.}\ \bibnamefont {Hastings}},\
  }\bibfield  {title} {\bibinfo {title} {Nontrivial quantum cellular automata
  in higher dimensions},\ }\href {https://doi.org/10.1007/s00220-022-04528-1}
  {\bibfield  {journal} {\bibinfo  {journal} {Commun. Math. Phys.}\ }\textbf
  {\bibinfo {volume} {398}},\ \bibinfo {pages} {469} (\bibinfo {year}
  {2023})},\ \Eprint {https://arxiv.org/abs/1812.01625} {arXiv:1812.01625}
  \BibitemShut {NoStop}%
\bibitem [{\citenamefont {Shirley}\ \emph {et~al.}(2018)\citenamefont
  {Shirley}, \citenamefont {Slagle}, \citenamefont {Wang},\ and\ \citenamefont
  {Chen}}]{Shirley2018}%
  \BibitemOpen
  \bibfield  {author} {\bibinfo {author} {\bibfnamefont {W.}~\bibnamefont
  {Shirley}}, \bibinfo {author} {\bibfnamefont {K.}~\bibnamefont {Slagle}},
  \bibinfo {author} {\bibfnamefont {Z.}~\bibnamefont {Wang}},\ and\ \bibinfo
  {author} {\bibfnamefont {X.}~\bibnamefont {Chen}},\ }\bibfield  {title}
  {\bibinfo {title} {Fracton models on general three-dimensional manifolds},\
  }\href {https://doi.org/10.1103/PhysRevX.8.031051} {\bibfield  {journal}
  {\bibinfo  {journal} {Phys. Rev. X}\ }\textbf {\bibinfo {volume} {8}},\
  \bibinfo {pages} {031051} (\bibinfo {year} {2018})},\ \Eprint
  {https://arxiv.org/abs/1712.05892} {arXiv:1712.05892} \BibitemShut {NoStop}%
\bibitem [{\citenamefont {Shirley}\ \emph {et~al.}(2019)\citenamefont
  {Shirley}, \citenamefont {Slagle},\ and\ \citenamefont {Chen}}]{Shirley2019}%
  \BibitemOpen
  \bibfield  {author} {\bibinfo {author} {\bibfnamefont {W.}~\bibnamefont
  {Shirley}}, \bibinfo {author} {\bibfnamefont {K.}~\bibnamefont {Slagle}},\
  and\ \bibinfo {author} {\bibfnamefont {X.}~\bibnamefont {Chen}},\ }\bibfield
  {title} {\bibinfo {title} {Fractional excitations in foliated fracton
  phases},\ }\href {https://doi.org/10.1016/j.aop.2019.167922} {\bibfield
  {journal} {\bibinfo  {journal} {Ann. Phys.}\ }\textbf {\bibinfo {volume}
  {410}},\ \bibinfo {pages} {167922} (\bibinfo {year} {2019})},\ \Eprint
  {https://arxiv.org/abs/1806.08625} {arXiv:1806.08625} \BibitemShut {NoStop}%
\bibitem [{\citenamefont {Eisenbud}(1995)}]{Eisenbud1995}%
  \BibitemOpen
  \bibfield  {author} {\bibinfo {author} {\bibfnamefont {D.}~\bibnamefont
  {Eisenbud}},\ }\href {https://doi.org/10.1007/978-1-4612-5350-1} {\emph
  {\bibinfo {title} {Commutative Algebra with a View Toward Algebraic
  Geometry}}},\ \bibinfo {series} {Graduate Texts in Mathematics}, Vol.\
  \bibinfo {volume} {150}\ (\bibinfo  {publisher} {Springer},\ \bibinfo
  {address} {New York},\ \bibinfo {year} {1995})\BibitemShut {NoStop}%
\bibitem [{\citenamefont {Swan}(1978)}]{Swan1978}%
  \BibitemOpen
  \bibfield  {author} {\bibinfo {author} {\bibfnamefont {R.~G.}\ \bibnamefont
  {Swan}},\ }\bibfield  {title} {\bibinfo {title} {Projective modules over
  {Laurent} polynomial rings},\ }\href
  {https://doi.org/10.1090/S0002-9947-1978-0469906-4} {\bibfield  {journal}
  {\bibinfo  {journal} {Trans. Amer. Math. Soc.}\ }\textbf {\bibinfo {volume}
  {237}},\ \bibinfo {pages} {111} (\bibinfo {year} {1978})}\BibitemShut
  {NoStop}%
\bibitem [{\citenamefont {Lam}(1999)}]{Lam1999}%
  \BibitemOpen
  \bibfield  {author} {\bibinfo {author} {\bibfnamefont {T.-Y.}\ \bibnamefont
  {Lam}},\ }\href {https://doi.org/10.1007/978-1-4612-0525-8} {\emph {\bibinfo
  {title} {Lectures on Modules and Rings}}},\ \bibinfo {series} {Graduate Texts
  in Mathematics}, Vol.\ \bibinfo {volume} {189}\ (\bibinfo  {publisher}
  {Springer},\ \bibinfo {address} {New York},\ \bibinfo {year}
  {1999})\BibitemShut {NoStop}%
\end{thebibliography}%
\end{document}